\documentclass[9pt,preprint,nocopyrightspace]{sigplanconf}

\usepackage{pifont}
\usepackage{amssymb}

\usepackage{multirow}
\usepackage{enumerate}
\usepackage{xspace}
\usepackage{afterpage}
\usepackage{float}
\usepackage{txfonts} 
\usepackage{latexsym} 
\usepackage{mdwlist} 
\usepackage{stmaryrd} 
\usepackage{mathrsfs}
\usepackage{amsfonts} 
\usepackage{amssymb} 
\usepackage{amstext}
\usepackage{enumitem}
\usepackage{color} 
\usepackage{xspace} 

\usepackage{bbm}
\usepackage{alltt}
\usepackage{verbdef}
\usepackage{titlesec}
\usepackage{parskip}
\usepackage{wrapfig}
\usepackage{code}
\usepackage{url} 
\usepackage[colorlinks=true,citecolor=blue,linkcolor=blue,urlcolor=black]{hyperref} 
\usepackage{balance}
\usepackage[pdftex]{graphicx}
\usepackage{listings}
\usepackage{filecontents}
\usepackage{supertabular}
\usepackage{cite}
\usepackage{comment}
\usepackage{flushend}

\usepackage{xspace}
\usepackage{xcolor}

\usepackage{denot}
\usepackage{prooftree}
\usepackage{abbrev}
\usepackage{graphviz}

\usepackage{etoolbox}
\makeatletter
\patchcmd{\lsthk@TextStyle}{\let\lst@DefEsc\@empty}{}{}{\errmessage{failed to patch}}
\makeatother

\setlist[itemize]{leftmargin=*}
\titlespacing*{\section}{0pt}{*1.5}{*1.5} 
\titlespacing*{\subsection}{0pt}{*0.6}{*0.5}
\titlespacing*{\subsubsection}{0pt}{*0.8}{*0.5}
\titlespacing*{\paragraph}{0pt}{*0.5}{*1.2}

\definecolor{shadecolor}{gray}{1.00}
\definecolor{ddarkgray}{gray}{0.75}
\definecolor{darkgray}{gray}{0.30}
\definecolor{light-gray}{gray}{0.85}

\usepackage[T1]{fontenc}

\DeclareFontFamily{OML}{zavm}{\skewchar\font=127 }
\DeclareFontShape{OML}{zavm}{m}{n}{<-> s*[.80] zavmr7t}{}
\DeclareFontShape{OML}{zavm}{b}{n}{<-> s*[.80] zavmb7t}{}
\DeclareFontShape{OML}{zavm}{m}{it}{<-> s*[.80] zavmri7m}{}
\DeclareFontShape{OML}{zavm}{b}{it}{<-> s*[.80] zavmbi7m}{}
\DeclareFontShape{OML}{zavm}{m}{sl}{<->ssub * zavm/m/it}{}
\DeclareFontShape{OML}{zavm}{bx}{it}{<->ssub * zavm/b/it}{}
\DeclareFontShape{OML}{zavm}{b}{sl}{<->ssub * zavm/b/it}{}
\DeclareFontShape{OML}{zavm}{bx}{sl}{<->ssub * zavm/b/sl}{}
\DeclareMathAlphabet{\mathsf}{OML}{zavm}{m}{n} 

\newcommand{\ifext}[2]{\ifdefined\extflag{#1}\else{#2}\fi}

\newcommand{\xmark}{\text{\ding{55}}}
\makeatletter
\newcommand{\oset}[3][0ex]{%
  \mathrel{\mathop{#3}\limits^{
    \vbox to#1{\kern-1\ex@
    \hbox{$\scriptstyle#2$}\vss}}}}
\makeatother

\makeatletter
\newcommand{\ojset}[3][0ex]{%
  \mathrel{\mathop{#3}\limits^{
    \vbox to#1{\kern-3\ex@
    \hbox{$\scriptstyle#2$}\vss}}}}
\makeatother

\newcommand{\angled}[1]{\langle {#1} \rangle}
\newcommand{\langled}[1]{\langle {#1}}
\newcommand{\rangled}[1]{{#1} \rangle}

\renewcommand{\phi}{\varphi} 

\DeclareFontFamily{U}{mathb}{\hyphenchar\font45}
\DeclareFontShape{U}{mathb}{m}{n}{
      <5> <6> <7> <8> <9> <10> gen * mathb
      <10.95> mathb10 <12> <14.4> <17.28> <20.74> <24.88> mathb12
}{}
\DeclareSymbolFont{mathb}{U}{mathb}{m}{n}
\DeclareMathSymbol{\blacktriangleleft} {2}{mathb}{"9E}
\DeclareMathSymbol{\blacktriangleright}{2}{mathb}{"9F}

\newcommand{\subst}[2]{[#1/#2]}

\newcommand{\etc}{\emph{etc}}
\newcommand{\ie}{\emph{i.e.}\xspace}

\newcommand{\eg}{\emph{e.g.}\xspace}

\newcommand{\etal}{\emph{et~al.}\xspace}

\definecolor{shadecolor}{gray}{1.00}
\definecolor{ddarkgray}{gray}{0.75}
\definecolor{darkgray}{gray}{0.30}
\definecolor{light-gray}{gray}{0.85}

\newcommand{\graybox}[1]{\colorbox{light-gray}{#1}}

\makeatletter
\newcommand*{\textalltt}{}
\DeclareRobustCommand*{\textalltt}{%
  \begingroup
    \let\do\@makeother
    \dospecials
    \catcode`\\=\z@
    \catcode`\{=\@ne
    \catcode`\}=\tw@
    \verbatim@font\@noligs
    \@vobeyspaces
    \frenchspacing
    \@textalltt
}
\newcommand*{\@textalltt}[1]{%
    #1%
  \endgroup
}
\makeatother

\newcommand{\seqCstM}{\lstmath{sc}}
\newcommand{\relAcqM}{\lstmath{relAcq}}
\newcommand{\relM}{\lstmath{rel}}
\newcommand{\acqM}{\lstmath{acq}}
\newcommand{\rlxM}{\lstmath{rlx}}
\newcommand{\naM}{\lstmath{na}}
\newcommand{\conM}{\lstmath{con}}

\newcommand{\StatePair}[2]{\angled{#1,~#2}}

\newcommand{\AST}{\mathsf{s}}
\newcommand{\RT}{\AST_{\text{\textsf{RT}}}}
\newcommand{\Expr}{\mathsf{e}}

\newcommand{\RM}{\mathsf{RM}}

\newcommand{\WM}{\mathsf{WM}}
\newcommand{\SM}{\mathsf{SM}}
\newcommand{\FM}{\mathsf{FM}}

\newcommand{\Write}[3]{[#2]_{\mathsf{#1}} := #3}
\newcommand{\Read}[2]{[#2]_{\mathsf{#1}}}

\newcommand{\Cas}[5]{\lstmath{cas}_{\mathsf{#1}, \mathsf{#2}}(#3, #4, #5)}

\newcommand{\lstmath}[1]{\text{\lstinline{#1}}}

\newcommand{\Ret}[1]{#1}

\newcommand{\hole}{[~]}
\newcommand{\IfThenElse}[3]{\lstmath{if}~#1~\lstmath{then}~#2~\lstmath{else}~#3~\lstmath{fi}}
\newcommand{\Repeat}[1]{\lstmath{repeat}~#1~\lstmath{end}}
\newcommand{\Par}[2]{\lstmath{par}~#1~#2}
\newcommand{\Spw}[2]{\lstmath{spw}~#1~#2}
\newcommand{\Bind}[3]{{#1} = {#2}\!;{#3}}

\newcommand{\Stuck}{\lstmath{stuck}}
\newcommand{\vName}{\mathsf{x}}
\newcommand{\op}{\lstmath{op}}
\newcommand{\Number}{\mathbb{Z}}
\newcommand{\Choice}[2]{\lstmath{choice}#1~#2}
\newcommand{\First}[1]{\lstmath{fst}~#1}
\newcommand{\Second}[1]{\lstmath{snd}~#1}
\newcommand{\EvalContext}{\mathsf{E}}
\newcommand{\EvalSpecContext}{\mathsf{E}\alpha}
\newcommand{\EvalEUContext}{\mathsf{EU}}
\newcommand{\Pair}[2]{(#1,~#2)}

\newcommand{\loc}{\ell}
\newcommand{\locVar}{\iota}
\newcommand{\auxX}{\xi}

\newcommand{\mval}{\mu}
\newcommand{\mvalSubst}{\text{\emph{v}}}

\newcommand{\stEta}{H}

\newcommand{\stPsiRead}{\psi^{\text{\scriptsize{\texttt{rd}}}}}
\newcommand{\stPsiWrite}{\psi^{\text{\scriptsize{\texttt{wr}}}}}
\newcommand{\Path}{\text{\textsf{path}}}
\newcommand{\stpath}{\pi}

\newcommand{\AppendAlpha}{\text{\textsf{append}}}

\newcommand{\updateDep}{\text{\textsf{updateDep}}}
\newcommand{\updateSync}{\text{\textsf{updateSync}}}
\newcommand{\promote}{\text{\textsf{promote}}}
\newcommand{\remove}{\text{\textsf{remove}}}

\newcommand{\stSigma}{\sigma}

\newcommand{\stSC}{\sigma^{\text{\scriptsize{\texttt{sc}}}}}
\newcommand{\stNA}{\sigma^{\text{\scriptsize{\texttt{na}}}}}
\newcommand{\stSigmaRead}{\sigma_{\text{\scriptsize{\texttt{rd}}}}}
\newcommand{\stSigmaWrite}{\sigma_{\text{\scriptsize{\texttt{wr}}}}}
\newcommand{\stSigmaWriteNew}{\sigma'_{\text{\scriptsize{\texttt{wr}}}}}
\newcommand{\stSigmaSync}{\sigma_{\text{\scriptsize{\texttt{sync}}}}}
\newcommand{\stSigmaEmpty}{()}

\newcommand{\stTau}{\tau}

\newcommand{\NextTau}[2]{Next\stTau(#1, #2)}
\newcommand{\lastt}{\text{\textsf{LastTS}}}
\newcommand{\LastTau}[2]{\lastt(#1, #2)}

\newcommand{\stPhi}{\phi} 
\newcommand{\stAlpha}{\alpha}
\newcommand{\stGamma}{\gamma} 

\newcommand{\stPostOp}{\beta}
\newcommand{\stObservedWrites}{\omega}
\newcommand{\arrayBlock}[1]{\begin{array}{c}#1\end{array}}

\newcommand{\tauFst}{$\tauFstM$}
\newcommand{\tauSnd}{$\tauSndM$}

\newcommand{\tauFstM}{\textbf{0}}
\newcommand{\tauSndM}{\textbf{1}}

\newcommand{\cntrd}[1]{\begin{center} #1 \end{center}}

\newcommand{\stateTblThree}[2][5pt]{%
\cntrd{
\begin{supertabular}{p{.03\linewidth} |@{\hskip #1}| p{.15\linewidth} | p{.15\linewidth} | p{.15\linewidth}}
$\stTau$ & \xloc & \yloc & \zloc \\ 
\hline
#2
\end{supertabular}%
}
}

\newcommand{\stateTbl}[4][5pt]{%
\begin{supertabular}{p{.03\linewidth} |@{\hskip #1}| p{.45\linewidth} | p{.25\linewidth}}

$\stTau$ & #2 & #3 \\ 
\hline

#4
\end{supertabular}%
}

\newcommand{\stateTblFrame}[1]
{{\small
\vspace{3pt}
#1
\vspace{3pt}
}}

\newcommand{\emptyCell}[1]{%
  \multicolumn{1}{c#1}{-}
}

\newcommand{\emptyFront}{$\bot$}

\newcommand{\smapsto}{{\mapsto}}

\newcommand\redcolor{red!30!white!100}
\newcommand\bluecolor{blue!30!white!100}

\newcommand\greencolor{green!30!white!100}

\newcommand\redsq[1][L]{\colorbox{\redcolor}{\ensuremath{#1}}}
\newcommand\bluesq[1][R]{\colorbox{\bluecolor}{\ensuremath{#1}}}
\newcommand\greensq[1][P]{\colorbox{\greencolor}{\ensuremath{#1}}}

\newcommand{\parentFrontMark}{\greensq\xspace}
\newcommand{\leftFrontMark}{\redsq\xspace}
\newcommand{\rightFrontMark}{\bluesq\xspace}

\newcommand{\floc}{\lstinline{f}\xspace}
\newcommand{\dloc}{\lstinline{d}\xspace}

\newcommand{\xloc}{\lstinline{x}}
\newcommand{\yloc}{\lstinline{y}}
\newcommand{\zloc}{\lstinline{z}}

\newcommand{\prarrow}{\rightharpoonup}

\newcommand{\tick}{\checkmark}%
\newcommand{\tickP}{\checkmark}%
\newcommand{\tickPP}{\checkmark}%
\newcommand{\fail}{\xmark}%

\newcommand{\spawn}[2]{\text{\textsf{spawn}}(#1,~#2)}
\newcommand{\joinP}[2]{\text{\textsf{join}}(#1,~#2)}

\newcommand{\ruleF}[1]{\textsc{#1}}

\newcommand{\litmusTestStart}[3]{
\begin{minipage}[t]{0.2\linewidth}
\textbf{#1} \\
Fully Supported: $#2$ \\
Requires: #3\\
\end{minipage}
}
\newcommand{\litmusTestEnd}{
\vspace{.2cm}
\hrule
\vspace{.2cm}
}

\newcommand{\MpName}{\textsf{MP\_rel+acq+na}}

\newcommand{\spProg}{\textsf{SE\_simple}\xspace}

\begin{document}

\def\denot#1{[\![ #1 ]\!]}

\clubpenalty=10000 
\widowpenalty = 10000 

\lstset{escapeinside=||}

\setlength{\pdfpageheight}{\paperheight}
\setlength{\pdfpagewidth}{\paperwidth}


\publicationrights{licensed}     


\setlength{\pdfpageheight}{\paperheight}
\setlength{\pdfpagewidth}{\paperwidth}
\newcounter{tags}
\def\extflag{}
\lstdefinelanguage{while}{
keywords={choice, end, repeat, read, write, if,
          then, else, fi, od, ret, stuck, par, spw,
          fst, snd, null, delete},
sensitive=true,
basicstyle=\small, 
commentstyle=\scriptsize\rmfamily,
keywordstyle=\ttfamily\bfseries,
identifierstyle=\ttfamily,
basewidth={0.5em,0.5em},
columns=fixed,
fontadjust=true
}
\lstset{language=while}
\title{Operational Aspects of C/C++ Concurrency\ifext{}{\vspace{0pt}}}
\ifext{\subtitle{Extended Version\vspace{-30pt}}}{}

\subtitle{Extended version}
\subtitle{\large{\color{red}\today}}

\authorinfo{Anton Podkopaev}
           {Saint Petersburg State University and JetBrains Inc., Russia}
           {a.podkopaev@2009.spbu.ru}

\authorinfo{Ilya Sergey}
           {University College London, UK}
           {i.sergey@ucl.ac.uk}

\authorinfo{Aleksandar Nanevski}
           {IMDEA Software Institute, Spain}
           {aleks.nanevski@imdea.org}

\authorinfo{}{}{}

\maketitle 

\begin{abstract}


%
%
%
  
%

  In this work, we present a family of operational semantics that
  gradually approximates the realistic program behaviors in the
  C/C++11 memory model.
  Each semantics in our framework is built by elaborating and
  combining two simple ingredients: \emph{viewfronts} and
  \emph{operation buffers}. Viewfronts allow us to express the
  \emph{spatial} aspect of thread interaction, \ie, which values a
  thread can read, while operation buffers enable manipulation with
  the \emph{temporal} execution aspect, \ie, determining the order in
  which the results of certain operations can be observed by
  concurrently running threads.
  
  Starting from a simple abstract state machine, through a series of
  gradual refinements of the abstract state, we capture such language
  aspects and synchronization primitives as \emph{release/acquire}
  atomics, \emph{sequentially-consistent} and \emph{non-atomic} memory
  accesses, also providing a semantics for \emph{relaxed} atomics,
  while avoiding the Out-of-Thin-Air problem.
  To the best of our knowledge, this is the first formal and
  \emph{executable} operational semantics of C11 capable of expressing
  all essential concurrent aspects of the standard.

  %
  %
  %

  We illustrate our approach via a number of characteristic examples,
  relating the observed behaviors to those of standard litmus test
  programs from the literature.
  We provide an executable implementation of the semantics in PLT
  Redex, along with a number of implemented litmus tests and examples,
  and showcase our prototype on a large case study: randomized testing
  and debugging of a realistic Read-Copy-Update data structure.
 \end{abstract}




\section{Introduction}
\label{sec:intro}

Memory models describe the behavior of multithreaded programs, which
might concurrently access shared memory locations. The best studied
memory model is \emph{sequential consistency}
(SC)~\cite{Lamport:TC79}, which assumes a total order on all memory
accesses (\ie, read and write operations) in a single run of a
concurrent program, therefore, ensuring that the result of each read
from a location is a value that was stored by the last preceding write
to the very same location.

However, sequential consistency falls short when describing the
phenomena, observed in concurrent programs running on modern processor
architectures, such x86, ARM, and PowerPC, and resulting from
{store buffering}~\cite{SPARC-manual} and CPU- and compiler-level
optimizations, \eg, rearranging independent reads and
writes~\cite{Hennessy-Patterson:BOOK}. 
\emph{Relaxed memory models} aim to capture the semantics of such
programs and provide suitable abstractions for the developers to
reason about their code, written in a higher-level language,
independently from the hardware architecture it is going to be
executed on.

The most prominent example of a relaxed memory model is the C11 model,
introduced by the C/C++ 2011 standards~\cite{C:11,CPP:11} and
describing the behavior of concurrent C/C++ programs. It defines a
number of memory accesses, implementing different synchronization
policies and having corresponding performance costs. For instance,
\emph{SC-atomics} provide the SC-style total ordering between reads
and writes to the corresponding memory locations, while
\emph{release/acquire} (RA) accesses implement only partial one-way
synchronization, but are cheaper to implement. Finally, \emph{relaxed}
accesses are the cheapest in terms of performance, but provide the
weakest synchronization guarantees.


Existing formalizations of the \emph{full} C11 memory model adopt an
axiomatic style, representing programs by sets of \emph{consistent
  executions}~\cite{Batty-al:POPL11,Batty-al:POPL12,Batty-al:ESOP15}. Each
execution can be thought of as a graph, whose nodes are
read/write-accesses to memory locations. The edges of the graph
represent various orders between operations (\eg, total orders between
SC-atomics and operations in a single thread), some of which might be
partial. Defined this way, the executions help one to answer questions
of the following kind: ``\emph{Can the value X be read from the memory
  location L at the point R of the program P?}''

This axiomatic whole-program representation makes it difficult to
think of C11 programs in terms of step-by-step executions of a program
on some abstract machine, making it non-trivial to employ these
semantic approaches for the purposes of testing, debugging and
compositional symbolic reasoning about programs, \eg, by means of type
systems and program logics.
Recently, several attempts have been made to provide a more
operational semantics for C/C++ concurrency, however, all the
approaches existing to date focus on a specific subset of C11, \eg,
release/acquire/SC
synchronization~\cite{Lahav-al:POPL16,Turon-al:OOPSLA14} or relaxed
atomics~\cite{PichonPharabod-Sewell:POPL16}, without providing a
uniform framework accommodating all features of the standard.





In this work, we make a step towards providing a simple, yet uniform
foundations for accommodating all of the essential aspects of the C11
concurrency, and describe a framework for defining operational
semantics capturing the expected behaviors of concurrent executions
observed in realistic C/C++ programs, while prohibiting unwelcome
outcomes, such as Thin-Air executions.
The paramount idea of our constructions is maintaining a \emph{rich
  program state}, which is a subject of manipulation by concurrent
threads, and is represented by a combination of the following two
ingredients.

\paragraph{Ingredient 1: Viewfronts for threads synchronization}

We observe that, assuming a total ordering of writes to each
particular shared memory location, we can consider a state to be a
collection of \emph{per-location histories}, representing
totally-ordered updates---an idea adopted from the recent works on
logics for SC concurrency~\cite{Sergey-al:ESOP15}. We introduce the
notion of \emph{viewfronts} as a way to account for the phenomenon of
particular threads having specific, yet consistent, views to the
global history of each shared location, similarly to the way vector
clocks are used for synchronization in distributed
systems~\cite{Mattern88virtualtime}.
We then consider various flavors of C11 atomicity as ways to
``partially align'' viewfronts of several threads.

\paragraph{Ingredient 2: Operation buffers for speculative executions}

The mechanism of relaxed atomic accesses in C11 allows for speculative
reordering or removing of operations, involving them, in particular
threads. In order to formally define the resulting temporal phenomena,
observed by concurrently running threads (which can see some values
appearing ``out-of-order''), we need to capture a speculative nature
of such computations. As an additional challenge, the semantics has to
prohibit so-called Out-of-Thin-Air executions, in which results appear
out of nowhere.
We solve both problems by adopting the notion of \emph{operation
  buffers} from earlier works on relaxed memory
models~\cite{Boudol-Petri:POPL09,Boudol-al:EXPRESS12,EffingerDean-Grossman:MM},
and enhancing it with \emph{nesting} structure as a way to account for
conditional speculations.

\vspace{5pt}

While simple conceptually, the two described ingredients, when
combined, allow us to capture precisely the behavior of standard C11
synchronization primitives (including \emph{consume}-reads), desired
semantics of relaxed atomics, as well as multiple aspects of their
interaction, by elaborating the treatment of viewfronts and buffers.


The C11 standard is intentionally designed to be very general and
allow for multiple behaviors. However, particular compilation schemes
into different target architectures might focus only on specific
subsets of the enumerated features. To account for this diversity, our
framework comes in an \emph{aspect-oriented} flavor: it allows one to
``switch on and off'' specific aspects of C11 standard and to deal
only with particular sets of allowed concurrent behaviors.




\subsection{Contributions and outline}
\label{sec:contr-outl}


We start by outlining the basic intuition and illustrating a way of
handling C11's RA-synchronization and speculative executions,
introducing the idea of thread-specific viewfronts and operation
buffers in Section~\ref{sec:overview}.
Section~\ref{sec:advanced} demonstrates more advanced aspects of C11
concurrency expressed in our framework.
Section~\ref{sec:semantics} gives a formal definition of the
operational model for C11, which is our central theoretical
contribution. 
%
%
Section~\ref{sec:summary} describes evaluation of our semantics
implemented in the PLT Redex framework~\cite{Klein-al:POPL12,Felleisen-al:Redex}.
We argue for the adequacy of our constructions with respect to the
actual aspects of C11 using a large corpus of \emph{litmus test}
programs, adopted from the earlier works on formalizing C11
concurrency. To do so, we summarize the described operational aspects
of concurrent program behavior in C11, relating them to outputs of
litmus tests.
In Section~\ref{sec:rcu-example}, we showcase our operational model by
tackling a large realistic example: testing and debugging several
instances of a concurrently used Read-Copy-Update data
structure~\cite{McKenney:PhD,McKenney-Slingwine:PDCS98}, implemented
under relaxed memory assumptions. Our approach successfully detects
bugs in the cases when the employed synchronization primitives are not
sufficient to enforce the atomicity requirements, providing an
execution trace, allowing the programmer to reproduce the problem.
%
%
We compare to the related approaches to formalizing operational
semantics for relaxed memory in general and for C11 in particular in
Section~\ref{sec:related}, and conclude with a discussion of the
future work in Section~\ref{sec:conclusion}.


\section{Overview and Intuition}
\label{sec:overview}

We start by building the intuition for the program behaviors one can
observe in the C11 relaxed memory model.


The code below implements the \emph{message passing} pattern, where
one of the two parallel threads {waits} for the {notification} from
another one, and upon receiving it proceeds further with execution.

{
\label{code:mp}
\hspace{2.45cm} \lstinline{|$[$|f|$]$| := 0; |$[$|d|$]$| := 0;}
\vspace{.1cm}

\begin{center}
\begin{tabular}{l||l}
\begin{lstlisting}
|$[$|d|$]$| := 5;
|$[$|f|$]$| := 1; 
\end{lstlisting}
\hspace{.5cm}
&
\begin{lstlisting}
repeat |$[$|f|$]$| end;
r = |$[$|d|$]$|
\end{lstlisting}
\end{tabular}
\end{center}
\vspace{.2cm}
}

The identifiers in square parentheses (\eg, \texttt{[f]}) denote
accesses (\ie, writes and reads) to \emph{shared mutable} memory
locations, subject to concurrent manipulation, whereas plain
identifiers (\eg, \texttt{r}) stand for \emph{thread-local}
variables. In a sequentially consistent setting, assuming that reads
and writes to shared locations happen atomically, the right thread
will not reach the last assignment to \texttt{r} until the left thread
sets the flag \texttt{f} to be $1$. This corresponds to the ``message
passing'' idiom, and, hence, by the moment \texttt{[f]} becomes 1,
\texttt{d} will be pointing to~$5$. so by the end of the execution,
\texttt{r} must be~$5$.

In a more realistic setting of C/C++ concurrent programming, it is not
sufficient to declare all accesses to \texttt{[f]} and \texttt{[d]} as
atomic: depending on particular \emph{ordering annotations} on
read/write accesses (\eg, \emph{relaxed}, \emph{SC},
\emph{release/acquire} \etc) the outcome of the program might be
different and, in fact, contradictory to the ``natural''
expectations. For instance, annotating all reads and writes to
\texttt{[f]} and \texttt{[d]} as \emph{relaxed} might lead to
\texttt{r} being 0 at the end, due to the compiler and CPU-level
optimizations, rearranging instructions of the left thread with no
explicit dataflow dependency or, alternatively, assigning the value of
\texttt{[d]} to \texttt{r} in the right thread speculatively.

{
\setlength{\belowcaptionskip}{-10pt} 
\begin{figure}[t]
{\label{mp_mod}
\hspace{2.5cm} \lstinline{|$[$|f|$]_{na}$| := 0; |$[$|d|$]_{na}$| := 0;}
\vspace{.1cm}

\begin{center}
\begin{tabular}{l||l}
\begin{lstlisting}
|$[$|d|$]_{na}$| := 5;
|$[$|f|$]_{rel}$| := 1;
\end{lstlisting}
\hspace{.5cm}
&
\begin{lstlisting}
repeat |$[$|f|$]_{acq}$| end;
r = |$[$|d|$]_{na}$|
\end{lstlisting}
\end{tabular}
\end{center}
\vspace{.2cm}
}

\caption{Release/acquire message passing (\MpName).}
\label{fig:mp}
\end{figure}
}






One way to avoid these spurious results is to enforce stronger
synchronization guarantees between specific reads and writes in a
program using {release/acquire} order annotations. For instance, to
ensure the ``natural'' behavior of the message-passing idiom, the
program from above can be annotated as in Figure~\ref{fig:mp}.

In the modified program, all accesses to the location \texttt{d} are
now annotated as \emph{non-atomic}, which means racy concurrent
manipulations with them are considered run-time errors.
What is more important, the write to \texttt{f} in the left thread is
now annotated with $\relM$ modifier, which ``publishes'' the effects
of the previous operations, making them \emph{observable} by
concurrently running threads after the assigned to \texttt{f}
value~$1$ is read by them.
Furthermore, the read from \texttt{f} in the right thread is now
annotated with $\acqM$, preventing the operations following it in the
same thread from taking effect before the read itself takes
place. 
Together, the release/acquire modifiers in Figure~\ref{fig:mp} create
a \emph{synchronization order} we are seeking for and ensure that the
second assignment to \texttt{r} will only take place after the
\lstinline{repeat}-loop terminates, hence the observed value of
\texttt{f} is $1$, which implies that the observed value of \texttt{d}
is~5, thanks to the release-write, so the final value of \texttt{r} is
also~5.
Notice that there is also \emph{no} race between the two concurrent
non-atomic accesses to \texttt{d}, as those are clearly separated in
time, thanks to the synchronization between release/acquire accesses
to~\texttt{f}.

\paragraph{Axiomatic semantics and execution orders}

The state-of-the-art formalization \cite{Batty-al:POPL11} of C11
defines semantics for program execution as a set of graphs, where
nodes denote memory accesses (\ie, reads and writes) with particular
input/result values, and edges indicate ordering relations between
them.\footnote{%
  For illustrative purposes, here we employ a version of execution
  graphs~\cite{Vafeiadis-Narayan:OOPSLA13} with additional explicit
  nodes for spawning and joining threads.}

One instance of an execution graph for \MpName~is shown in
Figure~\ref{fig:graphMP}.
The edges labelled by \textsf{sb} indicate a natural program order,
reconstructed from the program's syntax. The green edge marked
\textsf{sw} indicates the \emph{synchronizes-with} relation, which
arises dynamically between a release-write and acquire-read of the
same value from the same location.
The transitive closure of the union of the \textsf{sb} and \textsf{sw}
relations is called \emph{happens-before} relation (\textsf{hb}) and
is central for defining the observed behaviors. In particular, a value
\emph{X}, written at a point $R_1$ of a program, can be only read at a
point $R_2$ if there is {no} \textsf{hb}-ordering between the
corresponding read event in $R_2$ and write event in $R_1$.  That, the
\emph{read-from} order (\textsf{rf}) must not contradict the
\textsf{hb} order.
%

{
\setlength{\belowcaptionskip}{-15pt} 
\begin{figure}[t]
\centering
\includegraphics[width=5cm]{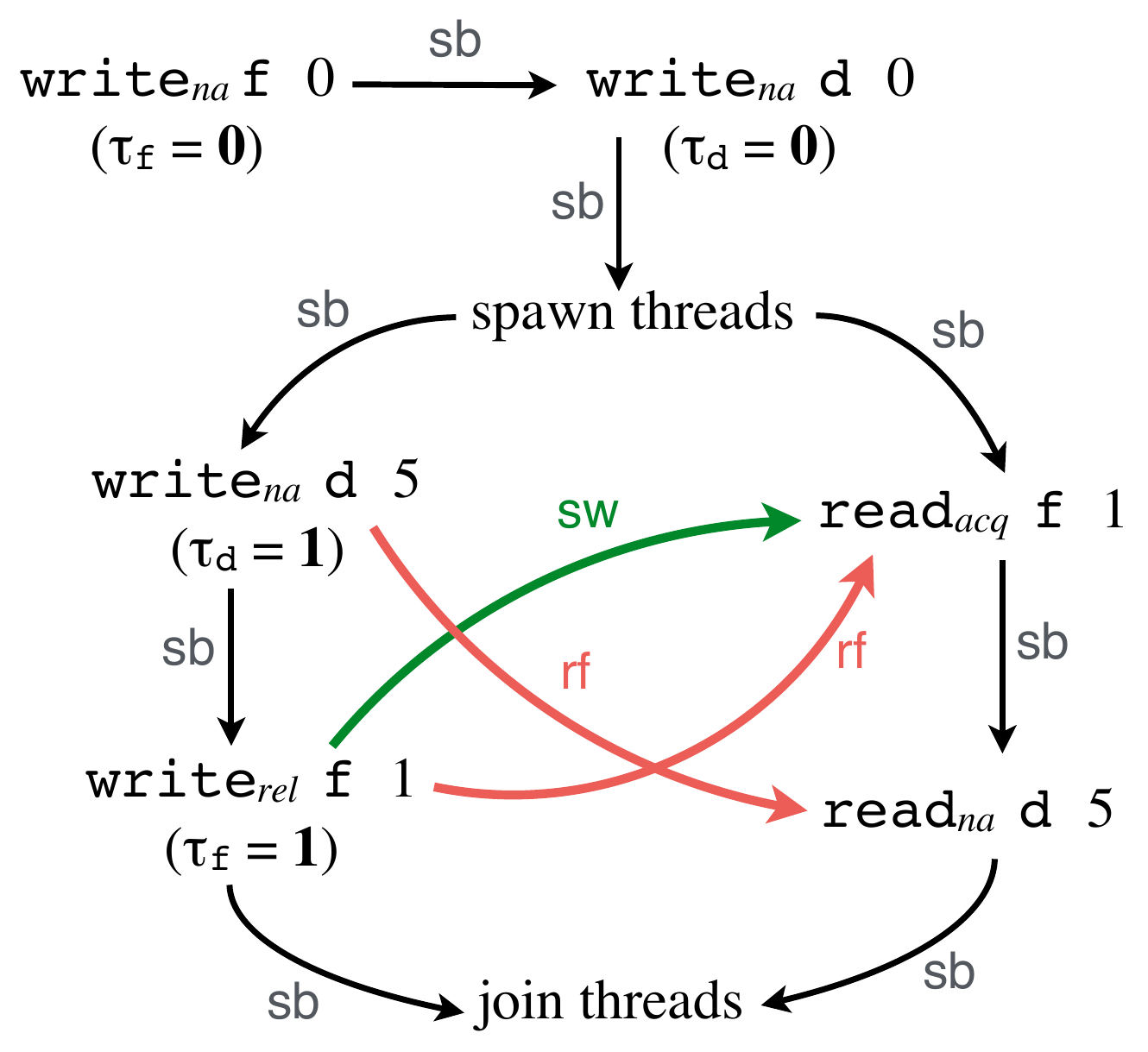}
\caption{Execution orders in the message passing example.}
\label{fig:graphMP}
\end{figure}
}

The C11 standard defines a number of additional axioms, specifying the
{consistent} execution graphs. In particular, all write actions on an
atomic location (\ie, such that it is not accessed non-atomically)
must be {totally ordered} via \emph{modification ordering} relation
(\textsf{mo}), which is consistent with \textsf{hb}, restricted to
this location: if a write operation $W_2$ to a location $\loc$ has a
greater timestamp than $W_1$, also writing to $\loc$, then $W_1$ cannot
be \textsf{hb}-ordered \emph{after} $W_2$.
The \textsf{mo} relation is shown in Figure~\ref{fig:graphMP} via
per-location {timestamps} $\tau_{-}$, incremented as new values are
stored.
Moreover, a read event $R_1$ cannot read from $W_1$, if $R_1$ is
\textsf{hb}-ordered after $W_2$, \ie, it's aware of the later write
$W_2$, as illustrated by \textsf{rf}-edges in
Figure~\ref{fig:graphMP}, preventing the case of reading of $0$ from
\texttt{d}.
%





\subsection{Synchronizing threads' knowledge via viewfronts}
\label{sec:hist}

\newcommand{\stateO}{%
  \tauFst & \emptyCell{|} & \emptyCell{} \\
}

\newcommand{\stateI}{%
  \tauFst & 0, \emptyFront \hfill \parentFrontMark
          & \emptyCell{} \\
}

\newcommand{\stateII}{%
  \tauFst & 0, \emptyFront \hfill \parentFrontMark
          & 0, \emptyFront \hfill \parentFrontMark \\
}

\newcommand{\stateIII}{%
  \tauFst & 0, \emptyFront \hfill \leftFrontMark \rightFrontMark
          & 0, \emptyFront                \hfill \leftFrontMark \rightFrontMark \\
}

\newcommand{\stateIV}{%
  \tauFst & 0, \emptyFront \hfill \leftFrontMark \rightFrontMark
          & 0, \emptyFront \hfill \rightFrontMark \\
  \tauSnd & \emptyCell{|}
          & 5, \emptyFront \hfill \leftFrontMark \\
}

\newcommand{\stateV}{%
  \tauFst & 0, \emptyFront \hfill \rightFrontMark 
          & 0, \emptyFront                \hfill \rightFrontMark \\
  \tauSnd & 1, (\floc\ $\smapsto$ \tauSnd, \dloc\ $\smapsto$ \tauSnd) \hfill \leftFrontMark
          & 5, \emptyFront                                          \hfill \leftFrontMark \\
}

\newcommand{\stateVI}{%
  \tauFst & 0, \emptyFront
          & 0, \emptyFront            \hfill \rightFrontMark \\
  \tauSnd & 1, (\floc\ $\smapsto$ \tauSnd, \dloc\ $\smapsto$ \tauSnd)
               \hfill \leftFrontMark \rightFrontMark
          & 5, \emptyFront                                  \hfill \leftFrontMark \\
}

\newcommand{\stateVII}{%
  \tauFst & 0, \emptyFront
          & 0, \emptyFront \\
  \tauSnd & 1, (\floc\ $\smapsto$ \tauSnd, \dloc\ $\smapsto$ \tauSnd)
               \hfill \leftFrontMark \rightFrontMark
          & 5, \emptyFront
               \hfill \leftFrontMark \rightFrontMark \\
}

\newcommand{\stateVIII}{%
  \tauFst & 0, \emptyFront
          & 0, \emptyFront \\
  \tauSnd & 1, (\floc\ $\smapsto$ \tauSnd, \dloc\ $\smapsto$ \tauSnd)
               \hfill \parentFrontMark
          & 5, \emptyFront
               \hfill \parentFrontMark \\
}

The \emph{read-from} relation determines the results of reads in
specific program locations depending on preceding or concurrent writes
by relying on the global \textsf{sb} and \textsf{sw} orderings, and
restricting them according to the C11 axioms.
To avoid the construction of global partial orders and provide an
incremental operational formalism for executing concurrent C11
programs, we focus on the \textsf{mo} relation for specific locations,
making it an inherent component of the program state.
We call this state component a \emph{history}: it contains totally
ordered ``logs'' of updates to every shared memory location, indexed
by timestamps (natural numbers). The history is objective (\ie,
\emph{global}): it contains information about {all} updates to shared
locations, as they took place during the execution.

However, the way threads ``see'' the history with respect to
particular locations is subjective (\ie, \emph{thread-local}): each
thread has its own knowledge of what is the ``latest'' written value
to each location. Thus, the value a thread actually reads can be
written \emph{no earlier} than what the thread considers to be the
location's latest value. To formalize this intuition, we define the
notion of \emph{viewfronts}.

A \emph{viewfront} is a partial function from memory locations to
natural numbers, representing timestamps in the corresponding
location's part of the history. A thread's viewfront represents its
knowledge of what were the timestamps of last written values to the
relevant locations that it is aware of.
When being a subject of a release-write, a location will {store} the
viewfront of a writing thread, which we will refer to as
\emph{synchronization front}, in addition to the actual value being
written. Symmetrically, another thread performing a synchronized load
(\eg, acquire-read) from the location will update its viewfront via
the one ``stored'' in the location. Viewfronts, when used for
  expressing release/acquire synchronization, are reminiscent to
  \emph{vector time frames}~\cite{Pozniansky-Schuster:PPoPP03}, but
  are used differently to express more advanced aspects of C11
  atomicity (see Section~\ref{sec:advanced} for details).



\begin{figure}[t]
\centering 
\vspace{-.2cm}
  \begin{tabular}{l@{\ \ \ }l}
    \begin{minipage}[l]{4.3cm} \small
\begin{lstlisting}
  |$[$|x|$]_{rlx}$| := 0; |$[$|y|$]_{rlx}$| := 0;
\end{lstlisting}
\vspace{-.2cm}
\begin{tabular}{l||l}
\begin{lstlisting}
r1 = |$[$|y|$]_{rlx}$|;
|$[$|x|$]_{rlx}$| := 1
\end{lstlisting}
\hspace{.6cm}
&
\begin{lstlisting}
r2 = |$[$|x|$]_{rlx}$|;
|$[$|y|$]_{rlx}$| := 1
\end{lstlisting}
\end{tabular}
    \end{minipage}
&
  \end{tabular}
\caption{An example with early reads (\textsf{LB\_rlx}).}
\label{fig:exampleEarlyReads}
\end{figure}

The following table represents the history and the threads' viewfronts
for the example from Figure~\ref{fig:mp} at the moment the left thread
has already written 1 and 5 to \texttt{f} and~\texttt{d}, but the
right thread has not yet exited the \lstinline{repeat}-loop.

\stateTblFrame{\stateTbl[2pt]{\floc}{\dloc}{ \stateV }}


\noindent
The values in the first column (\tauFst, \tauSnd) are
\emph{timestamps}, ascribing total order \textsf{mo} to the values
written to a certain location (\texttt{f} and \texttt{d}
correspondingly).
The remaining columns capture the sequence of updates of the
locations, each update represented as a pair of a value and a stored
synchronization front.
Viewfronts form a lattice, as partial maps from locations to
timestamps, with $\bot = \emptyset$.
%
The~\emptyFront\ fronts are used for location updates corresponding to
non-atomic stores. The front (\floc\ $\smapsto$ \tauSnd, \dloc\
$\smapsto$ \tauSnd) was stored to \texttt{f} upon executing the
corresponding release-write, capturing the actual viewfront
\leftFrontMark of the left thread.
The right thread's viewfront \rightFrontMark indicates that it still
considers \texttt{f} and \texttt{d} to be at least at timestamp
\textbf{0}, and, hence, can observe their values at timestamps larger
or equal than \textbf{0}.

Let us explore a complete execution trace of the program from
Figure~\ref{fig:mp}.
The initial state looks as follows:

\stateTblFrame{\stateTbl[2pt]{\floc}{\dloc}{ \stateO }}

\noindent
anf after the parent thread executes
`\lstinline{|$[$|f|$]_{na}$| := 0}' and
`\lstinline{|$[$|d|$]_{na}$| := 0}', it becomes:  

\noindent

\stateTblFrame{\stateTbl[2pt]{\floc}{\dloc}{ \stateII }}

\noindent Two subthreads are spawned, inheriting the parent's
viewfront:

\stateTblFrame{\stateTbl[2pt]{\floc}{\dloc}{ \stateIII }}

\noindent The left subthread performs `\lstinline{|$[$|d|$]_{na}$| := 5}', 
incrementing the timestamp $\tau$ of \texttt{d} and updating
its own viewfront \leftFrontMark:

\stateTblFrame{\stateTbl[2pt]{\floc}{\dloc}{ \stateIV }}

\noindent 
Next, the left thread executes `\lstinline{|$[$|f|$]_{rel}$| := 1}',
updating its viewfront and simultaneously storing it to the
\textbf{1}-entry~\texttt{f}:

\stateTblFrame{\stateTbl[2pt]{\floc}{\dloc}{ \stateV }}

\noindent 
The right thread can read the values of \texttt{f}, stored no later than
its viewfront \rightFrontMark indicates, thus, eventually it will
perform the acquire-read from \texttt{f} with
$\tau_{\text{\texttt{f}}} = $~\textbf{{1}}, updating its
\rightFrontMark correspondingly:

\stateTblFrame{\stateTbl[2pt]{\floc}{\dloc}{ \stateVII }}

\noindent Now the right thread's viewfront is updated with respect to
the latest store to \dloc, it reads $5$ from it, and the threads join:

\stateTblFrame{\stateTbl[2pt]{\floc}{\dloc}{ \stateVIII }}

\subsection{Speculating with operation buffers}
\label{sec:specs}



Relaxed atomics in C11 allow for speculative program optimizations,
which might result in out-of-order behaviors, observed during
concurrent executions under weak memory assumptions.

As a characteristic example of such a phenomenon, consider the program
in Figure~\ref{fig:exampleEarlyReads}. 
The C11 standard~\cite{C:11}, as well as its axiomatic formal
models~\cite{Batty-al:POPL11,Batty-al:ESOP15,Vafeiadis-Narayan:OOPSLA13},
allow for the outcome \lstinline{r1 = 1 |$ /\ $| r2 = 1} by the end of
its execution, as a result of rearranging instructions.
Alas, our viewfront-based semantics cannot account for such a
behavior: in order to be read from a location, a value should have
been first stored into the history by some thread!
However, in the example, it is either \texttt{x} or \texttt{y} that
stores $1$ (but not both) at the moments \texttt{r1} and \texttt{r2}
were assigned.
That is, while viewfront manipulation enables fine-grained control of
\emph{what} can be observed by threads, it does not provide enough
flexibility to specify \emph{when} effects of a particular thread's
operations should become visible to concurrent threads.


To account for such anomalies of relaxed behaviors, we introduce
\emph{per-thread operation buffers}, which allow a thread to postpone
an execution of an operation, ``resolving'' it later.
An operation buffer itself is a queue of records, each of which
contains an essential information for performing the corresponding
postponed operation.  For instance, for a postponed read action, a
thread allocates a fresh \emph{symbolic} value to substitute for a
not-yet-resolved read result, and adds a tuple, containing the
location and the symbolic value, to the buffer.
For a write action the thread puts an another tuple, the location and
the value to store to it, to the buffer.  
As it proceeds with the execution, the thread can
non-deterministically resolve an operation from the buffer \emph{if
  there is no operation before it}, which may affect its result,
\eg, a write to the same location, or an acquire-read changing the
local viewfront.
For instance, in Figure~\ref{fig:exampleEarlyReads}, buffering the
effects of the two relaxed reads, $[y]_{rlx}$ and $[x]_{rlx}$,
postpones their effects beyond the subsequent writes, enabling the
desired outcome, as by the moment the reads are resolved, the
corresponding 1's will be already stored to the history.

\paragraph{Nested buffers and speculative conditionals}

The idea of buffering operations for postponing their effects in a
relaxed concurrency settings is not novel and has previously appeared
in a number of related weak memory
frameworks~\cite{Boudol-Petri:POPL09,Boudol-al:EXPRESS12,Crary-Sullivan:POPL15,EffingerDean-Grossman:MM}.
However, in our case it comes with a twist, making it particularly
well suited for modelling C11 behaviors, while avoiding ``bad''
executions.

{
\setlength{\belowcaptionskip}{-10pt} 
\begin{figure}[t]

{
\hspace{2.05cm}
\lstinline{|$[$|x|$]_{rlx}$| := 0; |$[$|y|$]_{rlx}$| := 0; |$[$|z|$]_{rlx}$| := 0;}
\vspace{.1cm}

\begin{center}
\begin{tabular}{l||l}
\begin{lstlisting}
if |$[$|x|$]_{rlx}$|
then |$[$|z|$]_{rlx}$| := 1;
     |$[$|y|$]_{rlx}$| := 1
else |$[$|y|$]_{rlx}$| := 1 fi
\end{lstlisting}
\hspace{.5cm}
&
\begin{lstlisting}
if |$[$|y|$]_{rlx}$|
then |$[$|x|$]_{rlx}$| := 1;
else 0 fi
\end{lstlisting}
\end{tabular}
\end{center}
\vspace{.2cm}

\hspace{3.75cm}
\lstinline{res := |$[$|z|$]_{rlx}$|}
\vspace{.1cm}
}

\caption{A program allowing if-speculations (\spProg).}
\label{fig:sp}
\end{figure}
}

To illustrate this point, let us consider an example of a speculative
optimization involving a conditional statements. Such optimizations
are known to be difficult for modelling in relaxed
concurrency~\cite{Batty-al:ESOP15,PichonPharabod-Sewell:POPL16}.
For instance, in the program in Figure~\ref{fig:sp}, the assignment
$[y]_{rlx} := 1$ can be ``pulled out'' from both branches of the left
thread's conditional statement, as it will be executed anyway, and,
furthermore, it does not bear a data dependency with the possible
preceding assignment $[y]_{rlx} := 1$. Such an optimization will,
however, lead to interesting consequences: in the right thread, the
conditional statement might succeed assigning 1 to \texttt{x},
therefore leading to the overall result \texttt{res = 1}. This outcome
relies on the fact that the optimization, which made $[y]_{rlx} := 1$
unconditionally visible to the right thread, was done {speculatively},
yet it has been justified later, since the same assignment would have
been performed no matter which branch has been executed.

Luckily, to be able to express such a behavior, our buffer machinery
requires only a small enhancement: \emph{nesting}.
In the semantics, upon reaching an if-then-else statement, whose
condition's expression is a result of some preceding relaxed read,
which is not yet resolved, we create a \emph{tuple}, containing the
symbolic representation of the condition as well as two empty buffers,
to be filled with postponed operations of the left and the right
branches, correspondingly. The tuple is then added to the thread's
main operation buffer.


More specifically, in the program \spProg, the history after the three
initial relaxed writes is as follows:

\stateTblFrame{
  \stateTblThree[2pt]{
    \tauFst & 0, (\xloc\ $\smapsto$ \tauFst)
            & 0, (\yloc\ $\smapsto$ \tauFst)
            & 0, (\zloc\ $\smapsto$ \tauFst) \\}
}

\noindent
The left thread then postpones reading from \xloc~and start the
executing the \lstinline{if} statement speculatively, with the
following buffer:


\begin{center}
\begin{lstlisting}
|$\langle$| a = |$[$|x|$]_{rlx}$|; if a |$\angled{}$| |$\angled{}$| |$\rangle$|
\end{lstlisting}
\end{center}


\noindent
Proceeding to execute the two branches of the \texttt{if}-statement
with focusing on the corresponding nested buffers, the left thread
eventually fills them with the postponed commands:


\begin{center}
\begin{lstlisting}
|$\langle$| a = |$[$|x|$]_{rlx}$|; if a |$\langle$||$[$|z|$]_{rlx}$| := 1; |$[$|y|$]_{rlx}$| := 1|$\rangle$| |$\langle$||$[$|y|$]_{rlx}$| := 1|$\rangle$| |$\rangle$|
\end{lstlisting}
\end{center}


\noindent
At this point the two sub-buffers contain the postponed write
\lstinline{|$[$|y|$]_{rlx}$| := 1}, and no other postponed operations
in the same buffers are in conflict with them. This allows the
semantics to \emph{promote} this write to the upper-level buffer (\ie,
the main buffer of the thread):


\begin{center}
\begin{lstlisting}
|$\langle$| a = |$[$|x|$]_{rlx}$|; |$[$|y|$]_{rlx}$| := 1; if a |$\langle$||$[$|z|$]_{rlx}$| := 1|$\rangle$| |$\angled{}$| |$\rangle$|
\end{lstlisting}
\end{center}

\noindent
Next, the write is resolved, so its effect is visible to the right
thread:

\begin{center}
\begin{lstlisting}
|$\langle$| a = |$[$|x|$]_{rlx}$|; if a |$\langle$||$[$|z|$]_{rlx}$| := 1|$\rangle$| |$\angled{}$| |$\rangle$|
\end{lstlisting}
\end{center}

\noindent
At that moment, the overall history looks as follows:



\stateTblFrame{
  \stateTblThree[2pt]{
    \tauFst & 0, (\xloc\ $\smapsto$ \tauFst)
            & 0, (\yloc\ $\smapsto$ \tauFst)
            & 0, (\zloc\ $\smapsto$ \tauFst) \\
    \tauSnd & \emptyCell{|} & 1, (\yloc\ $\smapsto$ \tauSnd) & \emptyCell{} \\
  }
}


\noindent
Hence, the right thread can read 1 from the location \yloc, take the
\lstinline{then} branch of the \lstinline{if} statement, and perform
the write to \xloc:

\stateTblFrame{
  \stateTblThree[2pt]{
    \tauFst & 0, (\xloc\ $\smapsto$ \tauFst)
            & 0, (\yloc\ $\smapsto$ \tauFst)
            & 0, (\zloc\ $\smapsto$ \tauFst) \\
    \tauSnd & 1, (\xloc\ $\smapsto$ \tauSnd) & 1, (\yloc\ $\smapsto$ \tauSnd) & \emptyCell{} \\
  }
}

\noindent Now the left thread can resolve the postponed read
\lstinline{a = |$[$|x|$]_{rlx}$|} obtaining \lstinline{1} as its
result and reducing the operation buffer:

\begin{lstlisting}
|$\langle$| if 1 |$\langle$||$[$|z|$]_{rlx}$| := 1|$\rangle$| |$\angled{}$| |$\rangle$|
\end{lstlisting}

\noindent
By evaluating the buffered \lstinline{if} and resolving the write
\lstinline{|$[$|z|$]_{rlx}$| := 1}:

\stateTblFrame{
  \stateTblThree[2pt]{
    \tauFst & 0, (\xloc\ $\smapsto$ \tauFst)
            & 0, (\yloc\ $\smapsto$ \tauFst)
            & 0, (\zloc\ $\smapsto$ \tauFst) \\
    \tauSnd & 1, (\xloc\ $\smapsto$ \tauSnd) & 1, (\yloc\ $\smapsto$ \tauSnd) &  1, (\zloc\ $\smapsto$ \tauSnd) \\
  }
}

\noindent Reading from the latest record for \zloc~results in
\lstinline{res = 1}.

The idea of nested buffers with promoting duplicating records from a
lower to an upper level (under some dependency conditions) naturally
scales for the case of nested if-statements.




{
\setlength{\belowcaptionskip}{-10pt} 
\begin{figure}[t]

{
\hspace{2.85cm}
\lstinline{|$[$|x|$]_{rlx}$| := 0; |$[$|y|$]_{rlx}$| := 0;}
\vspace{.1cm}

\begin{center}
\begin{tabular}{l||l}
\begin{lstlisting}
if |$[$|x|$]_{rlx}$|
then |$[$|y|$]_{rlx}$| := 1
else 0 fi
\end{lstlisting}
\hspace{.5cm}
&
\begin{lstlisting}
if |$[$|y|$]_{rlx}$|
then |$[$|x|$]_{rlx}$| := 1
else 0 fi
\end{lstlisting}
\end{tabular}
\end{center}
\vspace{.1cm}
\hspace{2.85cm}
\lstinline{r1 = |$[$|x|$]_{rlx}$|; r2 = |$[$|y|$]_{rlx}$|}
\vspace{.2cm}
}

\caption{Program with C11-allowed Thin-Air behavior
  (\textsf{OTA\_if}).}
\label{fig:ota}
\end{figure}
}

\paragraph{On the Out-of-Thin-Air problem}

So what are the ``bad'' executions that should be prohibited by a
meaningful semantics? 

The C11 standard~\cite{C:11,CPP:11} and the axiomatic
semantics~\cite{Batty-al:POPL11} allow for so-called
\emph{Out-of-Thin-Air} (OTA) behaviors, witnessed by self-satisfying
conditionals, such as the one represented by the program in
Figure~\ref{fig:ota}, which, according to the standard is allowed to
end up with \lstinline{r1 = r2 = 1}.
Such behavior is, however, not observable on any of the major modern
architectures (x86, ARM, and POWER), and considered as a flaw of the
model~\cite{Boehm-Demsky:MSPC14, Batty-al:ESOP15}, with researchers
developing alternative semantics for relaxed atomics that avoid
OTA~\cite{PichonPharabod-Sewell:POPL16}.

Notice the only essential difference between the programs in
Figure~\ref{fig:sp} and~\ref{fig:ota} is that in the former the write
performed speculatively will \emph{always} take place, whereas in the
latter one the speculative writes in the then-branch might end up
unjustified.

As we have previously demonstrated, our semantics supports the weak behavior
of the program in Figure~\ref{fig:sp}, and outlaws it for the program
in Figure~\ref{fig:ota}, as the conditions for promoting buffered
operations in the if-branches will not be met in the latter case.


\section{Advanced Aspects of C11 Concurrency}
\label{sec:advanced}





In this section, we elaborate and employ the ideas of viewfronts and
operation buffers to adequately capture the remaining aspects of C11
concurrency.
In particular, we \emph{(i)} show how to extend the viewfront
mechanism to support {sequentially-consistent} (SC) and {non-atomic}
(NA) memory accesses, as well as consume-reads,
(\S\S\ref{sec:sc}--\ref{sec:consume-reads}); \emph{(ii)} employ
operation buffers to account for specific phenomena caused by
sequentialization optimization (\S\ref{sec:join}), and \emph{(iii)}
demonstrate the interplay between relaxed atomics and
RA-synchronization (\S\ref{sec:rlxAdvanced}).
%
%


\subsection{Sequentially-consistent memory accesses}
\label{sec:sc}

%
%

To see the difference between SC-style and
release/acquire-synchro-nization in C11, consider the program in
Figure~\ref{fig:exampleSC}.
All SC-operations are totally ordered with respect to each other, and
the last of them is either read from \lstinline{x}, or from
\lstinline{y}. Thus, the overall outcome 
\lstinline{r1 = 0 |$\land$| r2 = 0} is impossible.
Replacing any of the \emph{sc} modifiers by {release} or
acquire in the corresponding writes and reads, makes
\lstinline{r1 = 0 |$\land$| r2 = 0} a valid outcome, because there is
no more the total order on all operations.
In particular, the left subthread (with RA modifiers instead of SC)
could still read 0 from \texttt{y}, as by its viewfront, which at
that moment is (\texttt{x} $\smapsto$ \tauSnd, \texttt{y} $\smapsto$
\tauFst).
%

{
\setlength{\belowcaptionskip}{-10pt} 
\begin{figure}[t] 
\centering 
\vspace{-.2cm}
  \begin{tabular}{l@{\ \ \ }l}
    \begin{minipage}[l]{4.3cm} \small
\begin{lstlisting}
  |$[$|x|$]_{sc}$| := 0; |$[$|y|$]_{sc}$| := 0;
\end{lstlisting}
\vspace{-.2cm}
\begin{tabular}{l||l}
\begin{lstlisting}
|$[$|x|$]_{sc}$| := 1;
r1 = |$[$|y|$]_{sc}$|
\end{lstlisting}
\hspace{.6cm}
&
\begin{lstlisting}
|$[$|y|$]_{sc}$| := 1;
r2 = |$[$|x|$]_{sc}$|
\end{lstlisting}
\end{tabular}
    \end{minipage}
&
  \end{tabular}
  \caption{A program with SC synchronization (\textsf{SB\_sc}).}
\label{fig:exampleSC}
\end{figure}
}


In the axiomatic
model~\cite{Batty-al:POPL11,Vafeiadis-Narayan:OOPSLA13}, the
restricted set of SC behaviors is captured by introducing an
additional order \textsf{sc} and several axioms, requiring, in
particular, consistency of \textsf{sc} with respect to \textsf{hb} and
\textsf{mo}.
%
%
In our operational setting, it means that an SC-read cannot read from
a write with a \emph{smaller} timestamp than the \emph{greatest}
timestamp of SC-writes to this location.
To capture this requirement, we instrument the program state with an
additional component---a \emph{global} viewfront of sequentially
consistent memory operations ($\stSigma_{sc}$), which is being updated
at each SC-write.
%



\subsection{Non-atomic memory accesses and data races}
\label{sec:na}


Following the C11 standard \cite{C:11,CPP:11}, our semantics does not
draw a distinction between non-atomic and atomic locations (in
contrast with the axiomatic model~\cite{Batty-al:POPL11}). However,
data races involving non-atomic memory \emph{operations} (whose
purpose is data manipulation, not thread synchronization) might result
in an undefined behavior.

Consider the following two code fragments with data races on
non-atomics.
In the first case, a thread performs a $\naM$-read concurrently with a
write to the same location.


  \begin{tabular}{l@{\ \ \ }l}
    \begin{minipage}[l]{4.3cm} \small
\begin{lstlisting}
       |$[$|d|$]_{na}$| := 0;
\end{lstlisting}
\vspace{-.2cm}
\begin{tabular}{l||l}
\begin{lstlisting}
|$[$|d|$]_{rlx}$| := 1
\end{lstlisting}
\hspace{.6cm}
&
\begin{lstlisting}
r = |$[$|d|$]_{na}$|
\end{lstlisting}
\end{tabular}
    \end{minipage}
&
  \end{tabular}
\vspace{.2cm}

\noindent
We can detect the data race, when the right subthread is executed
after the left one, so it performs the $\naM$-read, while not being
``aware'' of the latest write to the same location. As our semantics
constructs the whole state-space for all possible program executions,
we will identify this data race on some execution path.


%


The second case is an opposite one: $\naM$-write and atomic read:

  \begin{tabular}{l@{\ \ \ }l}
    \begin{minipage}[l]{4.3cm} \small
\begin{lstlisting}
       |$[$|d|$]_{na}$| := 0;
\end{lstlisting}
\vspace{-.2cm}
\begin{tabular}{l||l}
\begin{lstlisting}
|$[$|d|$]_{na}$| := 1
\end{lstlisting}
\hspace{.6cm}
&
\begin{lstlisting}
r = |$[$|d|$]_{rlx}$|
\end{lstlisting}
\end{tabular}
    \end{minipage}
&
  \end{tabular}
\vspace{.2cm}

\noindent
It still has a data race involving a non-atomic access, which,
however, we cannot detect by comparing threads' viewfronts.
To identify data races of this kind, we extend the state with a global
$\naM$-front, storing a timestamp of the last $\naM$-write to a
location.
Now, if the left thread executes its $\naM$-write first, the atomic
read in the right one will not be aware of it, which will be
manifested as a data race, thanks to the $\naM$-front.
%

\subsection{Consume-reads}
\label{sec:consume-reads}

\begin{figure}[t]
\centering 
\vspace{-.2cm}
  \begin{tabular}{l@{\ \ \ }l}
    \begin{minipage}[l]{4.3cm} \small
\begin{lstlisting}
|$[$|p|$]_{na}$| := null; |$[$|d|$]_{na}$| := 0; |$[$|x|$]_{na}$| := 0;
\end{lstlisting}
\vspace{-.2cm}
\begin{tabular}{l||l}
\begin{lstlisting}
|$[$|x|$]_{rlx}$| := 1;
|$[$|d|$]_{na}$| := 1;
|$[$|p|$]_{rel}$| := d
\end{lstlisting}
\hspace{.6cm}
&
\begin{lstlisting}
r1 = |$[$|p|$]_{con}$|;
if   r1 != null
then r2 = |$[$|r1|$]_{na}$|;
     r3 = |$[$|x|$]_{rlx}$|
else r2 = 0; r3 = 0
fi
\end{lstlisting}
\end{tabular}
    \end{minipage}
&
  \end{tabular}
\caption{An example with a consume read (\textsf{MP\_con+na\_2}).}
\label{fig:exampleConsume}
\end{figure}


Unlike {acquire}-reads, {consume}-reads~\cite{consume:TR} do not
update a thread's viewfront, but provide a synchronization front only
for subsequent reads that are dereferencing their result.
 
Consider the code fragment in Figure~\ref{fig:exampleConsume}.  Here,
we have message-passing of data stored in \lstinline{d} via location
\lstinline{p}.  The right thread employs a consume-read from
\lstinline{p}.  In the case when it gets a non-\lstinline{null} value
(representing a pointer to~\lstinline{d}), it reads from it to
\lstinline{r2}, and after that from location \lstinline{x} to
\lstinline{r3}.  There might be three possible outcomes:
%
%
\lstinline{r1 = null} $ /\ $ \lstinline{r2 = 0} $ /\ $ \lstinline{r3 = 0},
\lstinline{r1 = d} $ /\ $ \lstinline{r2 = 1} $ /\ $ 
\lstinline{r3 = 1}, and
\lstinline{r1 = d} $ /\ $ \lstinline{r2 = 1} $ /\ $ \lstinline{r3 = 0}.
Changing the consume-read to an acquire one makes the last triple
forbidden, as the right thread's viewfront would become up to date
with both \lstinline{|$[$|x|$]_{rlx}$| := 1} and
\lstinline{|$[$|d|$]_{rlx}$| := 1} after acquire-reading a
non-\lstinline{null} pointer value from~\lstinline{p}.
At the same time, consume-read \lstinline{r1 = |$[$|p|$]_{con}$|}
provides synchronization only for \lstinline{r2 = |$[$|r1|$]_{rlx}$|},
which explicitly dereferences its result, but not for
\lstinline{r3}, which has no data-dependency with it.
%

%

Adding consume-reads to the semantics requires us to change the
program syntax to allow \emph{run-time annotations} on reads, which
might be affected by consume ones.  When a consume-read is
executed, it retrieves some value/front-entry $(\mvalSubst, \stSigma)$
from the history, as any other read.  Unlike an acquire-read,
it does not update the thread's viewfront by the retrieved $\stSigma$.
Instead, it annotates all subsequent data-dependent reads by the
front~$\stSigma$. Later, when these reads will be executed, they will
join the front $\stSigma$ from the annotation with the thread's
viewfront for computing the lower boundary on the relevant location's
timestamp.  The same process is applied to annotate data-dependent
postponed reads in buffers, which might refer to the symbolic result
of a consume-read.

\subsection{Threads joining and synchronization}
\label{sec:join}

{
\setlength{\belowcaptionskip}{-10pt} 
\begin{figure}[t] 
\centering 
\vspace{-.2cm}
\hspace{-2cm}
  \begin{tabular}{l@{\ \ \ }l}
    \begin{minipage}[l]{4.3cm} \small
\begin{lstlisting}
                |$[$|x|$]_{rlx}$| := 0; |$[$|y|$]_{rlx}$| := 0;
\end{lstlisting}
\vspace{-.2cm}
\begin{tabular}{l||l||l||l}
\begin{lstlisting}
r1 = |$[$|y|$]_{rlx}$|;
|$[$|z1|$]_{rlx}$| := r1
\end{lstlisting}
\hspace{.6cm}
&
\begin{lstlisting}
  0
\end{lstlisting}
\hspace{.6cm}
&
\begin{lstlisting}
r2 = |$[$|x|$]_{rlx}$|;
|$[$|z2|$]_{rlx}$| := r2
\end{lstlisting}
\hspace{.6cm}
&
\begin{lstlisting}
  0
\end{lstlisting}
\end{tabular}

\vspace{-1pt}
\begin{tabular}{l||l}
  \begin{lstlisting}
            |$[$|x|$]_{rlx}$| := 1
  \end{lstlisting}
\hspace{2.52em}
&
  \begin{lstlisting}
      |$[$|y|$]_{rlx}$| := 1
  \end{lstlisting}
\end{tabular}
    \end{minipage}
&
  \end{tabular}
  \caption{A program with non-flat thread joining (\textsf{LB\_rlx+join}).}
\label{fig:join}
\end{figure}
}

Once two threads join, it is natural to expect that all their
postponed memory operations are resolved, \ie, they have
empty operation buffers.
This is reflected in the axiomatic semantics~\cite{Batty-al:POPL11} by
an \emph{additional-synchronizes-with} relation, which is a part of
the \emph{happens-before} relation. Thus, every memory action of
joined threads \emph{happens-before} actions which are syntactically
after the join point.  Our semantics achieves this by merging
viewfronts at join and forcing resolution of all operations in the
buffers.

However, the C11 standard is intended to allow
\emph{sequentialization} optimization \cite{Vafeiadis-al:POPL15}, \ie,
$S_{1} ~\|~ S_{2} \rightsquigarrow S_{1}; S_{2}$, making the previous
assumption unsound.  
This is illustrated by the program in Figure~\ref{fig:join}, in which
the parallel compositions with ``idle'' threads might be optimized by
replacing them with non-idle parts.
After such an optimization it is possible to observe \lstinline{1}s as
result values of \lstinline{z1} and \lstinline{z2}.
To account for this, we allow an alternative instance of a
thread-joining policy, implemented as an aspect, which takes all
possible interleavings of the threads' operation buffers (with an idle
thread being a natural unit), thus achieving the behavior we seek for.



\subsection{Relaxed atomics and synchronization} 
\label{sec:rlxAdvanced}

Interaction between relaxed atomics and RA-synchronization is
particularly subtle, due to a number of ways they might affect the
outcomes of each other. We identify these points of interaction and
describe several design decisions, elaborating the structure of the
state, so the requirements imposed by the C11 standard are met.

\subsubsection{Relaxed writes and release sequences}
\label{sec:relax-writ-rele}

%
%
Since a relaxed read cannot be used for synchronization, in our
semantics it {does not} update the viewfront of the thread with
a synchronization front from the history (as an acquire read does).
However, when an {acquire}-read in thread $T_2$
reads a result of a relaxed write performed by thread $T_1$, it should
be synchronized with a \emph{preceding} {release}-write to the same
location performed by thread $T_1$, if there is one.
This observation follows the spirit of the axiomatic
model~\cite{Batty-al:POPL11}, which defines a notion of \emph{release
  sequence} between the writes of thread $T_1$.

%
%


\begin{figure}[t] 
\centering 
\vspace{-.2cm}
  \begin{tabular}{l@{\ \ \ }l}
    \begin{minipage}[l]{4.3cm} \small
\begin{lstlisting}
|$[$|f|$]_{na}$| := 0; |$[$|d|$]_{na}$| := 0; |$[$|x|$]_{na}$| := 0;
\end{lstlisting}
\vspace{-.2cm}
\begin{tabular}{l||l}
\begin{lstlisting}
|$[$|d|$]_{na}$| := 5;
|$[$|f|$]_{rel}$| := 1;
|$[$|x|$]_{rel}$| := 1;
|$[$|f|$]_{rlx}$| := 2
\end{lstlisting}
\hspace{.6cm}
&
\begin{lstlisting}
repeat |$[$|f|$]_{acq}$| == 2 end;
r1 := |$[$|d|$]_{na}$|;
r2 := |$[$|x|$]_{rlx}$|
\end{lstlisting}
\end{tabular}
    \end{minipage}
&
  \end{tabular}
  \caption{Example of release sequence
    (\textsf{MP\_rel+acq+na+rlx\_2}).}
\label{fig:rlxSync}
\end{figure}

For an example, let us take a look at Figure~\ref{fig:rlxSync}
presenting a modified version of the message-passing program.
The only possible outcome for \lstinline{r1} is \lstinline{5}, because
when \lstinline{|$[$|f|$]_{acq}$|} gets \lstinline{2}, it also becomes
synchronized with \lstinline{|$[$|f|$]_{rel}$| := 1}, which precedes
\lstinline{|$[$|f|$]_{rlx}$| := 2} in the left thread.  
At the same time \lstinline{r2} can be either \lstinline{0}, or
\lstinline{1}: \lstinline{0} is a possible outcome for \lstinline{r2},
because \lstinline{|$[$|f|$]_{acq}$|} synchronizes with
\lstinline{|$[$|f|$]_{rel}$| := 1}, which precedes the
\lstinline{|$[$|x|$]_{rel}$| := 1} write, which therefore might be
missed.

To express this synchronization pattern in our model, we instrument
the state with per-thread \emph{write-fronts}, containing information
about last {release}-writes to locations performed by the thread.
Upon a relaxed write, this information is used to retrieve a
synchronization front from the history record with a timestamp equal
to the write-front value, contributed by a preceding
{release}-write.


\subsubsection{Postponed relaxed operations and synchronization}
\label{sec:postponed}

{
\setlength{\belowcaptionskip}{-10pt} 
\begin{figure}[t]
\centering 
\vspace{-.2cm}
  \begin{tabular}{l@{\ \ \ }l}
    \begin{minipage}[l]{4.3cm} \small
\begin{lstlisting}
  |$[$|x|$]_{rlx}$| := 0; |$[$|y|$]_{rlx}$| := 0;
\end{lstlisting}
\vspace{-.2cm}
\begin{tabular}{l||l}
\begin{lstlisting}
r1 = |$[$|y|$]_{rlx}$|;
|$[$|x|$]_{rel}$| := 1
\end{lstlisting}
\hspace{.6cm}
&
\begin{lstlisting}
r2 = |$[$|x|$]_{rlx}$|;
|$[$|y|$]_{rel}$| := 1
\end{lstlisting}
\end{tabular}
    \end{minipage}
&
  \end{tabular}
  \caption{Postponed relaxed reads and release-writes
    (\textsf{LB\_rel+rlx}).}
\label{fig:earlyReadsRel}
\end{figure}
}

\begin{figure}[t]
\centering 
\vspace{-.2cm}
  \begin{tabular}{l@{\ \ \ }l}
    \begin{minipage}[l]{4.3cm} \small
\begin{lstlisting}
  |$[$|x|$]_{rlx}$| := 0; |$[$|y|$]_{rlx}$| := 0;
\end{lstlisting}
\vspace{-.2cm}
\begin{tabular}{l||l}
\begin{lstlisting}
r1 = |$[$|y|$]_{acq}$|;
|$[$|x|$]_{rlx}$| := 1
\end{lstlisting}
\hspace{.6cm}
&
\begin{lstlisting}
r2 = |$[$|x|$]_{rlx}$|;
|$[$|y|$]_{rel}$| := 1
\end{lstlisting}
\end{tabular}
    \end{minipage}
&
  \end{tabular}
\caption{Postponed relaxed reads and RA (\textsf{LB\_rel+acq+rlx}).}
\label{fig:earlyReadsRA}
\end{figure}




Consider the program in Figure~\ref{fig:earlyReadsRel}, which is
similar to the example from Figure~\ref{fig:exampleEarlyReads}, but
writes have {release}-modifiers.
Since a release-write does not impose any restriction without a related
acquire read, it is still possible to get the result
\lstinline{r1 = r2 = 1}. 
Therefore, our semantics allows to perform a release-write even if
there are postponed reads from other locations in the thread's buffer.

The program in Figure~\ref{fig:earlyReadsRA} is more problematic, as
it has {release}/{acquire} modifiers on accesses to \lstinline{y}, and
postponing the read \lstinline{|$[$|x|$]_{rlx}$|} in the right thread
might lead to \lstinline{r1 = r2 = 1}.
Notice that such an outcome is in conflict with the semantics of RA,
as assigning \lstinline{1} to \lstinline{r1} would imply
synchronization between \lstinline{|$[$|y|$]_{rel}$| = 1} and
\lstinline{r1 = |$[$|y|$]_{acq}$|}, thus the former happens
\emph{before} the latter one!
Following the rules of RA-synchronization, 
\lstinline{r2 = |$[$|x|$]_{rlx}$|} should happen {before}
\lstinline{|$[$|x|$]_{rlx}$| = 1} as well, making it impossible to
read \lstinline{1} into \lstinline{r2}.


The problem is clear now: we need to prevent from happening the
situations, when a read, postponed beyond a release-write $W$,
is resolved after some concurrent acquire-read gets
synchronized with $W$, as it might damage RA-synchronization
sequences.

To achieve this, we instrument the state with a global list $\stGamma$
of triples that consist of: \emph{(i)} a location~$\loc$, \emph{(ii)} a
timestamp~$\stTau$ of some executed write, and \emph{(iii)} a
symbolic value~$\vName$ of a read postponed beyond this write.
%
%
%
When executing a {release}-write $W$, which stores a value to a
location $\loc$ with a timestamp $\stTau$, for each thread-local
postponed read, we globally record a triple
$\angled{\loc, \stTau, \vName}$, where $\vName$ is the symbolic value
of the read. An {acquire}-read of the $(\loc, \stTau)$ history entry
by another thread succeeds only if there are no
$\angled{\loc, \stTau, \vName}$ left in $\stGamma$ for any symbolic
value $\vName$.
Resolving a postponed read with a symbolic value $\vName$ removes all
$\vName$-related entries from $\stGamma$.




\begin{figure}[t]
\centering 
\vspace{-.2cm}
  \begin{tabular}{l@{\ \ \ }l}
    \begin{minipage}[l]{4.3cm} \small
\begin{lstlisting}
  |$[$|x|$]_{rlx}$| := 0; |$[$|y|$]_{rlx}$| := 0;
\end{lstlisting}
\vspace{-.2cm}
\begin{tabular}{l||l}
\begin{lstlisting}
|$[$|x|$]_{rlx}$| := 1;
|$[$|y|$]_{rel}$| := 2
\end{lstlisting}
\hspace{.6cm}
&
\begin{lstlisting}
|$[$|y|$]_{rlx}$| := 1;
|$[$|x|$]_{rel}$| := 2
\end{lstlisting}
\end{tabular}
\begin{lstlisting}
  r1 = |$[$|x|$]_{rlx}$|; r2 = |$[$|y|$]_{rlx}$|
\end{lstlisting}
    \end{minipage}
&
  \end{tabular}
  \caption{Postponed writes and release-writes
    (\textsf{WR\_rlx+rel}).}
\label{fig:wrRel}
\end{figure}

The program in Figure~\ref{fig:wrRel} (\textsf{2+2W} from
\cite{Lahav-al:POPL16}) demonstrates another subtlety, caused by
interaction between postponed writes and RA-synchronization.
According to the standard, \lstinline{r1 = r2 = 1} is its valid
outcome, and to achieve that in our semantics the corresponding
relaxed writes should be committed to the history \emph{after} the
release ones.
However, if the another (third) thread performs an acquire read from
the history record of one of the release writes, it should become
aware of the previous corresponding relaxed write. 

We solve the problem using the same global list $\stGamma$ as with
postponed reads.  If a release-write is performed before a postponed
one, it adds a corresponding triple to $\stGamma$. 
Subsequently, a synchronizing acquire-read from a history record will
not be performed until the corresponding triple is in the
list. 
The only difference is that when the semantics resolves a postponed
write, it does not only delete the corresponding records from
$\stGamma$, but also updates the synchronization front in the history
record stored by the release write, therefore, ``bringing it
up-to-date'' with the globally published stored values.

\subsection{Putting it all together}

As one can notice, almost every aspect of the C11 standard, outlined
in Sections~\ref{sec:sc}--\ref{sec:rlxAdvanced} requires us to enhance
our semantics in one way or another. The good news are that almost all
of these enhancements are \emph{orthogonal}: they can be added to the
operational model independently. For instance, one can consider a
subset of C11 with RA-synchronization, relaxed and non-atomic
accesses, but without accounting for release-sequences or SC-accesses.
%

\section{Operational Semantics, Formally} 
\label{sec:semantics}

In this section, we formally describe main components of our
operational semantics for C11, starting from the definition of the
language, histories and viewfronts, followed by the advanced aspects.
The semantics of consume-reads is described in
\ifext{Appendix~\ref{sec:consume}}{the appendix of the accompanying
  extended version}.

\subsection{Language syntax and basic reduction rules}
\label{sec:syntax}

\begin{figure}[t]\small
\[\begin{array}{rcl}
\Expr   & ::= & \vName \mid z~(\in \Number) \mid \Expr_{1}~\op~\Expr_{2}
                \mid \Choice~\Expr_{1}~\Expr_{2} \\
        &     & \First{\Expr} \mid \Second{\Expr} \mid \Pair{\Expr_{1}}{\Expr_{2}} \mid \locVar\\
\op     & ::= & + \mid - \mid * \mid / \mid \% \mid \texttt{==} \mid \texttt{!=} \\
\locVar & ::= & \loc \mid \vName \\
\loc    & \text{---} & \text{location identifier} \\
\vName  & \text{---} & \text{local variable}\\ 
\mvalSubst & ::= & \loc \mid z \mid \Pair{\mvalSubst_{1}}{\mvalSubst_{2}} \\  
\\
\AST & ::=  & \Ret{\Expr} \mid \Bind{\vName}{\AST_{1}}{\AST_{2}} \mid
              \Spw{\AST_{1}}{\AST_{2}}  \mid\\
     &      & \IfThenElse{\Expr}{\AST_{1}}{\AST_{2}} \mid
              \Repeat{\AST} \mid \\
     &      & \Read{\RM}{\locVar} \mid
              \Write{\WM}{\locVar}{\Expr} 
              \mid \Cas{SM}{FM}{\locVar}{\Expr_{1}}{\Expr_{2}} \\
\RT & ::=  & \Stuck \mid \Par{\AST_{1}}{\AST_{2}}\\
              \\
\RM   & ::= & \seqCstM | \acqM | \conM | \rlxM | \naM \\
\WM   & ::= & \seqCstM | \relM | \rlxM | \naM \\
\SM   & ::= & \seqCstM | \relAcqM | \relM | \acqM | \conM | \rlxM \\
\FM   & ::= & \seqCstM | \acqM | \conM | \rlxM
\end{array}\]
\caption{Syntax of statements and expressions.}
\label{fig:syntax}
\end{figure}

The syntax of the core language is presented in
Figure~\ref{fig:syntax}.
The meta-variable $\Expr$ ranges over expressions, which might be
integer numbers $z$, location identifiers $\loc$, (immutable) local
variables $\vName$, pairs, selectors and binary operations. The random
\lstinline{choice} operator, which non-deterministically returns one
of its arguments.
At the moment, arrays or pointer arithmetics are not supported.

Programs are statements, represented by terms $\AST$, most of which
are standard.
As customary in operational semantics, the result of a fully evaluated
program is either a value $\mvalSubst$ or the \emph{run-time} $\Stuck$
statement, which denotes the result of a program that ``went wrong''.
For instance, it is used to indicate all kinds of undefined behavior,
\eg, data races on \emph{non-atomic} operations or reading from
non-initialized locations.
The $\Spw{\AST_{1}}{\AST_{2}}$, when reduced, spawns two threads with
subprograms $\AST_{1}$ and $\AST_{2}$ respectively, emitting the
\emph{run-time} statement $\Par{\AST_{1}}{\AST_{2}}$, which is
necessary for implementing dynamic viewfront allocation for the newly
forked threads, as will be described below. In our examples, we will
use the parallel composition operator \textsf{||} for both
\lstinline{spw} and \lstinline{par} statements.

A binding statement $\Bind{\vName}{\AST_{1}}{\AST_{2}}$ implements
sequential composition by means of substituting $\AST_{1}$ in
$\AST_{2}$ for all occurrences of $\vName$. Location-manipulating
statements include reading from a location $(\Read{RM}{\locVar})$,
writing $(\Write{WM}{\locVar}{\Expr})$, and compare-and-set on a
location $\locVar$
($\Cas{SM}{FM}{\locVar}{\Expr_{1}}{\Expr_{2}}$). These statements are
annotated with order modifiers. We will sometimes abbreviate
\lstinline{r1 = |$[$|x|$]_{\text{\textsf{RM}}}$|; |$[$|y|$]_{\text{\textsf{WM}}}$| := r2}
as
\lstinline{|$[$|y|$]_{\text{\textsf{WM}}}$| := |$[$|x|$]_{\text{\textsf{RM}}}$|}
and \lstinline{r1 = s; r1} as \lstinline{r1 = s}.






Meta-variable $\auxX$ ranges over dynamic environments, defined
further.
Evaluation of a program $\AST$ in the semantics starts with the
initial state $\StatePair{\AST}{\auxX_{init}}$, where $\auxX_{init}$
contains an empty history, and an empty viewfront for the only initial
thread.
The semantics is defined in reduction style
\cite{Felleisen-Hieb:TCS92}, with most of its rules of the form
\[\prooftree
...
----------------------------------{}
\StatePair{\EvalContext[\AST]}{\auxX} ==> \StatePair{\EvalContext[\AST']}{\auxX'}
\endprooftree \]
where $\EvalContext$ is a reduction context, defined as follows: 
{\small
{
\[\begin{array}{rcl}
\EvalContext   & ::= & \hole
                         \mid \Bind{\vName}{\EvalContext}{\AST}  \mid \Par{\EvalContext}{\AST}
                       \mid \Par{\AST}{\EvalContext} \\  
\end{array}\]
}}
\hspace{-5pt}
If there is more than one thread currently forked, \ie, the program
expression contains a \lstinline{par} node, its statement might be
matched against $\EvalContext[\AST]$ in multiple possible ways
non-deterministically.
%
%

\begin{figure}[t]
\centering
{\small{
\[
\begin{array}{c}
\prooftree
\auxX' = \spawn{\EvalContext}{\auxX}
----------------------------------{Spawn}
\angled{\EvalContext[\Spw{\AST_{1}}{\AST_{2}}], \auxX} ==>
\angled{\EvalContext[\Par{\AST_{1}}{\AST_{2}}], \auxX'}
\endprooftree
\\\\
\prooftree
\auxX' = \joinP{\EvalContext}{\auxX}
----------------------------------{Join}
\angled{\EvalContext[\Par{\Ret{\mvalSubst_{1}}}{\Ret{\mvalSubst_{2}}}], \auxX} ==>
\angled{\EvalContext[\Ret{\Pair{\mvalSubst_{1}}{\mvalSubst_{2}}}], \auxX'}
\endprooftree
\end{array}
\]
}}  
\caption{Generic rules for spawning and joining threads}
\label{fig:spawn}
\end{figure}

The core rules of our semantics, involving non-memory operations, are
standard and are presented in
\ifext{Appendix~\ref{sec:appSemanticsRules}}{the appendix of the
  extended version of the paper (available in supplementary
  material)}.  The only interesting rules are for spawning and joining
threads (Figure~\ref{fig:spawn}), as they alter the thread-related
information in the environment (\eg, the viewfronts). The exact shape
of these rules depends on the involved concurrency aspects, which
define the meta-functions \textsf{spawn} and \textsf{join}.


\subsection{Histories and Viewfronts}
\label{sec:histvf}

\begin{figure}[t]\small
\[\begin{array}{lrcl}
\text{State} & \auxX      & ::= & \angled{\stEta, \stPsiRead} \\
\text{History} &\stEta     & ::= & (\loc, \stTau) \prarrow \angled{\mvalSubst, \stSigma}\\ 
\text{Viewfront function~~~~~} & \stPsiRead & ::= & \stpath \prarrow \stSigma\\
\text{Viewfront} & \stSigma   & ::= & \loc \prarrow \stTau\\
\text{Thread path} & \stpath    & \text{---} & (l|r)^{*}\\
\text{Timestamp} & \stTau \in \mathbb{N}   & \text{---} & \text{timestamp}
\end{array}\]
\caption{States, histories and viewfronts.}
\label{fig:auxXrelAcq}
\end{figure}

In its simplest representation, the program environment $\auxX$ is a
pair, whose components are a history $\stEta$ and per-thread viewfront
function $\stPsiRead$, defined in Figure~\ref{fig:auxXrelAcq}.

A history $\stEta$ is a partial function from location identifiers
$\loc$ and timestamps $\stTau$ to pairs of a stored value
$\mvalSubst$ and a synchronization front $\stSigma$.
Further aspects of our semantics feature different kinds of fronts,
but all of them have the same shape, mapping memory locations to
timestamps.
Per-thread viewfront function $\stPsiRead$ maps \emph{thread paths}
($\stpath$) to viewfronts. A thread path is a list of directions
($l|r$), which shows how to get to the thread subexpression inside a
program statement tree through the \lstinline{par} nodes: it uniquely
identifies a thread in a program statement.  We use an auxiliary
function \textsf{path} in the rules to calculate a path from an
evaluation context $\EvalContext$.

Once threads are spawn, they inherit a viewfront of their parent
thread, hence the simplest \textsf{spawn} function is defined as
follows:
\[
{\small{
\spawn{\EvalContext}{\angled{\AST,\stPsiRead}} \triangleq
\angled{\AST, \stPsiRead
[\stpath\:l \mapsto \stSigmaRead,~\stpath\:r \mapsto \stSigmaRead]}
}}\]
where $\stpath = \Path(\EvalContext)$, and $\stSigmaRead = \stPsiRead(\stpath)$.
\noindent
When threads join, their parent thread gets a viewfront, which is the
least upper bound (join) of subthread viewfronts:
$$
\joinP{\EvalContext}{\angled{\AST,\stPsiRead}} =
\angled{\AST, \stPsiRead[\stpath \mapsto \stSigmaRead^{l} \join \stSigmaRead^{r}]}
$$
where $\stpath = \Path(\EvalContext)$,
$\stSigmaRead^{l} = \stPsiRead(\stpath\:l)$, and
$\stSigmaRead^{r} = \stPsiRead(\stpath\:r)$, thus, synchronizing the
children threads' views.

We can now define the first class of ``wrong'' behaviors,
corresponding to reading from non-initialized locations
(Figure~\ref{fig:uninit-stuckRules}).  
The rules are applicable in the case when a thread tries to read from
a location, which it knows nothing about, \ie, its viewfront is not yet
defined for the location, making it uninitialized from the thread's
point of view.  This condition is also satisfied if the location is not
initialized at all, \ie, it has no corresponding records in the
history.

\begin{figure}[t]\small
\centering
\[\arrayBlock{
\prooftree
\arrayBlock{
  \auxX = \angled{\stEta, \stPsiRead} \quad \stpath = \Path(\EvalContext) \quad
  \stSigmaRead = \stPsiRead(\stpath) \quad \stSigmaRead(\loc) = \bot
}
----------------------------------{Read-Uninit}
\angled{\EvalContext[\Read{\RM}{\loc}], \auxX} ==> 
\angled{\lstmath{stuck}, \auxX_{init}}
~~~~~
\arrayBlock{
  \auxX = \angled{\stEta, \stPsiRead} \quad \stpath = \Path(\EvalContext) 
  \quad
  \stSigmaRead = \stPsiRead(\stpath) \quad \stSigmaRead(\loc) = \bot
}
----------------------------------{CAS-Uninit}
\angled{\EvalContext[\Cas{\SM}{\FM}{\loc}{\Expr_{1}}{\Expr_{2}}], \auxX} ==> 
\angled{\lstmath{stuck}, \auxX_{init}}
\endprooftree
}\]
\caption{Rules for reading from an uninitialized location.}
\label{fig:uninit-stuckRules}
\end{figure}

\subsection{Release/Acquire synchronization}
\label{sem:relAcqOps}

\begin{figure}[t]
\centering
{\small{
\[\arrayBlock{
\prooftree
\arrayBlock{
  \auxX = \angled{\stEta, \stPsiRead} \quad \stpath = \Path(\EvalContext) \quad
  \stTau = \NextTau{\stEta}{\loc} \\
  \stSigmaRead = \stPsiRead(\stpath) \quad
  \stSigma = \stSigmaRead[\loc \mapsto \stTau] \\
  \auxX' = \angled{\stEta[(\loc, \stTau) \mapsto (\mvalSubst, \stSigma)],
                   \stPsiRead[\stpath \mapsto \stSigma]}
} 
----------------------------------{WriteRel}
\angled{\EvalContext[\Write{\relM}{\loc}{\mvalSubst}], \auxX} ==> 
\angled{\EvalContext[\Ret{\mvalSubst}], \auxX'}
~~~~~
\arrayBlock{
  \auxX = \angled{\stEta, \stPsiRead} \quad \stpath = \Path(\EvalContext) \quad 
  \stEta(\loc, \stTau) = (\mvalSubst, \stSigma) \\
  \stSigmaRead = \stPsiRead(\stpath) \quad
  \stSigmaRead(\loc) \leq \stTau \\
  \auxX' = \angled{\stEta, \stPsiRead[\stpath \mapsto \stSigmaRead \lub \stSigma]}
}
----------------------------------{ReadAcq}
\angled{\EvalContext[\Read{\acqM}{\loc}], \auxX} ==> 
\angled{\EvalContext[\Ret{\mvalSubst}], \auxX'}
\endprooftree
}\]
}}
\caption{Reduction rules for release/acquire atomics.}
\label{fig:rel/acq-sem}
\end{figure}

The reduction rules for release-write and acquire-read
are given in Figure~\ref{fig:rel/acq-sem}.  
A release-write augments the history with a new entry
$(\loc, \stTau) \mapsto (\mvalSubst, \stSigma)$,
where $\mvalSubst$ is a value argument of the write, and $\stSigma$ is
a synchronization front, which now might be retrieved by the threads
reading from the new history entry. 
The stored front $\stSigma$ is the same as the viewfront of the writer
thread after having stored the value, \ie, featuring updated
$\loc$-entry with the new timestamp $\stTau$.

An acquire-read is more interesting. It
\emph{non-deterministically} chooses, from the global history, an
entry
$(\loc, \stTau) \mapsto (\mvalSubst, \stSigma)$, with a timestamp
$\stTau$ which is at least a new as the timestamp $\stTau'$ for the
corresponding location $\loc$ in the thread's viewfront
$\stSigmaRead$ (\ie, $\stTau' \le \stTau$).
The value $\mvalSubst$ replaces the read expression inside the context
$\EvalContext$, and the thread's local viewfront $\stSigmaRead$ is
updated via the retrieved synchronization front $\stSigma$.
The RA-CAS operations (rules omitted for brevity) behave similarly
with only difference: a successful CAS reads from the \emph{latest}
entry in the history. 
%



\subsection{SC operations}
\label{sec:scop}

\begin{figure}[t]\small
\centering
\[\arrayBlock{
\prooftree
\arrayBlock{
  \auxX = \angled{..., \stSC} \quad ... \quad
  \auxX' = \angled{..., \stSC[\loc \mapsto \stTau]}
} 
----------------------------------{WriteSC}
\angled{\EvalContext[\Write{\seqCstM}{\loc}{\mvalSubst}], \auxX} ==> 
\angled{\EvalContext[\mvalSubst], \auxX'}
~~~~~
\arrayBlock{
  \auxX = \angled{..., \stSC} \quad ... \quad
  \mathsf{max}(\stSigmaRead(\iota), \stSC(\iota)) \leq \stTau
}
----------------------------------{ReadSC}
\angled{\EvalContext[\Read{\seqCstM}{\loc}], \auxX} ==> 
\angled{\EvalContext[\Ret{\mvalSubst}], \auxX'}
\endprooftree
}\]
\caption{Reduction rules for SC atomics.}
\label{fig:sc-sem}
\end{figure}


To account for SC operations, we augment $\auxX$ with $\seqCstM$-front:
%
%
\[
\begin{array}{rcl}
\auxX      & ::= & \angled{..., \stSC}
\end{array}\]
%
%
that maps each location to a timestamp of the \emph{latest} entry in
the location history, which has been added by a SC-write.

The SC operations update the history and local/stored fronts similarly
to RA atomics. In addition, an SC-write updates $\seqCstM$-front, and
an SC-read introduces an additional check for the timestamp $\stTau$,
taking $\max$ of two viewfronts, as defined in
Figure~\ref{fig:sc-sem}.\footnote{Read and write rules only depict 
  difference with the {release}/{acquire} ones.}
That is, the rule~\textsc{ReadSC} ensures that an SC-read gets a
history entry, which is not older than the one added by the last SC
write to the location.
%

\subsection{Non-atomic operations}

\begin{figure}[t]\small
\centering
\[\arrayBlock{
\prooftree
\arrayBlock{
  \auxX = \angled{\stEta, \stPsiRead, ..., \stNA} \quad ... \\
  \stSigmaRead(\loc) == \LastTau{\stEta}{\loc} \quad
  \stSigma = \stSigmaRead[\loc \mapsto \stTau] \\[2pt]
  \auxX' = \angled{\stEta[(\loc, \stTau) \mapsto (\mvalSubst, \stSigmaEmpty)],
                   \stPsiRead[\stpath \mapsto \stSigma], ...,
                   \stNA[\loc \mapsto \stTau]}
} 
----------------------------------{WriteNA}
\angled{\EvalContext[\Write{\naM}{\loc}{\mvalSubst}], \auxX} ==> 
\angled{\EvalContext[\Ret{\mvalSubst}], \auxX'}
~~~~~
\arrayBlock{
  \auxX = \angled{\stEta, \stPsiRead, ..., \stNA} \quad ... \\[2pt]
  \stTau = \LastTau{\stEta}{\loc} \quad
  \stTau == \stSigmaRead(\loc)  \quad
  \stEta(\loc, \stTau) = (\mvalSubst, \stSigma)\\[2pt]
}
----------------------------------{ReadNA}
\angled{\EvalContext[\Read{\naM}{\loc}], \auxX} ==> 
\angled{\EvalContext[\Ret{\mvalSubst}], \auxX}
~~~~~
\arrayBlock{
  \auxX = \angled{\stEta, \stPsiRead, ..., \stNA}
  \quad ... \quad 
  \stSigmaRead(\loc) \neq \LastTau{\stEta}{\loc}\\[2pt]
}
----------------------------------{ReadNA-stuck1}
\angled{\EvalContext[\Read{\naM}{\loc}], \auxX} ==> 
\angled{\Stuck, \auxX}
~~~~~
\arrayBlock{
  \auxX = \angled{\stEta, \stPsiRead, ..., \stNA}
  \quad ... \quad
  \stSigmaRead(\loc)~<~\stNA(\loc)
}
----------------------------------{ReadNA-stuck2}
\angled{\EvalContext[\Read{\RM}{\loc}], \auxX} ==> 
\angled{\Stuck, \auxX}
\endprooftree
}\]
\caption{Reduction rules for non-atomics.}
\label{fig:na-sem}
\end{figure}

For non-atomic accesses we augment $\auxX$ with $\naM$-front:
%
%
\[\begin{array}{rcl}
\auxX      & ::= & \angled{..., \stNA}
\end{array}\]
%
%
Similarly to $\seqCstM$-front, it maps a location to the latest
corresponding $\naM$-entry in the history, and it is updated by
NA-writes.

The real purpose of $\naM$-front is to detect \emph{data races}
involving NA-operations as defined in Figure~\ref{fig:na-sem}.  
When a thread performs NA-write or NA-read (see \ruleF{WriteNA},
\ruleF{ReadNA} rules), it must be aware of the latest stored record of
the location (\ie, it should match the timestamp in its local front
$\stSigmaRead$). Violating this side condition is condemned to be a
data race and leads to undefined behavior (\ruleF{ReadNA-stuck1}).
In addition, if a thread performs \emph{any} write to or read from a
location, it should be aware of the latest NA-record to the location
(see the side condition $\stSigmaRead(\loc)~<~\stNA(\loc)$ in the rule
\ruleF{ReadNA-stuck2}).%
\footnote{Rules \ruleF{WriteNA-stuck1} and \ruleF{WriteNA-stuck2} are
  similar and can be found in
  \ifext{Appendix~\ref{sec:appSemanticsRules}}{the appendix of the
    extended version of the paper}.}
%
The stuck-cases reflect the cases when a write or a read is in data
race with the last NA-write to the location.

Unlike release or \emph{SC-}writes, NA-writes do not store a
front to the history entry, as they cannot be used for
synchronization. A similar fact holds for NA-reads: they do not get a
stored front from the history entry, upon reading from~it.


\subsection{Release-sequences and write-fronts}
\label{sec:wrf}

\begin{figure}[t]
\centering
{\small{
\[\arrayBlock{
\prooftree
\arrayBlock{
  \auxX = \angled{\stEta, \stPsiRead, ...} \quad ... \\
  \auxX' = \angled{\stEta,
    \stPsiRead[\stpath \mapsto \stSigmaRead[\loc \mapsto \stTau]], ...}
}
----------------------------------{ReadRlx}
\angled{\EvalContext[\Read{\rlxM}{\loc}], \auxX} ==> 
\angled{\EvalContext[\Ret{\mvalSubst}], \auxX'}
~~~~~
\arrayBlock{
  \auxX = \angled{\stEta, \stPsiRead, ..., \stPsiWrite}
  \quad ... \\
  \stTau_{rel} = \stPsiWrite(\stpath)(\loc) \quad
  (\_, \stSigmaSync) = \stEta(\loc, \stTau_{rel}) \\
  \auxX' =
  \langled{\stEta[(\loc, \stTau) \mapsto
          (\mvalSubst, \stSigmaSync[\loc \mapsto \stTau])],}\\
   \rangled{\stPsiRead[\stpath \mapsto
                       \stSigmaRead[\loc \mapsto \stTau]],
     ..., \stPsiWrite}
} 
----------------------------------{WriteRlx}
\angled{\EvalContext[\Write{\rlxM}{\loc}{\mvalSubst}], \auxX} ==> 
\angled{\EvalContext[\Ret{\mvalSubst}], \auxX'}
~~~~~
\arrayBlock{
  \auxX = \angled{\stEta, \stPsiRead, ..., \stPsiWrite}
  \quad ... \\
  \stSigmaWrite = \stPsiWrite(\stpath) \quad
  \stSigmaWriteNew = \stSigmaWrite[\loc \mapsto \stTau] \\
  \auxX' = \angled{..., \stPsiWrite[\stpath \mapsto \stSigmaWriteNew]}
} 
----------------------------------{WriteRel'}
\angled{\EvalContext[\Write{\relM}{\loc}{\mvalSubst}], \auxX} ==> 
\angled{\EvalContext[\Ret{\mvalSubst}], \auxX'}
\endprooftree
}\]
}}
\caption{Reduction rules for relaxed atomics.}
\label{fig:rlx-sem}
\end{figure}

{Relaxed reads} do not update
their thread's viewfront with a synchronization front from the history
(see \ruleF{ReadRlx} rule in Figure~\ref{fig:rlx-sem}). At this stage,
their support does not require augmenting the state.


An additional instrumentation is required, though, to encode release
sequences.
As discussed in
Section \ref{sec:rlxAdvanced}, an acquire-read, when reading
the result of a relaxed write, might get synchronized with a
release-write to the same location performed earlier by the
same writer thread.
To account for this, we introduce per-thread write-front function
$\stPsiWrite$ as an environment component:
\[\begin{array}{rcl}
\auxX      & ::= & \angled{..., \stPsiWrite}
\end{array}\]
It is similar to the viewfront function $\stPsiRead$, but it stores a
timestamp of the last release-write to a location by the
thread.
Specifically, when a thread performs a relaxed write $W$ (see
rule \ruleF{WriteRlx} in Figure~\ref{fig:rlx-sem}), it checks if there
was a release-write $W'$ performed by it earlier, takes a
synchronization front $\stSigmaSync$ from the history entry, added by
$W'$, and stores it as the synchronization front in the \emph{new}
history entry.\footnote{If $H(\loc, \stTau_{rel}) = \bot$ then
  $\stSigmaSync = \bot$.} 
Additionally, we need to modify the old rules \ruleF{WriteRel},
\ruleF{WriteSC}, \ruleF{CAS-Rel}, \etc., so they update $\stPsiWrite$
correspondingly (\eg, see rule \ruleF{WriteRel'} in
Figure~\ref{fig:rlx-sem}).

We also need to change our meta-functions, in order to account for the
$\stPsiWrite$ component of the state environment:
{\small{
\[
\begin{array}{rcl}
\spawn{\EvalContext}{\angled{...,\stPsiWrite}} &\!\!=\!\!&
\angled{..., \stPsiWrite
[\stpath\:l \mapsto \bot,~\stpath\:r \mapsto \bot]} 
\\
\joinP{\EvalContext}{\angled{..., \stPsiWrite}} &\!\!=\!\!&
\angled{..., \stPsiWrite[\stpath \mapsto \bot]}
\end{array}
\]
}}

Subthreads do not inherit write-fronts upon spawning, and a parent
thread does not inherit the joined one, since the described
synchronization effects via relaxed writes and {release sequences} do
not propagate through spawn/join points according to the
model~\cite{Batty-al:POPL11}.

\subsection{Postponed operations and speculations}
\label{sec:postponed-sem}

To support postponed operations (or, equivalently, speculative
executions)
we instrument the state with two additional components:
\[\begin{array}{rcl}
\auxX             & ::=        & \angled{..., \stPhi, \stGamma} \\
\end{array}\]

The main one, $\stPhi$, is a function
that maps a thread path $\stpath$ to a per-thread hierarchical
{buffer} $\stAlpha$ of {postponed operations} $\stPostOp$:
{\small{
\[\begin{array}{rcl}
    \stPhi            & ::=        & \stpath \prarrow \stAlpha \\
    \stAlpha          & ::=        & \stPostOp^{*} \\
    \stPostOp         & ::=        & read\angled{\vName,\locVar,\RM} \mid write\angled{\vName,\locVar,\WM,\Expr} \mid bind\angled{\vName,\Expr} \mid if\angled{\vName,\Expr,\stAlpha,\stAlpha}\\
\end{array}\]
}}
\hspace{-5pt}
Each operation $\stPostOp$ is uniquely identified by its symbolic
value $\vName$.
Read entries contain a (possibly unresolved) location $\locVar$ to
read from as well as a read modifier $\RM$. Write entries additionally
contain an expression $\Expr$ to be stored to the
location. \emph{Bind} entries are used to postpone calculation of an
expression depending on a symbolic value, making it possible to
postpone the reads as follows:

{\small
\begin{center} \begin{tabular}{c}
\begin{lstlisting}
r1 = |$[$|x|$]_{rlx}$|; r2 = r1 + 1; ...
\end{lstlisting}
\end{tabular} \end{center}
}
\vspace{5pt}
\noindent
Both reads \lstinline{r1} and bind \lstinline{r2} might be postponed,
so the second statement will not ``trigger'' evaluation of the first
one.
\emph{If}-entries have a conditional expression $\Expr$ and two
\emph{subbuffers} $\stAlpha$ representing operations speculatively put
to the buffer under \emph{then} and \emph{else} branches.  To
represent speculation under (possibly nested) \lstinline{if} statement
we define an if-specialized reduction context $\EvalSpecContext$ as
follows:
{\small{
\[\begin{array}{rcl}
\EvalSpecContext & ::= & \hole \mid
                         \Bind{\vName}{\EvalSpecContext}{\AST} \mid
    \IfThenElse{\vName}{\EvalSpecContext}{\AST}~\mid \\
&& \IfThenElse{\vName}{\AST}{\EvalSpecContext}
\end{array}\]
}}
\hspace{-5pt}
where the symbolic value $\vName$ in the condition is the same as in
the corresponding buffer entry
$if\angled{\vName,\stAlpha_1, \stAlpha_2}$.  The list of symbolic
values from conditions of the context can be used to uniquely identify
an operation buffer inside an hierarchical per-thread buffer $\stAlpha$.
%


%

The list $\stGamma$ encodes \emph{Acquire-Read Restrictions} by
containing triples $\angled{\loc, \stTau, \vName}$, forbidding to
acquire-read from $(\loc, \stTau)$ until $\vName$ is not resolved.
%


During a thread execution, any read, write, or bind operation
(including those under not fully reduced if-branches) can be postponed
by the semantics by adding a corresponding record into the matching
subbuffer of the thread buffer $\stAlpha$.
%
For the sake of brevity we discuss only write rules here; other rules can
be found in the appendix%
\ifext{}{ of the extended version} 
and in our implementation.


The postpone-rules append operation records into the corresponding
buffer $\stPhi$. An operation to be postponed can be nested under a
not fully reduced if statement.

\begin{center} {\small{
\[\arrayBlock{
\prooftree
\arrayBlock{
  \auxX = \angled{..., \stPhi, \stGamma} \quad ... \\
  \vName \text{ is fresh symbolic variable} \quad
  \stAlpha = \stPhi(\stpath) \\
  \auxX' = \angled{...,
\stPhi[\stpath \mapsto \AppendAlpha(\stAlpha, \EvalSpecContext, write\angled{\vName, \locVar, \WM, \Expr})],
  \stGamma} \\
} 
----------------------------------{Write-Postpone}
\angled{\EvalContext[\EvalSpecContext[\Write{\WM}{\locVar}{\Expr}]], \auxX} ==> 
\angled{\EvalContext[\EvalSpecContext[\Ret{\vName}]], \auxX'}
\endprooftree
}\] }}
\end{center}

After postponing, a write-record has to be resolved eventually: 
\begin{center} {\small{
\[\arrayBlock{
\prooftree
\arrayBlock{
  \auxX = \angled{\stEta, ... ,\stPsiWrite, \stPhi, \stGamma} \quad ... \\
  write\angled{\vName, \loc, \WM, \mvalSubst} \in \stPhi(\stpath)\;\text{and not in conflict} \\
  \stPhi' = \remove(\stPhi, \stpath, \vName) \\ 
 \stGamma' = \updateDep(\vName, \WM, \stPsiWrite, \loc, \stTau, \stGamma, \stPhi(\stpath))\\ 
 \stEta' = \updateSync(\vName, \WM, \stSigma, \stGamma, \stEta) \\
 \auxX' = \angled{\stEta'[(\loc, \stTau) \smapsto (\mvalSubst, \stSigma)], ...,
\stPhi'[\mvalSubst/\vName], \stGamma'} \\
} 
----------------------------------{Write-Resolve}
\angled{\AST, \auxX} ==> 
\angled{\AST[\mvalSubst/\vName], \auxX'}
\endprooftree
}\]
}}
\end{center}

\noindent
We point out several important side conditions of the rule.
The write must be in the top-level buffer $\stPhi(\stpath)$,
and there must be no operation \emph{before} it in the
buffer, which is in conflict with it, \eg, an
acquire-read or a write to the same location {(line 2)}.
%
%
%
%
In line 4, $\updateDep$ updates $\stGamma$ as follows.
First, it removes from $\stGamma$ all entries that mention the
symbolic value $\vName$, $\angled{\loc',\stTau',\vName}$, since the
write is resolved.  
Second, for specific postponed operations $\vName'$, it adds
$\angled{\loc,\stTau,\vName'}$ entries to $\stGamma$, blocking
acquire-reads from the newly created history entry $(\loc, \stTau)$
until $\vName'$ are resolved.
These are the operations that can affect an acquire-read from
$(\loc, \stTau)$: \emph{(i)} ones related by $\stGamma$ to
$(\loc, \stPsiWrite(\stpath, \loc))$ (\ie, a record of the last
release-write), and \emph{(ii)}, if the resolved write $\vName$ is a
release one, operations, that precede the write in the thread buffer
$\stPhi(\stpath)$ or are unresolved writes observed by thread (\ie,
elements of $\stObservedWrites(\stpath)$).
In line 6, $\updateSync$ updates synchronization fronts in history
records $(\loc',\stTau')$ related to $\vName$ by $\stGamma$.

Duplicated non-conflicting writes from nested buffers can be
(non-deterministically) promoted to an upper-level:

\begin{center} {\small{
\[\arrayBlock{
\prooftree
\arrayBlock{
  \auxX = \angled{\stEta, \stPsiRead, ..., \stPhi, \stGamma} \quad ... \\
  if\angled{\vName'', \Expr, \stAlpha_1, \stAlpha_2 }\;\text{is inside}\;\stPhi(\stpath)\\
  write\angled{\vName , \loc, \WM, \mvalSubst} \in \stAlpha_1 \quad
  write\angled{\vName', \loc, \WM, \mvalSubst} \in \stAlpha_2 \\
  \vName, \vName'\;\text{are not in conflict in}\;\stAlpha_1, \stAlpha_2\\
  \stPhi' = \promote(\vName, \vName', \stPhi) \quad \stGamma' = \stGamma[\vName'/\vName]
} 
----------------------------------{Write-Promote}
\angled{\AST, \auxX} ==> \angled{\AST[\vName'/\vName], \auxX'}
\endprooftree
}\]
}}
\end{center}
Line 3 of the rule's premise requires two identical writes to be
present in the ``sibling'' buffers. 
In line 4, $\promote$ removes the writes from $\stAlpha_1$ and
$\stAlpha_2$, and puts $write\angled{\vName , \loc, \WM, \mvalSubst}$
in the parent buffer before
$if\angled{\vName'', \Expr, \stAlpha'_1, \stAlpha'_2}$.



The rule for initialization of speculative execution of branches of an
\lstinline{if}-statement adds
$if\angled{\vName, \Expr, \angled{}, \angled{}}$ into $\stAlpha$,
similarly to postponing a write, and replaces the condition with a
symbolic value:
{\small{
\[\arrayBlock{
\prooftree
\arrayBlock{
  \auxX = \angled{..., \stPhi, \stGamma} \quad ... \\
  \Expr\;\text{depends on an unresolved symbolic value} \\
  \vName \text{ --- fresh symbolic variable} \quad
  \stAlpha = \stPhi(\stpath) \\
  \auxX' = \angled{...,
\stPhi[\stpath \mapsto \AppendAlpha(\stAlpha, \EvalSpecContext,
  if\angled{\vName,\Expr, \angled{}, \angled{}})], \stGamma} \\
} 
----------------------------------{If-Speculation-Init}
\arrayBlock{
  \angled{\EvalContext[\EvalSpecContext[\IfThenElse{\Expr}{\AST_{1}}{\AST_{2}}]], \auxX} ==>\\ 
  \angled{\EvalContext[\EvalSpecContext[\IfThenElse{\vName}{\AST_{1}}{\AST_{2}}]], \auxX'} \\
} 
\endprooftree
}\] }}
%
\noindent
\hspace{-3pt}
Finally, upon resolving all symbolic values in the condition $\Expr$
of an \lstinline{if}-statement, the statement itself can be reduced:

{\small{
\[\arrayBlock{
\prooftree
\arrayBlock{
  \auxX = \angled{... , \stPhi, \stGamma} \quad ... \quad
  if\angled{\vName, z, \stAlpha_1, \stAlpha_2} \in \stPhi(\stpath) \quad
  z \neq 0\\
  \stPhi' = \stPhi[if\angled{\vName, z, \stAlpha_1, \stAlpha_2} / \stAlpha_1] \quad
  \auxX' = \angled{..., \stPhi', \stGamma}
} 
----------------------------------{If-Resolve-True}
\angled{\EvalContext[\EvalSpecContext[\IfThenElse{\vName}{\AST_{1}}{\AST_{2}}]], \auxX} ==> 
\angled{\EvalContext[\EvalSpecContext[\AST_{1}]], \auxX'}
\endprooftree
}\] }}

\section{Implementation and Evaluation}
\label{sec:summary}

We implemented our semantics in PLT Redex,\footnote{The sources are
  available as supplementary material for the paper.} a framework on
top of the Racket programming
language~\cite{Klein-al:POPL12,Felleisen-al:Redex}.
The implementation of the core language definitions
(Sections~\ref{sec:syntax}--\ref{sec:histvf}) is {2070~LOC}, with
various C11 concurrency aspects
(Sections~\ref{sem:relAcqOps}--\ref{sec:postponed-sem}) implemented on
top of them, in {1310~LOC}. Implementation of litmus tests
(Section~\ref{sec:summary}) and case studies
(Section~\ref{sec:rcu-example}) took {3130~LOC}.




\paragraph{Evaluation via Litmus Tests}
\label{sec:evaluation}

To ensure the adequacy of our semantics with respect to the C++11
standard~\cite{CPP:11} and gain confidence in its implementation, we
evaluated it on a number of litmus test programs from the
literature. For each test, we encoded the set of expected results and
checked, via extensive state-space enumeration, provided by PLT Redex,
that these are the only outcomes produced.

Figure~\ref{fig:litmusTbl} provides a table, relating specific litmus
tests from the
literature~\cite{Bornat-al:LACE,Batty-al:POPL11,Lahav-al:POPL16,Maranget-al:tutorial,Turon-al:OOPSLA14}
to the relevant aspects of our semantics from
Section~\ref{sec:semantics}, required in order to support their
desired behavior. All tests mentioned before in the paper are presented in the table.
%
%
Since there is no common naming conventions for litmus tests in a
{high-level language}, making consistent appearance in related papers,
we supplied ours with meaningful names, grouping them according to the
behavioral pattern they exercise (\eg, message-passing, store
buffering, \etc.). Exact definitions of the test program and
descriptions of their behaviors can be found in
\ifext{Appendix~\ref{sec:litmusTests}}{the appendix of the extended
  version of the paper in the supplementary material}.


All tests within the same group have a similar structure but differ in
memory access modifiers.
The columns \textsf{Hst}--\textsf{JN} in Figure~\ref{fig:litmusTbl}
show, which semantic aspects a test requires for its complete and
correct execution.  The last column indicates, whether the test's
behavior in our semantics matches fully its outcome according to the
C11 standard or not. Below, we discuss the tests that behave
differently in the C11 standard and in our semantics.

{
\setlength{\belowcaptionskip}{-20pt} 
\begin{figure}[h!]
\centering
{\small
\begin{tabular}{| l ||@{~}c@{~}|@{~}c@{~}|@{~}c@{~}|@{~}c@{~}|@{~}c@{~}|@{~}c@{~}|@{~}c@{~}|@{~}c@{~}||@{~}c@{~}|}
  \hline
  \textbf{Test name} & \textsf{VF} & \textsf{WF} & \textsf{SCF}
  & \textsf{NAF} & \textsf{PO} & \textsf{ARR} 
  & \textsf{CR} & \textsf{JN} & \textbf{C11} \\
%

\hline\hline
\multicolumn{10}{|c|}{Store Buffering (\textsf{SB})\ifext{, \S\ref{app:sb}}{}} \\
\hline
\textsf{rel+acq}   & \tick & &       & & & & & & \tick\\ 
\textsf{sc}        & \tick & & \tick & & & & & & \tick\\ 
\textsf{sc+rel}    & \tick & & \tick & & & & & & \tick\\ 
\textsf{sc+acq}    & \tick & & \tick & & & & & & \tick\\ 

\hline
\multicolumn{10}{|c|}{Load Buffering (\textsf{LB})\ifext{, \S\ref{app:lb}}{}} \\
\hline
\textsf{rlx}         & \tick & & & & \tick & & & & \tick\\ 
\textsf{rel+rlx}     & \tick & & & & \tick & & & & \tick\\ 
\textsf{acq+rlx}     & \tick & & & & \tick & & & & \fail\\ 
\textsf{rel+acq+rlx} & \tick & & & & \tick & \tick & & & \tick\\ 
\textsf{rlx+use}     & \tick & & & & \tick & & & & \tick\\ 
\textsf{rlx+let}     & \tick & & & & \tick & & & & \tick\\ 
\textsf{rlx+join}    & \tick & & & & \tick & & & \tick & \tickP\\ 
\textsf{rel+rlx+join} & \tick & & & & \tick & & & \tick & \tickP\\ 
\textsf{acq+rlx+join} & \tick & & & & \tick & & & \tick & \fail\\ 

\hline
\multicolumn{10}{|c|}{Message passing (\textsf{MP})\ifext{, \S\ref{app:mp}}{}} \\
\hline
\textsf{rlx+na}            & \tick &       & & \tick & & &       & & \tick\\ 
\textsf{rel+rlx+na}        & \tick &       & & \tick & & &       & & \tick\\ 
\textsf{rlx+acq+na}        & \tick &       & & \tick & & &       & & \tick\\ 
\textsf{rel+acq+na}        & \tick &       & & \tick & & \tick & & & \tick\\ 
\textsf{rel+acq+na+rlx(\_2)} & \tick & \tick & & \tick & & \tick & & & \tick\\ 
\textsf{con+na(\_2)}       & \tick &       & & \tick & & & \tick & & \tick\\ 
\textsf{cas+rel+acq+na}    & \tick &       & & \tick & & \tick & & & \tick\\ 
\textsf{cas+rel+rlx+na}    & \tick &       & & \tick & & &     & & \tick\\ 

\hline
\multicolumn{10}{|c|}{Coherence of Read-Read (\textsf{CoRR})\ifext{, \S\ref{app:corr}}{}} \\
\hline
\textsf{rlx}      & \tick & &       & & &  & & & \tick\\ 
\textsf{rel+acq}  & \tick & &       & & &  & & & \tick\\ 

\hline
\multicolumn{10}{|c|}{Independent Reads of Independent Writes (\textsf{IRIW})\ifext{, \S\ref{app:iriw}}{}} \\
\hline
\textsf{rlx}      & \tick & &       & & &  & & & \tick\\ 
\textsf{rel+acq}  & \tick & &       & & &  & & & \tick\\ 
\textsf{sc}       & \tick & & \tick & & &  & & & \tick\\ 

\hline
\multicolumn{10}{|c|}{Write-to-Read Causality (\textsf{WRC})\ifext{, \S\ref{app:wrc}}{}} \\
\hline
\textsf{rlx}      & \tick & &       & & & &  & & \tick\\ 
\textsf{rel+acq}  & \tick & &       & & & &  & & \tick\\ 
\textsf{cas+rel}  & \tick & &       & & & \tick & & & \tick\\ 
\textsf{cas+rlx}  & \tick & &       & & & &  & & \tick\\ 

\hline
\multicolumn{10}{|c|}{Out-of-Thin-Air (\textsf{OTA})\ifext{, \S\ref{app:ota}}{}} \\
\hline
\textsf{lb}       & \tick & &       & & \tick & & & & \fail\\ 
\textsf{if}       & \tick & &       & & \tick & & & & \fail\\ 

\hline
\multicolumn{10}{|c|}{Write Reorder (\textsf{WR})\ifext{, \S\ref{app:wr}}{}} \\
\hline
\textsf{rlx}      & \tick & &       & & \tick & & & & \tick\\ 
\textsf{rlx+rel}  & \tick & &       & & \tick & \tick & & & \tick\\ 
\textsf{rel}      & \tick & &       & & \tick & \tick & & & \tick\\ 


\hline
\multicolumn{10}{|c|}{Speculative Execution (\textsf{SE})\ifext{, \S\ref{app:se}}{}} \\
\hline
\textsf{simple}      & \tick & &       & & \tick & & & & \tick\\ 
\textsf{prop}        & \tick & &       & & \tick & & & & \tick\\ 
\textsf{nested}      & \tick & &       & & \tick & & & & \tick\\ 

  \hline
  \multicolumn{10}{|c|}{Locks\ifext{, \S\ref{app:locks}}{}} \\
  \hline
  Dekker & \tick & & & \tick & & & & & \tick\\ 
  Cohen~\cite{Turon-al:OOPSLA14}  & \tick & & & \tick & & & & & \tick\\ 


\hline

\end{tabular}
}
\caption{Litmus tests
  \ifext{(Appendix~\ref{sec:litmusTests})}{} and corresponding
  semantic aspects of our framework:
  {viewfronts}~(\textsf{VF},~\S\ref{sec:hist}),
  {write-fronts}~(\textsf{WF},~\S\ref{sec:wrf}),
  SC-fronts~(\textsf{SCF},~\S\ref{sec:sc}), {non-atomic
    fronts}~(\textsf{NAF},~\S\ref{sec:na}), {postponed
    operations}~(\textsf{PO},~\S\ref{sec:postponed-sem}),
  {acquire read restrictions
    ($\stGamma$)}~(\textsf{ARR},~\S\ref{sec:postponed-sem}),
  {consume-reads}~(\textsf{CR}), {joining threads with non-empty
    operation buffers}~(\textsf{JN},~\S\ref{sec:join}).  The column
  \textbf{C11} indicates whether the behavior is coherent with the C11
  standard.  }
\label{fig:litmusTbl}
\end{figure}
}

\paragraph{Discrepancies with the C11 standard}

The combining relaxed and acquire-writes,
\textsf{LB\_\{acq+rlx,acq+rlx+join\}}, in order to adhere to the
``canonical'' C11 behavior, require an ability to do an acquire-read
and a subsequent relaxed write out-of-order.
%
%
%
%
Even though the relaxed behavior of this kind is not supported by our
semantics, it is not observable under sound compilation schemes of
acquire-read to the major architectures~\cite{CompScheme}:
\textit{(i)} load buffering is not observable on x86 in
general~\cite{Sewell-al:CACM10}, \textit{(ii}) all barriers
(\emph{sync}, \emph{lwsync}, \emph{ctrl+isync}) forbid reorderings of
read and write on Power~\cite{Alglave-al:TOPLAS14},
\textit{(iii)} as well as barriers (\emph{dmb sy}, \emph{dmb ld},
\emph{ctrl+isb}) on ARM~\cite{Flur-al:POPL16}.

%


Our semantics rules out OTA behaviors (tests \textsf{OTA\_\{lb,
  if\}}), which are considered to be an issue of the
standard~\cite{Batty-al:ESOP15,Boehm-Demsky:MSPC14,PichonPharabod-Sewell:POPL16}.

\begin{figure*}[t]
{\small{
\textbf{Program:}
\vspace{-17pt}
\begin{lstlisting}[language=while]
                        |$[$|cw|$]_{na}$| := 0; |$[$|cr1|$]_{na}$| := 0; |$[$|cr2|$]_{na}$| := 0; |$[$|lhead|$]_{na}$| := null;
\end{lstlisting}
\begin{tabular}{l||l}
\begin{lstlisting}[language=while]
|$[$|a|$]_{rlx}$| := (1, null);
|$[$|ltail|$]_{na}$| := a;
|$[$|lhead|$]_{rel}$| := a;
append(b, 10 , ltail);
append(c, 100, ltail);
updateSecondNode(d, 1000)
\end{lstlisting}
\hspace{.6cm} &
\begin{lstlisting}[language=while]
\end{lstlisting}
\begin{tabular}{l||l}
\begin{lstlisting}[language=while]
|$[$|sum11|$]_{na}$| := 0;
rcuOnline (cw, cr1);
traverse  (lhead, cur1, sum11)
rcuOffline(cw, cr1);

|$[$|sum12|$]_{na}$| := 0;
rcuOnline (cw, cr1);
traverse  (lhead, cur1, sum12)
rcuOffline(cw, cr1);

r11 = |$[$|sum11|$]_{na}$|;
r12 = |$[$|sum12|$]_{na}$|
\end{lstlisting}
\hspace{.6cm} &
\begin{lstlisting}[language=while]
|$[$|sum21|$]_{na}$| := 0;
rcuOnline (cw, cr2);
traverse  (lhead, cur2, sum21)
rcuOffline(cw, cr2);

|$[$|sum22|$]_{na}$| := 0;
rcuOnline (cw, cr2);
traverse  (lhead, cur2, sum22);
rcuOffline(cw, cr2);

r21 = |$[$|sum21|$]_{na}$|;
r22 = |$[$|sum22|$]_{na}$|
\end{lstlisting}
\end{tabular}
\begin{lstlisting}[language=while]
\end{lstlisting}
\end{tabular}
\begin{lstlisting}[language=while]
\end{lstlisting}

\vspace{-5pt}
\hrule
\vspace{5pt}

\textbf{Functions:}

\begin{tabular}{l@{}l@{}l}

\begin{tabular}{@{}l@{}}
\begin{lstlisting}[language=while]
append(loc, value, ltail) |$\triangleq$| 
  |$[$|loc|$]_{rlx}$| := (value, null);
  rt  = |$[$|ltail|$]_{na}$|;
  rtc = |$[$|rt|$]_{rlx}$|;
  |$[$|rt|$]_{rel}$|    := (fst rtc, loc);
  |$[$|ltail|$]_{na}$| := loc
\end{lstlisting}
\\
\begin{lstlisting}[language=while]
updateSecondNode(loc, value) |$\triangleq$|
  r1  = |$[$|lhead|$]_{rlx}$|;
  r1c = |$[$|r1|$]_{rlx}$|;
  r2  = snd r1c;
  r2c = |$[$|r2|$]_{rlx}$|;
  r3  = snd r2c;
  |$[$|loc|$]_{rel}$| := (value, r3);
  |$[$|r1|$]_{rel}$|  := (fst r1c, loc);
  sync(cw, cr1, cr2);
  delete r2
\end{lstlisting}
\end{tabular}

&
\hspace{.6cm}

\begin{tabular}{@{}l@{}}
\begin{lstlisting}[language=while]
traverse(lhead, curNodeLoc, resLoc) |$\triangleq$| 
  rh = |$[$|lhead|$]_{acq}$|;
  |$[$|curNodeLoc|$]_{na}$| := rh;
  repeat
    rCurNode = |$[$|curNodeLoc|$]_{na}$|;
    if (rCurNode != null)
    then rNode = |$[$|rCurNode|$]_{acq}$|;
         rRes = |$[$|resLoc|$]_{na}$|;
         rVal = fst rNode;
         |$[$|resLoc|$]_{na}$| := rVal + rRes;
         |$[$|curNodeLoc|$]_{na}$| := snd rNode;
         0
    else 1
    fi
  end
\end{lstlisting}
\end{tabular}

&
\hspace{.6cm}

\begin{tabular}{@{}l@{}}
\begin{lstlisting}[language=while]
sync(cw, cr1, cr2) |$\triangleq$|
  rcw  = |$[$|cw|$]_{rlx}$|;
  rcwn = rcw + 2;
  |$[$|cw|$]_{rel}$| := rcwn;
  |\graybox{\texttt{syncWithReader(rcwn, cr1);}}|
  |\graybox{\texttt{syncWithReader(rcwn, cr2)}}|
\end{lstlisting}
\\
\\
\begin{lstlisting}[language=while]
syncWithReader(rcwn, cr) |$\triangleq$| 
  repeat |$[$|cr|$]_{acq}$| >= rcwn end
\end{lstlisting}
\\
\\
\begin{lstlisting}[language=while]
rcuOnline(cw, cr) |$\triangleq$| 
  |$[$|cr|$]_{rlx}$| := |$[$|cw|$]_{acq}$| + 1
\end{lstlisting}
\\
\\
\begin{lstlisting}[language=while]
rcuOffline(cw, cr) |$\triangleq$| 
  |$[$|cr|$]_{rel}$| := |$[$|cw|$]_{rlx}$|
\end{lstlisting}
\end{tabular}

\end{tabular}
}}

\caption{The QSBR RCU implementation. Fragments in gray boxes are later
  removed for testing purposes (Section~\ref{sec:testing}).}
\label{fig:rcuProg}
\end{figure*}

\section{Case Study: Read-Copy-Update}
\label{sec:rcu-example}

%
We showcase our implementation by testing and debugging a
Read-Copy-Update structure~(RCU)
~\cite{McKenney:PhD,McKenney-Slingwine:PDCS98} and its client
programs.
%
%

\subsection{RCU: background and implementation}

Read-Copy-Update is a standard way to implement non-blocking sharing
of a linked data structure (\eg, list or a tree) between single writer
and multiple readers, running concurrently.
For our purposes, we focus on RCU for a singly linked list,
implemented via \emph{Quiescent State Based Reclamation} (QSBR)
technique~\cite{Desnoyers-al:TPDS12}.  The central idea of RCU is the
way the writer treats nodes of the linked structure. Specifically,
instead of in-place modification of a list node, the writer creates a
\emph{copy} of it, modifies the copy, \emph{updates} the link to the
node, making the older version inaccessible, and then waits until
\emph{all} readers stop using the older version, so it could be
reclaimed.
The crux of the algorithm's correctness is a fine-grained
synchronization between the writer and the readers: the writer updates
the link via release-write, and the readers must traverse the list
using an acquire-read for dereferencing its nodes, ensuring that readers
will not observe {partially} modified nodes.


A QSBR RCU implementation and its client program are shown in
Figure~\ref{fig:rcuProg}.
The first line in the top of the figure initializes thread counters,
which are used by the reader threads to signal if they use the list or
not (\ie, they are in a quiescent state), and a pointer to the list
(\lstinline{lhead}), which is going to be shared between the threads.
Next, three threads are spawned: a writer and two readers.
The writer thread on the left appends \lstinline{1}, \lstinline{10},
and \lstinline{100} to the list.\footnote{In the absence of
  implemented allocation, the example uses the fixed locations
  \lstinline{a}, \lstinline{b}, and \lstinline{c} for storing nodes of
  the list.}
A call to \lstinline{append} creates a new node and adds a link from
the current last node to the new one via {relaxed} write to
\lstinline{ltail}. The updating write to the last node
\lstinline{[rt]} is a release one, guaranteeing that a reader thread,
which might be observing the added node pointer via an {acquire}-read
in concurrent \lstinline{traverse} call, will become aware of the
value and link data, stored to the node. This release/acquire
synchronization eliminates a potential data race.

At the end, the writer thread changes the second value in the list,
\lstinline{10}, to \lstinline{1000}.  In the corresponding
\lstinline{updateSecondNode} routine, first five commands get pointers
to first, second, and third list nodes. The next two commands create a
new node with \lstinline{value} stored in it, and update the
corresponding link in the previous node (\ie, the first one).  By
executing \lstinline{sync(cw, cr1, cr2)}, the writer checks that the
reader threads no longer use an older version of the list (with
\lstinline{10}), so, once the check succeeds, the old node can be
reclaimed.

The reader threads calculate {two times} the sum of the list's
elements by traversing the list.  Before and after each traversal they
call \lstinline{rcuOnline} and \lstinline{rcuOffline} routines
respectively to signal to the writer about their state.

\subsection{Additional infrastructure for testing RCU}
\label{sec:test-struct}

Having an executable operational semantics gives us a possibility to
run a dynamic analysis of the RCU and its client, exercising {all}
possible executions. Realistically, running such an analysis would
take forever, because of the size of the program state-space. The
state-space explosion is because of the following three reasons:

\begin{enumerate}
\item Non-determinism due to concurrent thread scheduling;
\item Resolution of postponed operations;
\item Loading any value, which is \emph{newer} than the one in the
  thread's viewfront representation of the reading location.
\end{enumerate}

\noindent
Indeed, our semantics accounts for all combinations of the factors
above, exploring {all} possible execution traces.

\paragraph{Randomized semantics}

In order to make dynamic analysis practically feasible, we implemented
a semantics, which non-determinis-tically chooses a random path in the
program state space of the original semantics.  It does so by applying
semantic rules to the current state getting a set of new states,
checks if there is a \lstinline{stuck} state, and randomly chooses the
next state from the set. The presence of the randomized semantics
makes possible to implement property-based of testing of
executions~\cite{Hritcu-al:ICFP13}.

\paragraph{Deallocation}

As an additional aspect, we added \lstinline{delete} operator to the
language for reclaiming retired nodes in RCU, and extended the state
by a global list of reclaimed locations. That is, if a location is
added to the ``retired'' list, any read or write on it will lead to
the \lstinline{stuck} state, indicating accessing a deallocated
pointer.





\subsection{Testing and debugging the RCU implementation}
\label{sec:testing}

We can now run some random tests on our RCU implementation and see
whether it meets the basic safety requirements. In particular, we can
check that no matter what path is being exercised by the randomized
semantics, the execution of the program does not get stuck. In our
experience, this correctness condition held for the implementation in
Figure~\ref{fig:rcuProg} for all test runs.

Next, we intentionally introduced a synchronization bug into our
implementation.
%
%
In particular, we removed from the implementation
\lstinline{syncWithReader} loops (see grayed code fragments), which
were used to synchronize the writer with the readers.
%
%
Additionally, we considered the following correctness criteria:
\emph{(i)} the values of \lstinline{r11}, \lstinline{r12},
\lstinline{r21}, and \lstinline{r22} must be in
$\{0, 1, 11, 111, 1101\}$: this guarantees that the list is read
correctly, and, \emph{(ii)} it should be the case that by the end
$\lstmath{r11} \leq \lstmath{r12} \land \lstmath{r21} \leq
\lstmath{r22}$,
\ie, second list traversals see the list at least as up to date as the
first ones.

We ran the test twenty times on Core i7 2.5GHz Linux machine with 8 GB
RAM.  Despite the large number of states to visit, all runs terminated
in less than 27 seconds, and did not violate the desired criteria of
correctness with respect to \lstinline{r*} invariants.\footnote{The
  table with run results is in the \ifext{appendix,
    Figure~\ref{fig:tblRun}}{appendix of the extended version}.}  In
the absence of intentionally removed active wait loops, non-guarded
deallocation has lead to \lstinline{stuck} state in four out of twenty
runs.


Can one implement the RCU with weaker order modifiers?
To check this hypothesis, we changed release write
\lstinline{|$[$|rt|$]_{rel}$| := } \lstinline{(fst rtc, loc)} in
\lstinline{append} to a relaxed one, which resulted in 8 out of 10
test runs getting \lstinline{stuck}, even without deallocation after
update in the writer.
We then changed only \lstinline{|$[$|lhead|$]_{rel}$| := a} in the
writer to a relaxed version, which led to 10 out of 10 test runs
ending up in the \lstinline{stuck} state, as these changes break
synchronization between the writer and the reader threads. The same
results are observed when changing acquire-reads to \emph{relaxed}
ones in \lstinline{traverse}.  Without the enforced
RA-synchronization, the reader threads do not get their viewfronts
updated with writes to locations \lstinline{a}--\lstinline{d}, so
attempts to read from them result in \lstinline{stuck} state,
according to \textsc{Read-Uninit} rule.

As our semantics is implemented in PLT Redex, these synchronization
bugs are easy reproduce: if a program gets stuck or delivers
unexpected results, one can retrieve the corresponding execution trace
via standard Redex machinery.

\paragraph{RCU via consume-reads}

The original implementation of the RCU used {consume}-reads instead of
acquire-reads.  Our version of RCU employs {release/acquire}
synchronization for the following reason.
Currently, our semantics does not support \emph{mutable} local
variables, as we did not need them for running litmus tests. In their
absence, the only possible way to transfer a pointer's value to the
next iteration of a loop is to store it in some location, as it is
currently done in the \lstinline{repeat}-loop of
\texttt{traverse}. The downside of using a proper memory location
instead of a local variable is that this breaks data-dependency chain,
which is required by {consume}-reads for synchronization.
There are no fundamental problems preventing us from adding mutable
variables to make proper use of consume-reads, and we plan to do it in
the future.

\section{Related Work}
\label{sec:related}

\paragraph{Existing semantics for C11 and their variations}

The axiomatic C11 semantics by Batty~\etal~\cite{Batty-al:POPL11} has
been adapted for establishing soundness of several program logics for
relaxed
memory~\cite{Vafeiadis-Narayan:OOPSLA13,Turon-al:OOPSLA14,Lahav-Vafeiadis:ICALP15,Doko-Vafeiadis:VMCAI16}.
While some of these adaptations bear a lot of similarity with
operational approach~\cite{Turon-al:OOPSLA14}, all of them are still
based on the notion of partial orders between reads and
writes. 
The recent operational semantics for C11 took steps to
\emph{incrementalize} Batty~\etal's model, constructing the partial
orders in a step-wise manner, checking the consistency axioms at every
execution step~\cite{Nienhuis-al:OSC11}.
This model does not follow the program execution order and allows OTA
behaviors.

The semantics for \emph{Strong Release-Acquire} (SRA) model by
Lahav~\etal~\cite{Lahav-al:POPL16} does not use graphs and consistency
axioms, relying instead on \emph{message buffers}, reminiscent to our
viewfronts in the way they are used for thread synchronization in our
approach.
However, Lahav~\etal's semantics only targets a (strengthened) subset
of C11, restricted to release/acquire-synchronization, thus,
sidestepping the intricacies of encoding the meaning of relaxed
atomics.

\paragraph{Operational semantics for relaxed memory models}

A low-level operational model for the total store order (TSO) memory
model, which is stronger than C11, has been defined by Owens~\etal,
targeting x86-TSO processors~\cite{Owens-al:TPHOL09}. A more complex
model is tackled by Sarkar~\etal, who provided an operational
semantics for the POWER architecture~\cite{Sarkar-al:PLDI11}.
Finally, the most recent work by Flur~\etal provides an operational
model for the ARMv8 architecture~\cite{Flur-al:POPL16}.
While in this work we are concerned with semantics of a high-level
language (\ie, C/C++), investigating compilation schemes to those
low-level models with respect to our semantics is our immediate future
work.

Related proposals with respect to operational semantics for relaxed
memory, inspired by the TSO model, are based on the idea of
write-buffers~\cite{Jagadeesan-al:ESOP10,EffingerDean-Grossman:MM,Boudol-Petri:POPL09,Boudol-al:EXPRESS12},
reminiscent to the buffers we use to define postponed operations in
our approach. The idea of write buffers and buffer pools fits
operational intuition naturally and was used to prove soundness of a
program logic~\cite{Sieczkowski-al:ESOP15}, but it is not trivial to
adapt for C11-style synchronization, especially for reconciling
RA-synchronization and relaxed atomics, as we demonstrated in
Sections~\ref{sec:advanced} and~\ref{sec:semantics}.
An alternative approach to define relaxed behavior is to allow the
programmer to manipulate with synchronization orders \emph{explicitly}
via program-level annotations~\cite{Crary-Sullivan:POPL15}. This
approach provides a highly generic way of modelling custom
synchronization patterns at atomic accesses, although, it does not
correspond to any specific standard and is not executable. Due to the
intentional possibility to use it for modelling very ``relaxed''
behaviors via arbitrary speculations, the approach allows OTA
behaviors.
In contrast, our semantics, tailored to allow for modelling all
essential concurrent features of C11, prevents OTA by careful
treatment of operation buffers, and is executable.

\paragraph{Semantics for relaxed atomics via event structures}

The OTA executions are considered a serious issue with the C++11
standard~\cite{Batty-al:ESOP15,Boehm-Demsky:MSPC14,McKenney-al:OTA},
in particular, because there is no well-stated and uniform definition
of this phenomenon, which is only characterized in the folklore as
``values appearing out of nowhere''.

At the moment, several proposals provide treatment for relaxed
atomics, avoiding the OTA behavior by presenting models for relaxed
memory accesses, which are defined in terms of \emph{event
  structures}~\cite{PichonPharabod-Sewell:POPL16,
  Jeffrey-Riely:LICS16}.
These models allow aggressive optimizations including value-range
speculations, without introducing classic out-of-thin-air behaviors.
Our semantics does not support all of these compiler transformations
yet, (\eg, speculative calculation of an arithmetic expression with
symbolic values), but they can be added as additional rules without
changing the underlying program state.

In contrast with our semantics, the model of Pichon-Pharabod and
Sewell~\cite{PichonPharabod-Sewell:POPL16} is not realistically
executable, as it requires for every read operation in a program
execution to consider $N$ events, where $N$ is a size of the value
domain. That is, for instance, for reading a 32-bit integer it
is~$2^{32}$. Furthermore, at the moment the
model~\cite{PichonPharabod-Sewell:POPL16} does not account for
release/acquire-synchronization.
%

The model by Jeffrey and Riely~\cite{Jeffrey-Riely:LICS16} does not
allow for reorderings of independent reads, which makes it too strong
to be efficiently implemented on such architectures Power and ARM.
The authors suggest a fix for it, which, however, invalidates some
other guarantees their initial proposal provides.
%
%











\paragraph{Reasoning about Read-Copy-Update}

RCU structures have been used recently to showcase program
logics~\cite{Tassarotti-al:PLDI15,Lahav-Vafeiadis:ICALP15}, semantic
frameworks~\cite{Lahav-al:POPL16}, and program repair/synthesis
methods~\cite{Meshman-al:FMCAD15} in the context of C11 concurrency.
To the best of our knowledge, no other existing approach provides a
way of efficiently debugging them by means of re-tracing executions,
exhibiting synchronization issues, as we demonstrated in
Section~\ref{sec:testing}.



\section{Conclusion and Future Work}
\label{sec:conclusion}

In this work, we presented a family of operational semantics for
modelling C/C++ concurrency features. The encoding of C11-style
semantics in our framework is based on the two main ideas:
\emph{viewfronts} and \emph{operation buffers}, with their various
combinations and elaborations allowing to express specific
synchronization mechanisms and language aspects from the C11 standard.
Our C11 semantics is executable, which we demonstrated by implementing
it in PLT Redex and showcasing with a number of examples.

As our future work, we plan to extend the defined formalism for C11
fences~\cite{Doko-Vafeiadis:VMCAI16} and establish formal results
relating executions in our semantics to executions in low-level
languages via standard compilation
schemes~\cite{Sewell-al:CACM10,Sarkar-al:PLDI11,Flur-al:POPL16}, thus,
proving that our semantics is weak enough to accommodate them.
Next, we are going to employ it as a basis for developing a
higher-order program logic for establishing Hoare specifications and
program refinement in the C11 model~\cite{Vafeiadis:CPP15}, proving
the logic's soundness with respect to our semantics, lifted to {sets
  of traces}~\cite{Brookes:TCS07} or {logical
  relations}~\cite{Turon-al:POPL13}.
Finally, we plan to use our operational framework for exploring the
ideas of efficiently synthesizing synchronization primitives via
bounded model checking~\cite{Meshman-al:FMCAD15} and partial order
reduction~\cite{Godefroid:PhD95}.





\bibliographystyle{abbrv}
\bibliography{references,proceedings}

\ifext{
\newpage
\onecolumn
\appendix

\section{The Catalogue of Litmus Tests}
\label{sec:litmusTests}

\subsection{Store Buffering (SB)}
\label{app:sb}

\litmusTestStart{SB\_rel+acq}{\tick}{History + Viewfronts}
\begin{minipage}[t]{0.3\linewidth}
Possible outcomes:\\
\lstinline{  r1 = 0 |$\land$| r2 = 0}\\
\lstinline{  r1 = 0 |$\land$| r2 = 1}\\
\lstinline{  r1 = 1 |$\land$| r2 = 0}\\
\lstinline{  r1 = 1 |$\land$| r2 = 1}\\
\end{minipage}
\begin{minipage}[t]{0.3\linewidth}
\vspace{-.2cm}
  \begin{tabular}{l@{\ \ \ }l}
    \begin{minipage}[l]{4.3cm} \small
\begin{lstlisting}
  |$[$|x|$]_{rel}$| := 0; |$[$|y|$]_{rel}$| := 0;
\end{lstlisting}
\vspace{-.2cm}
\begin{tabular}{l||l}
\begin{lstlisting}
|$[$|x|$]_{rel}$| := 1;
r1 = |$[$|y|$]_{acq}$|
\end{lstlisting}
\hspace{.6cm}
&
\begin{lstlisting}
|$[$|y|$]_{rel}$| := 1;
r2 = |$[$|x|$]_{acq}$|
\end{lstlisting}
\end{tabular}
    \end{minipage}
&
  \end{tabular}
\end{minipage}
\litmusTestEnd

\litmusTestStart{SB\_sc}{\tick}{SC + History + Viewfronts}
\begin{minipage}[t]{0.3\linewidth}
Forbidden outcomes:\\
\lstinline{  r1 = 0 |$\land$| r2 = 0}\\
\end{minipage}
\begin{minipage}[t]{0.3\linewidth}
\vspace{-.2cm}
  \begin{tabular}{l@{\ \ \ }l}
    \begin{minipage}[l]{4.3cm} \small
\begin{lstlisting}
  |$[$|x|$]_{sc}$| := 0; |$[$|y|$]_{sc}$| := 0;
\end{lstlisting}
\vspace{-.2cm}
\begin{tabular}{l||l}
\begin{lstlisting}
|$[$|x|$]_{sc}$| := 1;
r1 = |$[$|y|$]_{sc}$|
\end{lstlisting}
\hspace{.6cm}
&
\begin{lstlisting}
|$[$|y|$]_{sc}$| := 1;
r2 = |$[$|x|$]_{sc}$|
\end{lstlisting}
\end{tabular}
    \end{minipage}
&
  \end{tabular}
\end{minipage}
\litmusTestEnd

\litmusTestStart{SB\_sc+rel}{\tick}{SC + History + Viewfronts}
\begin{minipage}[t]{0.3\linewidth}
Possible outcomes:\\
\lstinline{  r1 = 0 |$\land$| r2 = 0}\\
\lstinline{  r1 = 0 |$\land$| r2 = 1}\\
\lstinline{  r1 = 1 |$\land$| r2 = 0}\\
\lstinline{  r1 = 1 |$\land$| r2 = 1}\\
\end{minipage}
\begin{minipage}[t]{0.3\linewidth}
\vspace{-.2cm}
  \begin{tabular}{l@{\ \ \ }l}
    \begin{minipage}[l]{4.3cm} \small
\begin{lstlisting}
  |$[$|x|$]_{sc}$| := 0; |$[$|y|$]_{sc}$| := 0;
\end{lstlisting}
\vspace{-.2cm}
\begin{tabular}{l||l}
\begin{lstlisting}
|$[$|x|$]_{rel}$| := 1;
r1 = |$[$|y|$]_{sc}$|
\end{lstlisting}
\hspace{.6cm}
&
\begin{lstlisting}
|$[$|y|$]_{sc}$| := 1;
r2 = |$[$|x|$]_{sc}$|
\end{lstlisting}
\end{tabular}
    \end{minipage}
&
  \end{tabular}
\end{minipage}
\litmusTestEnd

\litmusTestStart{SB\_sc+acq}{\tick}{SC + History + Viewfronts}
\begin{minipage}[t]{0.3\linewidth}
Possible outcomes:\\
\lstinline{  r1 = 0 |$\land$| r2 = 0}\\
\lstinline{  r1 = 0 |$\land$| r2 = 1}\\
\lstinline{  r1 = 1 |$\land$| r2 = 0}\\
\lstinline{  r1 = 1 |$\land$| r2 = 1}\\
\end{minipage}
\begin{minipage}[t]{0.3\linewidth}
\vspace{-.2cm}
  \begin{tabular}{l@{\ \ \ }l}
    \begin{minipage}[l]{4.3cm} \small
\begin{lstlisting}
  |$[$|x|$]_{sc}$| := 0; |$[$|y|$]_{sc}$| := 0;
\end{lstlisting}
\vspace{-.2cm}
\begin{tabular}{l||l}
\begin{lstlisting}
|$[$|x|$]_{sc}$| := 1;
r1 = |$[$|y|$]_{acq}$|
\end{lstlisting}
\hspace{.6cm}
&
\begin{lstlisting}
|$[$|y|$]_{sc}$| := 1;
r2 = |$[$|x|$]_{sc}$|
\end{lstlisting}
\end{tabular}
    \end{minipage}
&
  \end{tabular}
\end{minipage}
\litmusTestEnd


\subsection{Load Buffering (LB)}
\label{app:lb}

\litmusTestStart{LB\_rlx}{\tick}{Postponed Reads + History + Viewfronts}
\begin{minipage}[t]{0.3\linewidth}
Possible outcomes:\\
\lstinline{  r1 = 0 |$\land$| r2 = 0}\\
\lstinline{  r1 = 0 |$\land$| r2 = 1}\\
\lstinline{  r1 = 1 |$\land$| r2 = 0}\\
\lstinline{  r1 = 1 |$\land$| r2 = 1}\\
\end{minipage}
\begin{minipage}[t]{0.3\linewidth}
\vspace{-.2cm}
  \begin{tabular}{l@{\ \ \ }l}
    \begin{minipage}[l]{4.3cm} \small
\begin{lstlisting}
  |$[$|x|$]_{rlx}$| := 0; |$[$|y|$]_{rlx}$| := 0;
\end{lstlisting}
\vspace{-.2cm}
\begin{tabular}{l||l}
\begin{lstlisting}
r1 = |$[$|y|$]_{rlx}$|;
|$[$|x|$]_{rlx}$| := 1
\end{lstlisting}
\hspace{.6cm}
&
\begin{lstlisting}
r2 = |$[$|x|$]_{rlx}$|;
|$[$|y|$]_{rlx}$| := 1
\end{lstlisting}
\end{tabular}
    \end{minipage}
&
  \end{tabular}
\end{minipage}
\litmusTestEnd

\litmusTestStart{LB\_rel+rlx}{\tick}{Postponed Reads + History + Viewfronts}
\begin{minipage}[t]{0.3\linewidth}
Possible outcomes:\\
\lstinline{  r1 = 0 |$\land$| r2 = 0}\\
\lstinline{  r1 = 0 |$\land$| r2 = 1}\\
\lstinline{  r1 = 1 |$\land$| r2 = 0}\\
\lstinline{  r1 = 1 |$\land$| r2 = 1}\\
\end{minipage}
\begin{minipage}[t]{0.3\linewidth}
\vspace{-.2cm}
  \begin{tabular}{l@{\ \ \ }l}
    \begin{minipage}[l]{4.3cm} \small
\begin{lstlisting}
  |$[$|x|$]_{rlx}$| := 0; |$[$|y|$]_{rlx}$| := 0;
\end{lstlisting}
\vspace{-.2cm}
\begin{tabular}{l||l}
\begin{lstlisting}
r1 = |$[$|y|$]_{rlx}$|;
|$[$|x|$]_{rel}$| := 1
\end{lstlisting}
\hspace{.6cm}
&
\begin{lstlisting}
r2 = |$[$|x|$]_{rlx}$|;
|$[$|y|$]_{rel}$| := 1
\end{lstlisting}
\end{tabular}
    \end{minipage}
&
  \end{tabular}
\end{minipage}
\litmusTestEnd

\litmusTestStart{LB\_rel+rlx}{\tick}{Postponed Reads + History + Viewfronts}
\begin{minipage}[t]{0.3\linewidth}
Possible outcomes:\\
\lstinline{  r1 = 0 |$\land$| r2 = 0}\\
\lstinline{  r1 = 0 |$\land$| r2 = 1}\\
\lstinline{  r1 = 1 |$\land$| r2 = 0}\\
\lstinline{  r1 = 1 |$\land$| r2 = 1}\\
\end{minipage}
\begin{minipage}[t]{0.3\linewidth}
\vspace{-.2cm}
  \begin{tabular}{l@{\ \ \ }l}
    \begin{minipage}[l]{4.3cm} \small
\begin{lstlisting}
  |$[$|x|$]_{rlx}$| := 0; |$[$|y|$]_{rlx}$| := 0;
\end{lstlisting}
\vspace{-.2cm}
\begin{tabular}{l||l}
\begin{lstlisting}
r1 = |$[$|y|$]_{rlx}$|;
|$[$|x|$]_{rel}$| := 1
\end{lstlisting}
\hspace{.6cm}
&
\begin{lstlisting}
r2 = |$[$|x|$]_{rlx}$|;
|$[$|y|$]_{rel}$| := 1
\end{lstlisting}
\end{tabular}
    \end{minipage}
&
  \end{tabular}
\end{minipage}
\litmusTestEnd

\litmusTestStart{LB\_acq+rlx}{\fail}{Postponed Reads + History + Viewfronts}
\begin{minipage}[t]{0.3\linewidth}
Possible outcomes:\\
\lstinline{  r1 = 0 |$\land$| r2 = 0}\\
\lstinline{  r1 = 0 |$\land$| r2 = 1}\\
\lstinline{  r1 = 1 |$\land$| r2 = 0}\\
\lstinline{  r1 = 1 |$\land$| r2 = 1}\\
\end{minipage}
\begin{minipage}[t]{0.3\linewidth}
\vspace{-.2cm}
  \begin{tabular}{l@{\ \ \ }l}
    \begin{minipage}[l]{4.3cm} \small
\begin{lstlisting}
  |$[$|x|$]_{rlx}$| := 0; |$[$|y|$]_{rlx}$| := 0;
\end{lstlisting}
\vspace{-.2cm}
\begin{tabular}{l||l}
\begin{lstlisting}
r1 = |$[$|y|$]_{acq}$|;
|$[$|x|$]_{rlx}$| := 1
\end{lstlisting}
\hspace{.6cm}
&
\begin{lstlisting}
r2 = |$[$|x|$]_{acq}$|;
|$[$|y|$]_{rlx}$| := 1
\end{lstlisting}
\end{tabular}
    \end{minipage}
&
  \end{tabular}
\end{minipage}

Our semantics doesn't allow the \lstinline{  r1 = 1 |$\land$| r2 = 1} outcome for the program.
It doesn't allow reordering of an acquire read with a subsequent write.
The known sound compilation schemes of acquire read to major platforms (x86, ARM, Power) don't
allow the behavior either. 

\litmusTestEnd

\litmusTestStart{LB\_rel+acq+rlx}{\tick}{Postponed Reads + History + Viewfronts}
\begin{minipage}[t]{0.3\linewidth}
Forbidden outcomes:\\
\lstinline{  r1 = 1 |$\land$| r2 = 1}\\
\end{minipage}
\begin{minipage}[t]{0.3\linewidth}
\vspace{-.2cm}
  \begin{tabular}{l@{\ \ \ }l}
    \begin{minipage}[l]{4.3cm} \small
\begin{lstlisting}
  |$[$|x|$]_{rlx}$| := 0; |$[$|y|$]_{rlx}$| := 0;
\end{lstlisting}
\vspace{-.2cm}
\begin{tabular}{l||l}
\begin{lstlisting}
r1 = |$[$|y|$]_{acq}$|;
|$[$|x|$]_{rlx}$| := 1
\end{lstlisting}
\hspace{.6cm}
&
\begin{lstlisting}
r2 = |$[$|x|$]_{rlx}$|;
|$[$|y|$]_{rel}$| := 1
\end{lstlisting}
\end{tabular}
    \end{minipage}
&
  \end{tabular}
\end{minipage}
\litmusTestEnd

\litmusTestStart{LB\_rlx+use}{\tick}{Postponed Reads + History + Viewfronts}
\begin{minipage}[t]{0.3\linewidth}
Allowed outcome:\\
\lstinline{  r1 = 1 |$\land$| r2 = 1}\\
\end{minipage}
\begin{minipage}[t]{0.3\linewidth}
\vspace{-.2cm}
  \begin{tabular}{l@{\ \ \ }l}
    \begin{minipage}[l]{4.3cm} \small
\begin{lstlisting}
  |$[$|x|$]_{rlx}$| := 0; |$[$|y|$]_{rlx}$| := 0;
\end{lstlisting}
\vspace{-.2cm}
\begin{tabular}{l||l}
\begin{lstlisting}
r1 = |$[$|y|$]_{rlx}$|;
|$[$|z1|$]_{rlx}$| := r1;
|$[$|x|$]_{rlx}$| := 1
\end{lstlisting}
\hspace{.6cm}
&
\begin{lstlisting}
r2 = |$[$|x|$]_{rlx}$|;
|$[$|z2|$]_{rlx}$| := r2;
|$[$|y|$]_{rlx}$| := 1
\end{lstlisting}
\end{tabular}
    \end{minipage}
&
  \end{tabular}
\end{minipage}

\litmusTestEnd

\litmusTestStart{LB\_rlx+let}{\tick}{Postponed Reads + History + Viewfronts}
\begin{minipage}[t]{0.3\linewidth}
Allowed outcome:\\
\lstinline{  r1 = 1 |$\land$| r'1 = 2 |$\land$| r2 = 1 |$\land$| r'2 = 2}\\
\end{minipage}
\begin{minipage}[t]{0.3\linewidth}
\vspace{-.2cm}
  \begin{tabular}{l@{\ \ \ }l}
    \begin{minipage}[l]{4.3cm} \small
\begin{lstlisting}
  |$[$|x|$]_{rlx}$| := 0; |$[$|y|$]_{rlx}$| := 0;
\end{lstlisting}
\vspace{-.2cm}
\begin{tabular}{l||l}
\begin{lstlisting}
r1 = |$[$|y|$]_{rlx}$|;
r'1 = r1 + 1;
|$[$|x|$]_{rlx}$| := 1
\end{lstlisting}
\hspace{.6cm}
&
\begin{lstlisting}
r2 = |$[$|x|$]_{rlx}$|;
r'2 = r2 + 1;
|$[$|y|$]_{rlx}$| := 1
\end{lstlisting}
\end{tabular}
    \end{minipage}
&
  \end{tabular}
\end{minipage}
\litmusTestEnd

\litmusTestStart{LB\_rlx+join}{\tickPP}{Postponed Reads + History + Viewfronts + JN}
\begin{minipage}[t]{0.2\linewidth}
Allowed outcomes:\\
\lstinline{  r1 = 1 |$\land$| r2 = 1}\\
\end{minipage}
\begin{minipage}[t]{0.4\linewidth}
\vspace{-.2cm}
  \begin{tabular}{l@{\ \ \ }l}
    \begin{minipage}[l]{4.3cm} \small
\begin{lstlisting}
            |$[$|x|$]_{rlx}$| := 0; |$[$|y|$]_{rlx}$| := 0;
\end{lstlisting}
\vspace{-.2cm}
\begin{tabular}{l||l||l||l}
\begin{lstlisting}
r1 = |$[$|y|$]_{rlx}$|;
|$[$|z1|$]_{rlx}$| := r1
\end{lstlisting}
\hspace{.6cm}
&
\begin{lstlisting}
  0
\end{lstlisting}
\hspace{.6cm}
&
\begin{lstlisting}
r2 = |$[$|x|$]_{rlx}$|;
|$[$|z2|$]_{rlx}$| := r2
\end{lstlisting}
\hspace{.6cm}
&
\begin{lstlisting}
  0
\end{lstlisting}
\end{tabular}

\vspace{-1pt}
\begin{tabular}{l||l}
  \begin{lstlisting}
            |$[$|x|$]_{rlx}$| := 1
  \end{lstlisting}
\hspace{2.52em}
&
  \begin{lstlisting}
      |$[$|y|$]_{rlx}$| := 1
  \end{lstlisting}
\end{tabular}
    \end{minipage}
&
  \end{tabular}
\end{minipage}

\litmusTestEnd

\litmusTestStart{LB\_rel+rlx+join}{\tickPP}{Postponed Reads + History + Viewfronts + JN}
\begin{minipage}[t]{0.2\linewidth}
Allowed outcomes:\\
\lstinline{  r1 = 1 |$\land$| r2 = 1}\\
\end{minipage}
\begin{minipage}[t]{0.4\linewidth}
\vspace{-.2cm}
  \begin{tabular}{l@{\ \ \ }l}
    \begin{minipage}[l]{4.3cm} \small
\begin{lstlisting}
            |$[$|x|$]_{rlx}$| := 0; |$[$|y|$]_{rlx}$| := 0;
\end{lstlisting}
\vspace{-.2cm}
\begin{tabular}{l||l||l||l}
\begin{lstlisting}
r1 = |$[$|y|$]_{rlx}$|;
|$[$|z1|$]_{rlx}$| := r1
\end{lstlisting}
\hspace{.6cm}
&
\begin{lstlisting}
  0
\end{lstlisting}
\hspace{.6cm}
&
\begin{lstlisting}
r2 = |$[$|x|$]_{rlx}$|;
|$[$|z2|$]_{rlx}$| := r2
\end{lstlisting}
\hspace{.6cm}
&
\begin{lstlisting}
  0
\end{lstlisting}
\end{tabular}

\vspace{-1pt}
\begin{tabular}{l||l}
  \begin{lstlisting}
            |$[$|x|$]_{rel}$| := 1
  \end{lstlisting}
\hspace{2.52em}
&
  \begin{lstlisting}
      |$[$|y|$]_{rel}$| := 1
  \end{lstlisting}
\end{tabular}
    \end{minipage}
&
  \end{tabular}
\end{minipage}
\litmusTestEnd

\litmusTestStart{LB\_acq+rlx+join}{\fail}{Postponed Reads + History + Viewfronts + JN}
\begin{minipage}[t]{0.2\linewidth}
Allowed outcomes:\\
\lstinline{  r1 = 1 |$\land$| r2 = 1}\\
\end{minipage}
\begin{minipage}[t]{0.4\linewidth}
\vspace{-.2cm}
  \begin{tabular}{l@{\ \ \ }l}
    \begin{minipage}[l]{4.3cm} \small
\begin{lstlisting}
            |$[$|x|$]_{rlx}$| := 0; |$[$|y|$]_{rlx}$| := 0;
\end{lstlisting}
\vspace{-.2cm}
\begin{tabular}{l||l||l||l}
\begin{lstlisting}
r1 = |$[$|y|$]_{acq}$|;
|$[$|z1|$]_{rlx}$| := r1
\end{lstlisting}
\hspace{.6cm}
&
\begin{lstlisting}
  0
\end{lstlisting}
\hspace{.6cm}
&
\begin{lstlisting}
r2 = |$[$|x|$]_{acq}$|;
|$[$|z2|$]_{rlx}$| := r2
\end{lstlisting}
\hspace{.6cm}
&
\begin{lstlisting}
  0
\end{lstlisting}
\end{tabular}

\vspace{-1pt}
\begin{tabular}{l||l}
  \begin{lstlisting}
            |$[$|x|$]_{rlx}$| := 1
  \end{lstlisting}
\hspace{2.52em}
&
  \begin{lstlisting}
      |$[$|y|$]_{rlx}$| := 1
  \end{lstlisting}
\end{tabular}
    \end{minipage}
&
  \end{tabular}
\end{minipage}
\litmusTestEnd

\subsection{Message Passing (MP)}
\label{app:mp}

\litmusTestStart{MP\_rlx+na}{\tick}{NA + History + Viewfronts}
\begin{minipage}[t]{0.3\linewidth}
Possible outcomes:\\
\lstinline{  r1 = 0}\\
\lstinline{  r1 = 5}\\
\lstinline{  stuck}\\
\end{minipage}
\begin{minipage}[t]{0.3\linewidth}
\vspace{-.2cm}
  \begin{tabular}{l@{\ \ \ }l}
    \begin{minipage}[l]{4.3cm} \small
\begin{lstlisting}
        |$[$|f|$]_{rlx}$| := 0; |$[$|d|$]_{na}$| := 0;
\end{lstlisting}
\vspace{-.2cm}
\begin{tabular}{l||l}
\begin{lstlisting}
|$[$|d|$]_{na}$| := 5;
|$[$|f|$]_{rlx}$| := 1
\end{lstlisting}
\hspace{.6cm}
&
\begin{lstlisting}
repeat |$[$|f|$]_{rlx}$| end;
r1 = |$[$|d|$]_{na}$|
\end{lstlisting}
\end{tabular}
    \end{minipage}
&
  \end{tabular}
\end{minipage}
\litmusTestEnd

\litmusTestStart{MP\_rel+rlx+na}{\tick}{NA + History + Viewfronts}
\begin{minipage}[t]{0.3\linewidth}
Possible outcomes:\\
\lstinline{  r1 = 0}\\
\lstinline{  r1 = 5}\\
\lstinline{  stuck}\\
\end{minipage}
\begin{minipage}[t]{0.3\linewidth}
\vspace{-.2cm}
  \begin{tabular}{l@{\ \ \ }l}
    \begin{minipage}[l]{4.3cm} \small
\begin{lstlisting}
        |$[$|f|$]_{rlx}$| := 0; |$[$|d|$]_{na}$| := 0;
\end{lstlisting}
\vspace{-.2cm}
\begin{tabular}{l||l}
\begin{lstlisting}
|$[$|d|$]_{na}$| := 5;
|$[$|f|$]_{rel}$| := 1
\end{lstlisting}
\hspace{.6cm}
&
\begin{lstlisting}
repeat |$[$|f|$]_{rlx}$| end;
r1 = |$[$|d|$]_{na}$|
\end{lstlisting}
\end{tabular}
    \end{minipage}
&
  \end{tabular}
\end{minipage}
\litmusTestEnd

\litmusTestStart{MP\_rlx+acq+na}{\tick}{NA + History + Viewfronts}
\begin{minipage}[t]{0.3\linewidth}
Possible outcomes:\\
\lstinline{  r1 = 0}\\
\lstinline{  r1 = 5}\\
\lstinline{  stuck}\\
\end{minipage}
\begin{minipage}[t]{0.4\linewidth}
\vspace{-.2cm}
  \begin{tabular}{l@{\ \ \ }l}
    \begin{minipage}[l]{4.3cm} \small
\begin{lstlisting}
        |$[$|f|$]_{rlx}$| := 0; |$[$|d|$]_{na}$| := 0;
\end{lstlisting}
\vspace{-.2cm}
\begin{tabular}{l||l}
\begin{lstlisting}
|$[$|d|$]_{na}$| := 5;
|$[$|f|$]_{rlx}$| := 1
\end{lstlisting}
\hspace{.6cm}
&
\begin{lstlisting}
repeat |$[$|f|$]_{acq}$| end;
r1 = |$[$|d|$]_{na}$|
\end{lstlisting}
\end{tabular}
    \end{minipage}
&
  \end{tabular}
\end{minipage}
\litmusTestEnd

\litmusTestStart{MP\_rel+acq+na}{\tick}{NA + History + Viewfronts}
\begin{minipage}[t]{0.3\linewidth}
Possible outcomes:\\
\lstinline{  r1 = 5}\\
\end{minipage}
\begin{minipage}[t]{0.3\linewidth}
\vspace{-.2cm}
  \begin{tabular}{l@{\ \ \ }l}
    \begin{minipage}[l]{4.3cm} \small
\begin{lstlisting}
        |$[$|f|$]_{rel}$| := 0; |$[$|d|$]_{na}$| := 0;
\end{lstlisting}
\vspace{-.2cm}
\begin{tabular}{l||l}
\begin{lstlisting}
|$[$|d|$]_{na}$| := 5;
|$[$|f|$]_{rel}$| := 1
\end{lstlisting}
\hspace{.6cm}
&
\begin{lstlisting}
repeat |$[$|f|$]_{acq}$| end;
r1 = |$[$|d|$]_{na}$|
\end{lstlisting}
\end{tabular}
    \end{minipage}
&
  \end{tabular}
\end{minipage}
\litmusTestEnd

\litmusTestStart{MP\_rel+acq+na+rlx}{\tick}{Write-fronts + NA + History + Viewfronts}
\begin{minipage}[t]{0.3\linewidth}
Possible outcomes:\\
\lstinline{  r1 = 5}\\
\end{minipage}
\begin{minipage}[t]{0.3\linewidth}
\vspace{-.2cm}
  \begin{tabular}{l@{\ \ \ }l}
    \begin{minipage}[l]{4.3cm} \small
\begin{lstlisting}
        |$[$|f|$]_{rel}$| := 0; |$[$|d|$]_{na}$| := 0;
\end{lstlisting}
\vspace{-.2cm}
\begin{tabular}{l||l}
\begin{lstlisting}
|$[$|d|$]_{na}$| := 5;
|$[$|f|$]_{rel}$| := 1;
|$[$|f|$]_{rlx}$| := 2
\end{lstlisting}
\hspace{.6cm}
&
\begin{lstlisting}
repeat |$[$|f|$]_{acq}$| == 2 end;
r1 = |$[$|d|$]_{na}$|
\end{lstlisting}
\end{tabular}
    \end{minipage}
&
  \end{tabular}
\end{minipage}
\litmusTestEnd

\litmusTestStart{MP\_rel+acq+na+rlx\_2}{\tick}{Write-fronts + NA + History + Viewfronts}
\begin{minipage}[t]{0.3\linewidth}
Possible outcomes:\\
\lstinline{  r1 = 5 /\ r2 = <0, 1>}\\
\end{minipage}
\begin{minipage}[t]{0.3\linewidth}
\vspace{-.2cm}
  \begin{tabular}{l@{\ \ \ }l}
    \begin{minipage}[l]{4.3cm} \small
\begin{lstlisting}
|$[$|f|$]_{na}$| := 0; |$[$|d|$]_{na}$| := 0; |$[$|x|$]_{na}$| := 0;
\end{lstlisting}
\vspace{-.2cm}
\begin{tabular}{l||l}
\begin{lstlisting}
|$[$|d|$]_{na}$| := 5;
|$[$|f|$]_{rel}$| := 1;
|$[$|x|$]_{rel}$| := 1;
|$[$|f|$]_{rlx}$| := 2
\end{lstlisting}
\hspace{.6cm}
&
\begin{lstlisting}
repeat |$[$|f|$]_{acq}$| == 2 end;
r1 := |$[$|d|$]_{na}$|;
r2 := |$[$|x|$]_{rlx}$|
\end{lstlisting}
\end{tabular}
    \end{minipage}
&
  \end{tabular}
\end{minipage}
\litmusTestEnd

\litmusTestStart{MP\_con+na}{\tick}{Consume + NA + History + Viewfronts}
\begin{minipage}[t]{0.3\linewidth}
Possible outcomes:\\
\lstinline{  r1 = 0}\\
\lstinline{  r1 = 5}\\
\end{minipage}
\begin{minipage}[t]{0.3\linewidth}
\vspace{-.2cm}
  \begin{tabular}{l@{\ \ \ }l}
    \begin{minipage}[l]{4.3cm} \small
\begin{lstlisting}
        |$[$|f|$]_{con}$| := null; |$[$|d|$]_{na}$| := 0;
\end{lstlisting}
\vspace{-.2cm}
\begin{tabular}{l||l}
\begin{lstlisting}
|$[$|d|$]_{na}$| := 5;
|$[$|f|$]_{rel}$| := d
\end{lstlisting}
\hspace{.6cm}
&
\begin{lstlisting}
r0 := |$[$|f|$]_{con}$|;
if r0 != null
then r1 = |$[$|r0|$]_{na}$|
else r1 = 0
fi
\end{lstlisting}
\end{tabular}
    \end{minipage}
&
  \end{tabular}
\end{minipage}
\litmusTestEnd

\litmusTestStart{MP\_con+na\_2}{\tick}{Consume + NA + History + Viewfronts}
\begin{minipage}[t]{0.3\linewidth}
Possible outcomes:\\
\lstinline{  r2 = 0 /\ r3 = <0, 1>}\\
\lstinline{  r2 = 5 /\ r3 = <0, 1>}\\
\end{minipage}
\begin{minipage}[t]{0.3\linewidth}
\vspace{-.2cm}
  \begin{tabular}{l@{\ \ \ }l}
    \begin{minipage}[l]{4.3cm} \small
\begin{lstlisting}
|$[$|p|$]_{na}$| := null; |$[$|d|$]_{na}$| := 0; |$[$|x|$]_{na}$| := 0;
\end{lstlisting}
\vspace{-.2cm}
\begin{tabular}{l||l}
\begin{lstlisting}
|$[$|x|$]_{rlx}$| := 1;
|$[$|d|$]_{na}$| := 1;
|$[$|p|$]_{rel}$| := d
\end{lstlisting}
\hspace{.6cm}
&
\begin{lstlisting}
r1 = |$[$|p|$]_{con}$|;
if   r1 != null
then r2 = |$[$|r1|$]_{na}$|;
     r3 = |$[$|x|$]_{rlx}$|
else r2 = 0; r3 = 0
fi
\end{lstlisting}
\end{tabular}
    \end{minipage}
&
  \end{tabular}
\end{minipage}
\litmusTestEnd

\litmusTestStart{MP\_cas+rel+acq+na from \cite{Vafeiadis-Narayan:OOPSLA13}}{\tick}{NA + History + Viewfronts}
\begin{minipage}[t]{0.2\linewidth}
Impossible outcomes:\\
\lstinline{  stuck}\\
\end{minipage}
\begin{minipage}[t]{0.4\linewidth}
\vspace{-.2cm}
  \begin{tabular}{l@{\ \ \ }l}
    \begin{minipage}[l]{4.3cm} \small
\begin{lstlisting}
        |$[$|f|$]_{rlx}$| := 1; |$[$|d|$]_{na}$| := 0;
\end{lstlisting}
\vspace{-.2cm}
\begin{tabular}{l||l@{\ \ \ \ }||l}
\begin{lstlisting}
|$[$|d|$]_{na}$| := 5;
|$[$|f|$]_{rel}$| := 0
\end{lstlisting}
\hspace{.6cm}
&
\begin{lstlisting}
r1 = cas|$_{acq,rlx}$|(f, 0, 1);
if r1 == 0
then |$[$|d|$]_{rlx}$| := 6
else 0
fi
\end{lstlisting}
\hspace{.6cm}
&
\begin{lstlisting}
r2 = cas|$_{acq,rlx}$|(f, 0, 1);
if r2 == 0
then |$[$|d|$]_{rlx}$| := 7
else 0
fi
\end{lstlisting}
\end{tabular}
    \end{minipage}
&
  \end{tabular}
\end{minipage}
\litmusTestEnd

\litmusTestStart{MP\_cas+rel+rlx+na}{\tick}{NA + History + Viewfronts}
\begin{minipage}[t]{0.2\linewidth}
Possible outcomes:\\
\lstinline{  stuck}\\
\end{minipage}
\begin{minipage}[t]{0.4\linewidth}
\vspace{-.2cm}
  \begin{tabular}{l@{\ \ \ }l}
    \begin{minipage}[l]{4.3cm} \small
\begin{lstlisting}
        |$[$|f|$]_{rlx}$| := 1; |$[$|d|$]_{na}$| := 0;
\end{lstlisting}
\vspace{-.2cm}
\begin{tabular}{l||l||l}
\begin{lstlisting}
|$[$|d|$]_{na}$| := 5
|$[$|f|$]_{rel}$| := 0;
\end{lstlisting}
\hspace{.6cm}
&
\begin{lstlisting}
r1 = cas|$_{rlx,rlx}$|(f, 0, 1);
if r1 == 0
then |$[$|d|$]_{rlx}$| := 6
else 0
fi
\end{lstlisting}
\hspace{.6cm}
&
\begin{lstlisting}
r2 = cas|$_{rlx,rlx}$|(f, 0, 1);
if r2 == 0
then |$[$|d|$]_{rlx}$| := 7
else 0
fi
\end{lstlisting}
\end{tabular}
    \end{minipage}
&
  \end{tabular}
\end{minipage}
\litmusTestEnd

\subsection{Coherence of Read-Read (CoRR)}
\label{app:corr}

\litmusTestStart{CoRR\_rlx}{\tick}{History + Viewfronts}
\begin{minipage}[t]{0.3\linewidth}
Impossible outcomes:\\
\lstinline{  r1 = 1 |$\land$| r2 = 2 |$\land$| r3 = 2 |$\land$| r4 = 1}\\
\lstinline{  r1 = 2 |$\land$| r2 = 1 |$\land$| r3 = 1 |$\land$| r4 = 2}\\
\end{minipage}
\begin{minipage}[t]{0.3\linewidth}
\vspace{-.2cm}
  \begin{tabular}{l@{\ \ \ }l}
    \begin{minipage}[l]{4.3cm} \small
\begin{lstlisting}
                    |$[$|x|$]_{rlx}$| := 0;
\end{lstlisting}
\vspace{-.2cm}
\begin{tabular}{l||l||l||l}
\begin{lstlisting}
|$[$|x|$]_{rlx}$| := 1
\end{lstlisting}
\hspace{.6cm}
&
\begin{lstlisting}
|$[$|x|$]_{rlx}$| := 2
\end{lstlisting}
\hspace{.6cm}
&
\begin{lstlisting}
r1 = |$[$|x|$]_{rlx}$|;
r2 = |$[$|x|$]_{rlx}$|
\end{lstlisting}
\hspace{.6cm}
&
\begin{lstlisting}
r3 = |$[$|x|$]_{rlx}$|;
r4 = |$[$|x|$]_{rlx}$|
\end{lstlisting}
\end{tabular}
    \end{minipage}
&
  \end{tabular}
\end{minipage}
\litmusTestEnd

\litmusTestStart{CoRR\_rel+acq}{\tick}{History + Viewfronts}
\begin{minipage}[t]{0.3\linewidth}
Impossible outcomes:\\
\lstinline{  r1 = 1 |$\land$| r2 = 2 |$\land$| r3 = 2 |$\land$| r4 = 1}\\
\lstinline{  r1 = 2 |$\land$| r2 = 1 |$\land$| r3 = 1 |$\land$| r4 = 2}\\
\end{minipage}
\begin{minipage}[t]{0.3\linewidth}
\vspace{-.2cm}
  \begin{tabular}{l@{\ \ \ }l}
    \begin{minipage}[l]{4.3cm} \small
\begin{lstlisting}
                   |$[$|x|$]_{rel}$| := 0;
\end{lstlisting}
\vspace{-.2cm}
\begin{tabular}{l||l||l||l}
\begin{lstlisting}
|$[$|x|$]_{rel}$| := 1
\end{lstlisting}
\hspace{.6cm}
&
\begin{lstlisting}
|$[$|x|$]_{rel}$| := 2
\end{lstlisting}
\hspace{.6cm}
&
\begin{lstlisting}
r1 = |$[$|x|$]_{acq}$|;
r2 = |$[$|x|$]_{acq}$|
\end{lstlisting}
\hspace{.6cm}
&
\begin{lstlisting}
r3 = |$[$|x|$]_{acq}$|;
r4 = |$[$|x|$]_{acq}$|
\end{lstlisting}
\end{tabular}
    \end{minipage}
&
  \end{tabular}
\end{minipage}
\litmusTestEnd

\subsection{Independent Reads of Independent Writes (IRIW)}
\label{app:iriw}

\litmusTestStart{IRIW\_rlx}{\tick}{History + Viewfronts}
\begin{minipage}[t]{0.3\linewidth}
Possible outcomes:\\
\lstinline{  r1 = <0, 1>; r2 = <0, 1>;}\\
\lstinline{  r3 = <0, 1>; r4 = <0, 1>}\\
\end{minipage}
\begin{minipage}[t]{0.5\linewidth}
\vspace{-.2cm}
  \begin{tabular}{l@{\ \ \ }l}
    \begin{minipage}[l]{4.3cm} \small
\begin{lstlisting}
               |$[$|x|$]_{rlx}$| := 0; |$[$|y|$]_{rlx}$| := 0;
\end{lstlisting}
\vspace{-.2cm}
\begin{tabular}{l||l||l||l}
\begin{lstlisting}
|$[$|x|$]_{rlx}$| := 1
\end{lstlisting}
\hspace{.6cm}
&
\begin{lstlisting}
|$[$|y|$]_{rlx}$| := 1
\end{lstlisting}
\hspace{.6cm}
&
\begin{lstlisting}
r1 = |$[$|x|$]_{rlx}$|;
r2 = |$[$|y|$]_{rlx}$|
\end{lstlisting}
\hspace{.6cm}
&
\begin{lstlisting}
r3 = |$[$|y|$]_{rlx}$|;
r4 = |$[$|x|$]_{rlx}$|
\end{lstlisting}
\end{tabular}
    \end{minipage}
&
  \end{tabular}
\end{minipage}

Comment:
It is possible to get
\lstinline{r1 = 1; r2 = 0; r3 = 1; r4 = 0}
\litmusTestEnd

\litmusTestStart{IRIW\_rel+acq}{\tick}{History + Viewfronts}
\begin{minipage}[t]{0.3\linewidth}
Possible outcomes:\\
\lstinline{  r1 = <0, 1>; r2 = <0, 1>;}\\
\lstinline{  r3 = <0, 1>; r4 = <0, 1>}\\
\end{minipage}
\begin{minipage}[t]{0.5\linewidth}
\vspace{-.2cm}
  \begin{tabular}{l@{\ \ \ }l}
    \begin{minipage}[l]{4.3cm} \small
\begin{lstlisting}
               |$[$|x|$]_{rel}$| := 0; |$[$|y|$]_{rel}$| := 0;
\end{lstlisting}
\vspace{-.2cm}
\begin{tabular}{l||l||l||l}
\begin{lstlisting}
|$[$|x|$]_{rel}$| := 1
\end{lstlisting}
\hspace{.6cm}
&
\begin{lstlisting}
|$[$|y|$]_{rel}$| := 1
\end{lstlisting}
\hspace{.6cm}
&
\begin{lstlisting}
r1 = |$[$|x|$]_{acq}$|;
r2 = |$[$|y|$]_{acq}$|
\end{lstlisting}
\hspace{.6cm}
&
\begin{lstlisting}
r3 = |$[$|y|$]_{acq}$|;
r4 = |$[$|x|$]_{acq}$|
\end{lstlisting}
\end{tabular}
    \end{minipage}
&
  \end{tabular}
\end{minipage}

Comment:
It is possible to get
\lstinline{r1 = 1; r2 = 0; r3 = 1; r4 = 0}
\litmusTestEnd

\litmusTestStart{IRIW\_sc}{\tick}{SC + History + Viewfronts}
\begin{minipage}[t]{0.3\linewidth}
Forbidden outcomes:\\
\lstinline{  r1 = 1 |$\land$| r2 = 0 |$\land$| r3 = 1 |$\land$| r4 = 0}\\
\end{minipage}
\begin{minipage}[t]{0.5\linewidth}
\vspace{-.2cm}
  \begin{tabular}{l@{\ \ \ }l}
    \begin{minipage}[l]{4.3cm} \small
\begin{lstlisting}
               |$[$|x|$]_{sc}$| := 0; |$[$|y|$]_{sc}$| := 0;
\end{lstlisting}
\vspace{-.2cm}
\begin{tabular}{l||l||l||l}
\begin{lstlisting}
|$[$|x|$]_{sc}$| := 1
\end{lstlisting}
\hspace{.6cm}
&
\begin{lstlisting}
|$[$|y|$]_{sc}$| := 1
\end{lstlisting}
\hspace{.6cm}
&
\begin{lstlisting}
r1 = |$[$|x|$]_{sc}$|;
r2 = |$[$|y|$]_{sc}$|
\end{lstlisting}
\hspace{.6cm}
&
\begin{lstlisting}
r3 = |$[$|y|$]_{sc}$|;
r4 = |$[$|x|$]_{sc}$|
\end{lstlisting}
\end{tabular}
    \end{minipage}
&
  \end{tabular}
\end{minipage}
\litmusTestEnd

\subsection{Write-to-Read Causality (WRC)}
\label{app:wrc}

\litmusTestStart{WRC\_rel+acq}{\tick}{History + Viewfronts}
\begin{minipage}[t]{0.3\linewidth}
Forbidden outcomes:\\
\lstinline{  r2 = 1 |$\land$| r3 = 0}\\
\end{minipage}
\begin{minipage}[t]{0.3\linewidth}
\vspace{-.2cm}
  \begin{tabular}{l@{\ \ \ }l}
    \begin{minipage}[l]{4.3cm} \small
\begin{lstlisting}
        |$[$|x|$]_{rel}$| := 0; |$[$|y|$]_{rel}$| := 0;
\end{lstlisting}
\vspace{-.2cm}
\begin{tabular}{l||l||l}
\begin{lstlisting}
|$[$|x|$]_{rel}$| := 1
\end{lstlisting}
\hspace{.6cm}
&
\begin{lstlisting}
r1 = |$[$|x|$]_{acq}$|;
|$[$|y|$]_{rel}$| := r1
\end{lstlisting}
\hspace{.6cm}
&
\begin{lstlisting}
r2 = |$[$|y|$]_{acq}$|;
r3 = |$[$|x|$]_{acq}$|
\end{lstlisting}
\end{tabular}
    \end{minipage}
&
  \end{tabular}
\end{minipage}
\litmusTestEnd

\litmusTestStart{WRC\_rlx}{\tick}{History + Viewfronts}
\begin{minipage}[t]{0.3\linewidth}
Possible outcomes:\\
\lstinline{  r2 = 0 |$\land$| r3 = 0}\\
\lstinline{  r2 = 0 |$\land$| r3 = 1}\\
\lstinline{  r2 = 1 |$\land$| r3 = 0}\\
\lstinline{  r2 = 1 |$\land$| r3 = 1}\\
\end{minipage}
\begin{minipage}[t]{0.3\linewidth}
\vspace{-.2cm}
  \begin{tabular}{l@{\ \ \ }l}
    \begin{minipage}[l]{4.3cm} \small
\begin{lstlisting}
        |$[$|x|$]_{rlx}$| := 0; |$[$|y|$]_{rlx}$| := 0;
\end{lstlisting}
\vspace{-.2cm}
\begin{tabular}{l||l||l}
\begin{lstlisting}
|$[$|x|$]_{rlx}$| := 1
\end{lstlisting}
\hspace{.6cm}
&
\begin{lstlisting}
r1 = |$[$|x|$]_{rlx}$|;
|$[$|y|$]_{rlx}$| := r1
\end{lstlisting}
\hspace{.6cm}
&
\begin{lstlisting}
r2 = |$[$|y|$]_{rlx}$|;
r3 = |$[$|x|$]_{rlx}$|
\end{lstlisting}
\end{tabular}
    \end{minipage}
&
  \end{tabular}
\end{minipage}

\lstinline{r2 = 1; r3 = 0}
\litmusTestEnd

\litmusTestStart{WRC\_cas+rel}{\tick}{History + Viewfronts}
\begin{minipage}[t]{0.3\linewidth}
Impossible outcomes:\\
\lstinline{  r2 = 2 |$\land$| r3 = 0}\\
\end{minipage}
\begin{minipage}[t]{0.3\linewidth}
\vspace{-.2cm}
  \begin{tabular}{l@{\ \ \ }l}
    \begin{minipage}[l]{4.3cm} \small
\begin{lstlisting}
        |$[$|x|$]_{rel}$| := 0; |$[$|y|$]_{rel}$| := 0;
\end{lstlisting}
\vspace{-.2cm}
\begin{tabular}{l||l||l}
\begin{lstlisting}
|$[$|x|$]_{rel}$| := 1;
|$[$|y|$]_{rel}$| := 1
\end{lstlisting}
\hspace{.6cm}
&
\begin{lstlisting}
cas|$_{rel,acq}$|(y, 1, 2)
\end{lstlisting}
\hspace{.6cm}
&
\begin{lstlisting}
r1 = |$[$|y|$]_{rel}$|;
r2 = |$[$|x|$]_{rel}$|
\end{lstlisting}
\end{tabular}
    \end{minipage}
&
  \end{tabular}
\end{minipage}
\vspace{.2cm}
\hrule
\vspace{.2cm}

\litmusTestStart{WRC\_cas+rlx}{\tick}{History + Viewfronts}
\begin{minipage}[t]{0.3\linewidth}
Impossible outcomes:\\
\lstinline{  r2 = 2 |$\land$| r3 = 0}\\
\end{minipage}
\begin{minipage}[t]{0.3\linewidth}
\vspace{-.2cm}
  \begin{tabular}{l@{\ \ \ }l}
    \begin{minipage}[l]{4.3cm} \small
\begin{lstlisting}
        |$[$|x|$]_{rlx}$| := 0; |$[$|y|$]_{rlx}$| := 0;
\end{lstlisting}
\vspace{-.2cm}
\begin{tabular}{l||l||l}
\begin{lstlisting}
|$[$|x|$]_{rlx}$| := 1;
|$[$|y|$]_{rel}$| := 1
\end{lstlisting}
\hspace{.6cm}
&
\begin{lstlisting}
cas|$_{rlx,rlx}$|(y, 1, 2)
\end{lstlisting}
\hspace{.6cm}
&
\begin{lstlisting}
r1 = |$[$|y|$]_{rlx}$|;
r2 = |$[$|x|$]_{rlx}$|
\end{lstlisting}
\end{tabular}
    \end{minipage}
&
  \end{tabular}
\end{minipage}
\litmusTestEnd


\subsection{Out-of-Thin-Air reads}
\label{app:ota}

In our semantics it is not possible to get out-of-thin-air results,
unlike the C11 standard. But such reads are considered to be an
undesirable behavior by most of the standard's
clients~\cite{Batty-al:ESOP15}.

\litmusTestStart{OTA\_lb}{\fail}{Postponed reads + History + Viewfronts}
\begin{minipage}[t]{0.3\linewidth}
Possible outcomes:\\
\lstinline{  r1 = 0 |$\land$| r2 = 0}\\
\end{minipage}
\begin{minipage}[t]{0.3\linewidth}
\vspace{-.2cm}
  \begin{tabular}{l@{\ \ \ }l}
    \begin{minipage}[l]{4.3cm} \small
\begin{lstlisting}
  |$[$|x|$]_{rlx}$| := 0; |$[$|y|$]_{rlx}$| := 0;
\end{lstlisting}
\vspace{-.2cm}
\begin{tabular}{l||l}
\begin{lstlisting}
r1 = |$[$|y|$]_{rlx}$|;
|$[$|x|$]_{rlx}$| := r1
\end{lstlisting}
\hspace{.6cm}
&
\begin{lstlisting}
r2 = |$[$|x|$]_{rlx}$|;
|$[$|y|$]_{rlx}$| := r2
\end{lstlisting}
\end{tabular}
    \end{minipage}
&
  \end{tabular}
\end{minipage}

Comment: According to the C11 standard \cite{C:11,CPP:11},
\lstinline{r1} and \lstinline{r2} can get arbitrary values.
\litmusTestEnd

\litmusTestStart{OTA\_if}{\fail}{Postponed reads + History + Viewfronts}
\begin{minipage}[t]{0.4\linewidth}
Possible outcomes:\\
\lstinline{  r1 = 0 |$\land$| r2 = 0}\\
\end{minipage}
\begin{minipage}[t]{0.4\linewidth}
\vspace{-.2cm}
  \begin{tabular}{l@{\ \ \ }l}
    \begin{minipage}[l]{4.3cm} \small
\begin{lstlisting}
  |$[$|x|$]_{rlx}$| := 0; |$[$|y|$]_{rlx}$| := 0;
\end{lstlisting}
\vspace{-.2cm}
\begin{tabular}{l||l}
\begin{lstlisting}
r1 = |$[$|y|$]_{rlx}$|;
if r1
then |$[$|x|$]_{rlx}$| := 1
else r1 = 0 
fi
\end{lstlisting}
\hspace{.6cm}
&
\begin{lstlisting}
r2 = |$[$|x|$]_{rlx}$|;
if r2
then |$[$|y|$]_{rlx}$| := 1
else r2 = 0
fi
\end{lstlisting}
\end{tabular}
    \end{minipage}
&
  \end{tabular}
\end{minipage}

Comment: According to the C11 standard \cite{C:11,CPP:11},
\lstinline{r1} and \lstinline{r2} can be 1s at the end of execution.
\litmusTestEnd

\subsection{Write Reorder (WR), or 2+2W from \cite{Lahav-al:POPL16}}
\label{app:wr}

\litmusTestStart{WR\_rlx}{\tick}{History + Viewfronts + Operational Buffers}
\begin{minipage}[t]{0.3\linewidth}
Possible outcomes:\\
\lstinline{  r1 = 1 |$\land$| r2 = 2}\\
\lstinline{  r1 = 2 |$\land$| r2 = 1}\\
\lstinline{  r1 = 2 |$\land$| r2 = 2}\\
\end{minipage}
\begin{minipage}[t]{0.3\linewidth}
\vspace{-.2cm}
  \begin{tabular}{l@{\ \ \ }l}
    \begin{minipage}[l]{4.3cm} \small
\begin{lstlisting}
  |$[$|x|$]_{rlx}$| := 0; |$[$|y|$]_{rlx}$| := 0;
\end{lstlisting}
\vspace{-.2cm}
\begin{tabular}{l||l}
\begin{lstlisting}
|$[$|x|$]_{rlx}$| := 1;
|$[$|y|$]_{rlx}$| := 2
\end{lstlisting}
\hspace{.6cm}
&
\begin{lstlisting}
|$[$|y|$]_{rlx}$| := 1;
|$[$|x|$]_{rlx}$| := 2
\end{lstlisting}
\end{tabular}
\begin{lstlisting}
  r1 = |$[$|x|$]_{rlx}$|; r2 = |$[$|y|$]_{rlx}$|
\end{lstlisting}
    \end{minipage}
&
  \end{tabular}
\end{minipage}
\litmusTestEnd

\litmusTestStart{WR\_rlx+rel}{\tick}{History + Viewfronts + Operational Buffers}
\begin{minipage}[t]{0.3\linewidth}
Possible outcomes:\\
\lstinline{  r1 = 1 |$\land$| r2 = 2}\\
\lstinline{  r1 = 2 |$\land$| r2 = 1}\\
\lstinline{  r1 = 2 |$\land$| r2 = 2}\\
\end{minipage}
\begin{minipage}[t]{0.3\linewidth}
\vspace{-.2cm}
  \begin{tabular}{l@{\ \ \ }l}
    \begin{minipage}[l]{4.3cm} \small
\begin{lstlisting}
  |$[$|x|$]_{rlx}$| := 0; |$[$|y|$]_{rlx}$| := 0;
\end{lstlisting}
\vspace{-.2cm}
\begin{tabular}{l||l}
\begin{lstlisting}
|$[$|x|$]_{rlx}$| := 1;
|$[$|y|$]_{rel}$| := 2
\end{lstlisting}
\hspace{.6cm}
&
\begin{lstlisting}
|$[$|y|$]_{rlx}$| := 1;
|$[$|x|$]_{rel}$| := 2
\end{lstlisting}
\end{tabular}
\begin{lstlisting}
  r1 = |$[$|x|$]_{rlx}$|; r2 = |$[$|y|$]_{rlx}$|
\end{lstlisting}
    \end{minipage}
&
  \end{tabular}
\end{minipage}
\litmusTestEnd

\litmusTestStart{WR\_rel}{\tick}{History + Viewfronts + Operational Buffers}
\begin{minipage}[t]{0.3\linewidth}
Possible outcomes:\\
\lstinline{  r1 = 1 |$\land$| r2 = 2}\\
\lstinline{  r1 = 2 |$\land$| r2 = 1}\\
\lstinline{  r1 = 2 |$\land$| r2 = 2}\\
\end{minipage}
\begin{minipage}[t]{0.3\linewidth}
\vspace{-.2cm}
  \begin{tabular}{l@{\ \ \ }l}
    \begin{minipage}[l]{4.3cm} \small
\begin{lstlisting}
  |$[$|x|$]_{rel}$| := 0; |$[$|y|$]_{rel}$| := 0;
\end{lstlisting}
\vspace{-.2cm}
\begin{tabular}{l||l}
\begin{lstlisting}
|$[$|x|$]_{rel}$| := 1;
|$[$|y|$]_{rel}$| := 2
\end{lstlisting}
\hspace{.6cm}
&
\begin{lstlisting}
|$[$|y|$]_{rel}$| := 1;
|$[$|x|$]_{rel}$| := 2
\end{lstlisting}
\end{tabular}
\begin{lstlisting}
  r1 = |$[$|x|$]_{acq}$|; r2 = |$[$|y|$]_{acq}$|
\end{lstlisting}
    \end{minipage}
&
  \end{tabular}
\end{minipage}

\vspace{10pt}


\litmusTestEnd




\subsection{Speculative Execution}
\label{app:se}

\litmusTestStart{SE\_simple}{\tick}{}
\begin{minipage}[t]{0.3\linewidth}
Possible outcomes:\\
\lstinline{  r0 = 0}\\
\lstinline{  r0 = 1}\\
\end{minipage}
\begin{minipage}[t]{0.4\linewidth}
\vspace{-.2cm}
  \begin{tabular}{l@{\ \ \ }l}
    \begin{minipage}[l]{4.3cm} \small
\begin{lstlisting}
|$[$|x|$]_{rlx}$| := 0; |$[$|y|$]_{rlx}$| := 0; |$[$|z|$]_{rlx}$| := 0;
\end{lstlisting}
\vspace{-.2cm}
\begin{tabular}{l||l}
\begin{lstlisting}
r1 = |$[$|x|$]_{rlx}$|;
if r1
then |$[$|z|$]_{rlx}$| := 1;
     |$[$|y|$]_{rlx}$| := 1
else |$[$|y|$]_{rlx}$| := 1
fi
\end{lstlisting}
\hspace{.6cm}
&
\begin{lstlisting}
r2 = |$[$|y|$]_{rlx}$|;
if r2
then |$[$|x|$]_{rlx}$| := 1
else 0 
fi
\end{lstlisting}
\end{tabular}
    \end{minipage}
&
  \end{tabular}

\begin{lstlisting}
                  r0 = |$[$|z|$]_{rlx}$|
\end{lstlisting}
\end{minipage}

\litmusTestEnd

\litmusTestStart{SE\_prop}{\tick}{}
\begin{minipage}[t]{0.3\linewidth}
Possible outcomes:\\
\lstinline{  r0 = 0}\\
\lstinline{  r0 = 1}\\
\end{minipage}
\begin{minipage}[t]{0.4\linewidth}
\vspace{-.2cm}
  \begin{tabular}{l@{\ \ \ }l}
    \begin{minipage}[l]{4.3cm} \small
\begin{lstlisting}
  |$[$|x|$]_{rlx}$| := 0; |$[$|y|$]_{rlx}$| := 0; |$[$|z|$]_{rlx}$| := 0;
\end{lstlisting}
\vspace{-.2cm}
\begin{tabular}{l||l}
\begin{lstlisting}
r1 = |$[$|x|$]_{rlx}$|;
if r1
then |$[$|z|$]_{rlx}$| := 1;
     r1 = |$[$|z|$]_{rlx}$|;
     |$[$|y|$]_{rlx}$| := r1
else |$[$|y|$]_{rlx}$| := 1
fi
\end{lstlisting}
\hspace{.6cm}
&
\begin{lstlisting}
r2 = |$[$|y|$]_{rlx}$|;
if r2
then |$[$|x|$]_{rlx}$| := 1
else 0 
fi
\end{lstlisting}
\end{tabular}
    \end{minipage}
&
  \end{tabular}

\begin{lstlisting}
                  r0 = |$[$|z|$]_{rlx}$|
\end{lstlisting}
\end{minipage}

\litmusTestEnd

\litmusTestStart{SE\_nested}{\tick}{}
\begin{minipage}[t]{0.3\linewidth}
Possible outcomes:\\
\lstinline{  r0 = 0}\\
\lstinline{  r0 = 1}\\
\end{minipage}
\begin{minipage}[t]{0.4\linewidth}
\vspace{-.2cm}
  \begin{tabular}{l@{\ \ \ }l}
    \begin{minipage}[l]{4.3cm} \small
\begin{lstlisting}
|$[$|x|$]_{rlx}$| := 0; |$[$|y|$]_{rlx}$| := 0; |$[$|z|$]_{rlx}$| := 0; |$[$|f|$]_{rlx}$| := 0;
\end{lstlisting}
\vspace{-.2cm}
\begin{tabular}{l||l}
\begin{lstlisting}
r1 = |$[$|x|$]_{rlx}$|;
if r1
then r2 = |$[$|f|$]_{rlx}$|;
     if r2
     then |$[$|z|$]_{rlx}$| := 1;
          |$[$|y|$]_{rlx}$| := 1
     else |$[$|y|$]_{rlx}$| := 1
     fi
else |$[$|y|$]_{rlx}$| := 1
fi
\end{lstlisting}
\hspace{.6cm}
&
\begin{lstlisting}
r3 = |$[$|y|$]_{rlx}$|;
if r3
then |$[$|f|$]_{rlx}$| := 1;
     |$[$|x|$]_{rlx}$| := 1
else 0 
fi
\end{lstlisting}
\end{tabular}
    \end{minipage}
&
  \end{tabular}

\begin{lstlisting}
                  r0 = |$[$|z|$]_{rlx}$|
\end{lstlisting}
\end{minipage}

\litmusTestEnd

\subsection{Locks}
\label{app:locks}

\begin{minipage}[t]{0.4\linewidth}
\textbf{Dekker's lock}\\\\
Possible outcomes:\\
\lstinline{  stuck}\\
Requires: RA + na\\
Fully Supported: $\tick$\\
\end{minipage}
\begin{minipage}[t]{0.4\linewidth}
\vspace{-.2cm}
  \begin{tabular}{l@{\ \ \ }l}
    \begin{minipage}[l]{4.3cm} \small
\begin{lstlisting}
  |$[$|x|$]_{rel}$| := 0; |$[$|y|$]_{rel}$| := 0; |$[$|d|$]_{na}$| := 0;
\end{lstlisting}
\vspace{-.2cm}
\begin{tabular}{l||l}
\begin{lstlisting}
|$[$|x|$]_{rel}$| := 1;
r1 = |$[$|y|$]_{acq}$|
if r1 == 0
then |$[$|d|$]_{na}$| := 5
else 0 
fi
\end{lstlisting}
\hspace{.6cm}
&
\begin{lstlisting}
|$[$|y|$]_{rel}$| := 1;
r2 = |$[$|x|$]_{acq}$|
if r2 == 0
then |$[$|d|$]_{na}$| := 6
else 0 
fi
\end{lstlisting}
\end{tabular}
    \end{minipage}
&
  \end{tabular}
\end{minipage}
\litmusTestEnd

\begin{minipage}[t]{0.4\linewidth}
\textbf{Cohen's lock}\\\\
Impossible outcomes (according to \cite{Turon-al:OOPSLA14}):\\
\lstinline{  stuck}\\
Requires: RA + na\\
Fully Supported: $\tick$\\
\end{minipage}
\begin{minipage}[t]{0.4\linewidth}
\vspace{-.2cm}
  \begin{tabular}{l@{\ \ \ }l}
    \begin{minipage}[l]{4.3cm} \small
\begin{lstlisting}
  |$[$|x|$]_{rel}$| := 0; |$[$|y|$]_{rel}$| := 0; |$[$|d|$]_{na}$| := 0;
\end{lstlisting}
\vspace{-.2cm}
\begin{tabular}{l||l}
\begin{lstlisting}
|$[$|x|$]_{rel}$| := choice 1 2;
repeat |$[$|y|$]_{acq}$| end;
r1 = |$[$|x|$]_{acq}$|
r2 = |$[$|y|$]_{acq}$|
if r1 == r2
then |$[$|d|$]_{na}$| := 5
else 0 
fi
\end{lstlisting}
\hspace{.6cm}
&
\begin{lstlisting}
|$[$|y|$]_{rel}$| := choice 1 2;
repeat |$[$|x|$]_{acq}$| end;
r3 = |$[$|x|$]_{acq}$|
r4 = |$[$|y|$]_{acq}$|
if r3 != r4
then |$[$|d|$]_{na}$| := 6
else 0 
fi
\end{lstlisting}
\end{tabular}
    \end{minipage}
&
  \end{tabular}
\end{minipage}
\litmusTestEnd

\section{Additional Semantic Rules}
\label{sec:appSemanticsRules}

\begin{figure}[h]
{\small
{
\[\begin{array}{rcl}
\EvalContext   & ::= & \hole
                         \mid \Bind{\vName}{\EvalContext}{\AST} \\
               &     & \mid \Par{\EvalContext}{\AST}
                       \mid \Par{\AST}{\EvalContext} \\  
\EvalEUContext & ::= & \hole \mid (\EvalEUContext) \mid
                         \EvalEUContext~\op~\Expr \mid
                         \Expr~\op~\EvalEUContext \\
               &     &   \mid \Pair{\mval}{\EvalEUContext}
                         \mid \Pair{\EvalEUContext}{\mval} \\ 
               &     &   \mid \First{\EvalEUContext}
                         \mid \Second{\EvalEUContext} \\ 
               &     &   \mid \Choice{~\EvalEUContext}{\Expr} \mid
                         \Choice{~\Expr}{\EvalEUContext} \\
               &     &   \mid \Bind{\vName}{\EvalEUContext}{\AST} \\
               &     &   \mid \IfThenElse{\EvalEUContext}{\AST_{1}}{\AST_{2}} \\
               &     &   \mid \Write{\WM}{\locVar}{\EvalEUContext} \\
               &     &   \mid \Cas{SM}{FM}{\locVar}{\EvalEUContext}{\mval}
                         \mid \Cas{SM}{FM}{\locVar}{\mval}{\EvalEUContext} \\
\end{array}\]
}}
\caption{Syntax of evaluation contexts.}
\end{figure}

\begin{figure*}[h]\small
\centering
{
\small{
\[\begin{array}{c}
\prooftree
----------------------------------{Subst}
\angled{\EvalContext[\Bind{\vName}{\Ret{\mvalSubst}}{\AST}], \auxX} ==> 
\angled{\EvalContext[\AST\subst{\vName}{\mvalSubst}], \auxX}
~~~~~
n~\neq~0
----------------------------------{If-True}
\arrayBlock{
  \angled{\EvalContext[\IfThenElse{n}{\AST_{1}}{\AST_{2}}], \auxX} ==>
  \angled{\EvalContext[\AST_{1}]], \auxX}
}
~~~~~
----------------------------------{If-False}
\arrayBlock{
  \angled{\EvalContext[\IfThenElse{0}{\AST_{1}}{\AST_{2}}], \auxX} ==>
  \angled{\EvalContext[\AST_{2}]], \auxX}
}
~~~~~
\vName \text{ -- fresh invariable}
----------------------------------{Repeat-Unroll}
\arrayBlock{
  \angled{\EvalContext[\Repeat{\AST}], \auxX} ==>
  \angled{\EvalContext[\Bind{\vName}{\AST}
                 {\IfThenElse{\vName}{\Ret{\vName}}
                              {\Repeat{\AST}}}],
            \auxX}
}
~~~~~
\auxX' = \spawn{\EvalContext}{\auxX}
----------------------------------{Spawn}
\angled{\EvalContext[\Spw{\AST_{1}}{\AST_{2}}], \auxX} ==>
\angled{\EvalContext[\Par{\AST_{1}}{\AST_{2}}], \auxX'}
~~~~
\auxX' = \joinP{\EvalContext}{\auxX}
----------------------------------{Join}
\angled{\EvalContext[\Par{\Ret{\mvalSubst_{1}}}{\Ret{\mvalSubst_{2}}}], \auxX} ==>
\angled{\EvalContext[\Ret{\Pair{\mvalSubst_{1}}{\mvalSubst_{2}}}], \auxX'}
~~~~~
---------------------------------{Choice-Fst}
\arrayBlock{
  \angled{\EvalContext[\EvalEUContext[\Choice{~\Expr_{1}}{\Expr_{2}}]], \auxX} ==> 
  \angled{\EvalContext[\EvalEUContext[\Expr_{1}]], \auxX}
}
~~~~
----------------------------------{Choice-Snd}
\arrayBlock{
  \angled{\EvalContext[\EvalEUContext[\Choice{~\Expr_{1}}{\Expr_{2}}]], \auxX} ==> 
  \angled{\EvalContext[\EvalEUContext[\Expr_{2}]], \auxX}
}
\endprooftree
\end{array}\]
}}
\caption{The core rules of the semantics.}
\label{fig:basic-sem}
\end{figure*}

\begin{figure}[t]\small
\centering
\[\arrayBlock{
\prooftree
\arrayBlock{
  \auxX = \angled{\stEta, \stPsiRead, ..., \stNA} \quad ... \\
  \stSigmaRead(\loc) == \LastTau{\stEta}{\loc} \quad
  \stSigma = \stSigmaRead[\loc \mapsto \stTau] \\[2pt]
  \auxX' = \angled{\stEta[(\loc, \stTau) \mapsto (\mvalSubst, \stSigmaEmpty)],
                   \stPsiRead[\stpath \mapsto \stSigma], ...,
                   \stNA[\loc \mapsto \stTau]}
} 
----------------------------------{WriteNA}
\angled{\EvalContext[\Write{\naM}{\loc}{\mvalSubst}], \auxX} ==> 
\angled{\EvalContext[\Ret{\mvalSubst}], \auxX'}
~~~~~
\arrayBlock{
  \auxX = \angled{\stEta, \stPsiRead, ..., \stNA} \quad ... \\[2pt]
  \stTau = \LastTau{\stEta}{\loc} \quad
  \stTau == \stSigmaRead(\loc)  \quad
  \stEta(\loc, \stTau) = (\mvalSubst, \stSigma)\\[2pt]
}
----------------------------------{ReadNA}
\angled{\EvalContext[\Read{\naM}{\loc}], \auxX} ==> 
\angled{\EvalContext[\Ret{\mvalSubst}], \auxX}
~~~~~
\arrayBlock{
  \auxX = \angled{\stEta, \stPsiRead, ..., \stNA}
  \quad ... \\  
  \stSigmaRead(\loc) \neq \LastTau{\stEta}{\loc}\\[2pt]
}
----------------------------------{WriteNA-stuck1}
\angled{\EvalContext[\Write{\naM}{\loc}{\mvalSubst}], \auxX} ==> 
\angled{\Stuck, \auxX}
~~~~~
\arrayBlock{
  \auxX = \angled{\stEta, \stPsiRead, ..., \stNA}
  \quad ... \\  
  \stSigmaRead(\loc) \neq \LastTau{\stEta}{\loc}\\[2pt]
}
----------------------------------{ReadNA-stuck1}
\angled{\EvalContext[\Read{\naM}{\loc}], \auxX} ==> 
\angled{\Stuck, \auxX}
~~~~~
\arrayBlock{
  \auxX = \angled{\stEta, \stPsiRead, ..., \stNA}
  \quad ... \quad
  \stSigmaRead(\loc)~<~\stNA(\loc)
}
----------------------------------{WriteNA-stuck2}
\angled{\EvalContext[\Write{\RM}{\loc}{\mvalSubst}], \auxX} ==> 
\angled{\Stuck, \auxX}
~~~~~
\arrayBlock{
  \auxX = \angled{\stEta, \stPsiRead, ..., \stNA}
  \quad ... \quad
  \stSigmaRead(\loc)~<~\stNA(\loc)
}
----------------------------------{ReadNA-stuck2}
\angled{\EvalContext[\Read{\RM}{\loc}], \auxX} ==> 
\angled{\Stuck, \auxX}
\endprooftree
}\]
\caption{Reduction rules for non-atomics.}
\label{fig:na-semFull}
\end{figure}

\begin{figure}
\begin{center} {\small{
\[\arrayBlock{
\prooftree
\arrayBlock{
  \auxX = \angled{..., \stPhi, \stGamma} \quad ... \\
  \vName \text{ --- fresh symbolic variable} \quad
  \stAlpha = \stPhi(\stpath) \\
  \auxX' = \angled{...,
\stPhi[\stpath \mapsto \AppendAlpha(\stAlpha, \EvalSpecContext, read\angled{\vName, \locVar, \RM})],
  \stGamma} \\
} 
----------------------------------{Read-Postpone}
\angled{\EvalContext[\EvalSpecContext[\Read{\RM}{\locVar}]], \auxX} ==> 
\angled{\EvalContext[\EvalSpecContext[\Ret{\vName}]], \auxX'}
\endprooftree
}\] }}
\end{center}

\begin{center} {\small{
\[\arrayBlock{
\prooftree
\arrayBlock{
  \auxX = \angled{..., \stPhi, \stGamma} \quad ... \\
  \vName \text{ is fresh symbolic variable} \quad
  \stAlpha = \stPhi(\stpath) \\
  \auxX' = \angled{...,
\stPhi[\stpath \mapsto \AppendAlpha(\stAlpha, \EvalSpecContext, write\angled{\vName, \locVar, \WM, \Expr})],
  \stGamma} \\
} 
----------------------------------{Write-Postpone}
\angled{\EvalContext[\EvalSpecContext[\Write{\WM}{\locVar}{\Expr}]], \auxX} ==> 
\angled{\EvalContext[\EvalSpecContext[\Ret{\vName}]], \auxX'}
\endprooftree
}\] }}
\end{center}

\begin{center} {\small{
\[\arrayBlock{
\prooftree
\arrayBlock{
  \auxX = \angled{..., \stPhi, \stGamma} \quad ... \quad \Expr~\text{can't be substituted immediately}\\
  \vName \text{ --- fresh symbolic variable} \quad
  \stAlpha = \stPhi(\stpath) \\
  \auxX' = \angled{...,
\stPhi[\stpath \mapsto \AppendAlpha(\stAlpha, \EvalSpecContext, let\angled{\vName,\Expr})],
   \stGamma} \\
} 
----------------------------------{Let-Postpone}
\angled{\EvalContext[\EvalSpecContext[\Bind{\vName'}{\Expr}{\AST}]], \auxX} ==> 
\angled{\EvalContext[\EvalSpecContext\subst{\vName'}{\vName}], \auxX'}
\endprooftree
}\] }}
\end{center}

\begin{center} {\small{
\[\arrayBlock{
\prooftree
\arrayBlock{
  \auxX = \angled{\stEta, \stPsiRead, ..., \stPhi, \stGamma} \quad ... \quad
  read\angled{\vName, \loc, \RM}\;\text{is inside}\;\stPhi(\stpath)\\
  \text{and there is no conflicting operation before} \\
  \stPhi' = \text{remove}(\stPhi, \stpath, read\angled{\vName, \loc, \RM}) \\
  \stGamma' = \stGamma \setminus \{\vName\} \quad\quad
  \auxX' = \angled{...,
\stPhi'[\mvalSubst/\vName], \stGamma'} \\
} 
----------------------------------{Read-Resolve}
\angled{\AST, \auxX} ==> 
\angled{\AST[\mvalSubst/\vName], \auxX'}
\endprooftree
}\]
}}
\end{center}

\caption{Rules for work with postponing of operations.}
\end{figure}

\newpage
\section{RCU testing}
{
\setlength{\belowcaptionskip}{-10pt} 
\begin{figure}[h!]
\centering
{\small{
\begin{tabular}{|@{\ }r@{\ }|| c | c | c | c | c || c |}
  \hline
  \textbf{\#}
  & \texttt{r11} & \texttt{r12}
  & \texttt{r21} & \texttt{r22} & \textbf{Stuck} & \textbf{Runtime} (sec)\\
  \hline
  \hline

1  & 0   & 111  & 111 & 1101 & \tick & 25.2 \\
2  & 0   & 1    & 111 & 1101 &       & 21.4 \\
3  & 0   & 0    & 0   & 0    &       & 12.9 \\
4  & 0   & 1101 & 11  & 11   &       & 25.4 \\
5  & 0   & 1101 & 0   & 0    &       & 16.3 \\
6  & 0   & 11   & 0   & 0    & \tick & 17.5 \\
7  & 0   & 0    & 0   & 1101 &       & 22.1 \\
8  & 1   & 1    & 0   & 0    &       & 16.5 \\
9  & 0   & 1101 & 1   & 1101 &       & 19.2 \\
10 & 0   & 1101 & 111 & 1101 &       & 26.4 \\
11 & 1   & 1101 & 1   & 1    &       & 23.8 \\
12 & 0   & 0    & 111 & 1101 & \tick & 20.5 \\
13 & 11  & 1101 & 0   & 0    &       & 21.5 \\
14 & 0   & 111  & 0   & 111  & \tick & 21.5 \\
15 & 0   & 0    & 11  & 1101 &       & 22.1 \\
16 & 0   & 0    & 0   & 0    &       & 16.0 \\
17 & 0   & 0    & 0   & 1101 &       & 18.1 \\
18 & 1   & 1    & 0   & 0    &       & 22.2 \\
19 & 1   & 1101 & 1   & 1    &       & 26.0 \\
20 & 1   & 1101 & 0   & 0    &       & 20.1 \\

  \hline
  \end{tabular}
}}
\caption{Test results and runtimes for modified RCU.}
\label{fig:tblRun}
\end{figure}
}

}{}

\end{document}